\documentclass[aps,pre,reprint,onecolumn,superscriptaddress,showpacs,notitlepage,a4]{revtex4-2}
\usepackage{amssymb,amsmath}
\usepackage{graphicx}
\usepackage{comment}
\usepackage{mathtools}
\usepackage[export]{adjustbox}
\usepackage{overpic}
\usepackage{bm}
\usepackage{amsthm}
\usepackage{nameref,hyperref}
\usepackage[right]{lineno}
\usepackage{longtable}
\usepackage[table]{xcolor}
\usepackage[normalem]{ulem}
\usepackage{array}
\usepackage{float}
\usepackage{dcolumn}
\usepackage{bigints}
\usepackage{physics}

\begin{document}

\title{Forecasting emergency department visits in the reference hospital of the Balearic Islands: the role of tourist and weather data}

\author{Paride Crisafulli}
\affiliation{Instituto de F\'isica Interdisciplinar y Sistemas Complejos IFISC (CSIC-UIB), Campus UIB, 07122 Palma de Mallorca, Spain.}
\email{paride@ifisc.uib-csic.es}
\author{Angel del R\'{i}o Mangada}
\affiliation{Servicio de Cirugía Ortopédica y Traumatología, Hospital Son Llàtzer, Crta. Manacor km 4.3, 07198 Palma, Mallorca, Spain}
\author{Juan Jos\'{e} Segura Sampedro}
\affiliation{ General \& Digestive Surgery Service, Hospital Universitario La Paz, IdiPAZ, 28046 Madrid, Spain}
\author{Claudio R. Mirasso}
\affiliation{Instituto de F\'isica Interdisciplinar y Sistemas Complejos IFISC (CSIC-UIB), Campus UIB, 07122 Palma de Mallorca, Spain.}
\email{claudio@ifisc.uib-csic.es}
\author{Ra\'{u}l Toral}
\affiliation{Instituto de F\'isica Interdisciplinar y Sistemas Complejos IFISC (CSIC-UIB), Campus UIB, 07122 Palma de Mallorca, Spain.}
\email{raul@ifisc.uib-csic.es}
\author{Tobias Galla}
\affiliation{Instituto de F\'isica Interdisciplinar y Sistemas Complejos IFISC (CSIC-UIB), Campus UIB, 07122 Palma de Mallorca, Spain.}
\email{tobias.galla@ifisc.uib-csic.es}


\begin{abstract}
Accurate forecasting of patient arrivals at emergency departments (EDs) is vital for efficient resource allocation and high-quality patient care. In this study we investigate the relevance of exogenous variables, namely tourism, weather, calendar and demographic variables, in forecasting ED visits in the reference hospital in Palma de Mallorca, a city with significant seasonal population fluctuations due to tourism. Using a machine learning approach, we develop a model that predicts ED visits based solely on these exogenous variables. We test different machine learning algorithms (random forests, support vector machines, and feedforward neural networks) with different combinations of input variables and compare their symmetric mean average percentage errors (SMAPEs).  Our findings reveal that calendar information, resident, and tourist population data are statistically significant for the accuracy of the predictions, while the addition of weather data does not provide any further improvement. Comparison of non-time-series with time-series prediction models reveals that the latter provide better accuracy for short prediction horizons (e.g. shorter than a week). Furthermore, time-series models become less or equally accurate to models relying only on exogenous variables for long prediction horizons (e.g. fortnight or month). Our study highlights the importance of carefully selecting predictive variables to ensure short- and long-term, robust and reliable forecasts. This demonstrates that, despite their lower complexity, non-time-series models with well-chosen input variables can be as effective as time-series models when predicting for long time horizons.
\end{abstract}
\maketitle

\section{Introduction}
 
Accurate forecasting of the volume of patient arrivals in hospital emergency departments (EDs) is of vital importance in ensuring effective resource allocation and timely patient care. Despite the considerable attention this issue has received worldwide through numerous studies \cite{1, 2, 3, 4, 5, 6, 7, 8, 9, 10, 11, 12, 13, 14, 15, 16, 17, 18, 19, 20, 21, 22, 23, 24, 25}, it remains evident that there is no universally superior prediction algorithm \cite{1, 2}. 

All predictive models are trained using the existing ED data at a given location. Once the training has been carried out, the prediction of the number of arrivals on a particular date may or may not rely on patient numbers previous to the prediction date. We refer to models that, after training, continue to use past data to make predictions as “time-series models”. For example, to make a prediction for a particular date, a fully trained time-series model requires actual data for patient arrivals on the days before the target date. Time-series models include auto-regressive integrated moving average (ARIMA) models and variants \cite{1, 2, 3, 4, 5, 6, 7, 8, 9, 10, 11, 12, 13, 14, 15, 16, 17, 18}, seasonal exponential smoothing \cite{1, 2, 3, 4, 5, 6, 7, 8, 9, 10, 11, 12, 13, 14, 15, 16, 17, 18}, Holt-Winters methods \cite{1, 2, 3, 4, 5}, and recurrent or convolutional neural networks \cite{1, 2, 5, 11, 12, 16, 19}. Time-series models typically perform well for short-term forecasting but degrade over longer horizons; while exogenous variables can be incorporated \cite{1}, this does not necessarily lead to improved predictive performance.

Other, simpler, approaches use existing arrival data only during training, and subsequently rely solely on exogenous variables to make predictions. The most commonly used exogenous variables are calendar information (day of the week, month, holidays) \cite{1, 2, 4, 6, 8, 12, 13, 14, 15, 18, 20, 21, 22, 23}, weather data \cite{1, 2, 8, 10, 12, 14, 15, 18, 20, 21, 22}, online searches for relevant keywords such as “flu” \cite{15, 24}, and variables specific to a particular location (such as the timing of the Oktoberfest \cite{25}). We refer to models that, after training, do not use past ED data for future predictions as “non-time-series models". We note that these models might still require temporal information (e.g., the target date) as input, but the crucial difference to time-series models is that no patient arrival numbers, for example on days immediately preceding the target date, are required once the training of the model is complete. In that sense, these models do not extrapolate an existing time series. Algorithms used for such models include feedforward neural networks \cite{1, 2}, random forests \cite{1, 15}, and Poisson regression \cite{20}. Although generally less accurate than time-series models for short prediction horizons, non-time-series approaches maintain constant accuracy across different prediction horizons (as far as the exogenous variables allow). This makes these models more suitable for long-term, i.e. fortnightly or monthly, resource allocation than time-series models.

In this paper, we use a non-time-series machine-learning approach to predict ED visits at the major hospital on the island of Mallorca (Balearic Islands, Spain), located in its capital, Palma. The island attracts a considerable number of visitors (tourists as well as temporary workers), and during the summer months, the population approximately doubles. Therefore, it seems prudent to include the floating population (including tourists) as an exogenous variable, alongside calendar variables (including local holidays), resident population, and weather data. In order to characterize the impact of the different variables on prediction accuracy, we run the models with different combinations of input variables.

We find that, despite its simplicity, non-time-series models can yield symmetric mean average percentage errors (SMAPE) that are comparable to those obtained from more complex models in previous studies \cite{1, 2, 3, 4, 5, 6, 7, 8, 9, 10, 11, 12, 13, 14, 15, 16, 17, 18, 19, 20, 21, 22, 13, 24, 25}. To assess whether the differences in prediction errors between time-series models and non-time-series models are statistically meaningful, we conducted a series of Diebold–Mariano tests. The results show that, although models incorporating tourist population as an exogenous variable yield prediction errors that are statistically smaller than those of models without tourist information, the resulting change in the predicted number of incoming patients (NIP) remains small, below two patients for a typical hospital shift. By contrast, our analysis indicates that weather variables can be omitted, with calendar and population data providing sufficient exogenous information.

\

\section{Materials and methods}
\subsection{Data preprocessing and behavior}

We used a dataset comprising all ED visits at Hospital Universitari Son Espases (HUSE) in Palma, Mallorca, from December 26, 2015, to December 31, 2022. HUSE is the largest public hospital in the Balearic Islands. The study falls under the category of ``human subjects research'' in the PLOS classification. The Ethics Committee of the Balearic Islands (CEIm-IB) is the relevant review board, and the committee has confirmed that this study requires neither approval by the Ethics Committee nor patient consent, since the data used in this study are of an aggregated, anonymous and non-sensitive nature. The official document by the CEIm-IB (text in Spanish and English language) can be consulted online \cite{ethics}; as no direct intervention on patients or collection of personally identifiable data was involved, the study raised no issues regarding individual privacy.

The dataset was first accessed on the 19th of January 2023. Patients from pediatrics and gynecology were not included in the dataset. Each entry describes one patient arrival and subsequent processing in the ED (see below). Entries with an unrealistic length of stay were eliminated, reducing the number of entries from 824,718 to 824,695. We classify a duration of stay as unrealistic if it is either negative or longer than 50 days. This latter cutoff was chosen after discussion with medical practitioners from Son Espases Hospital. Instead of the commonly used 24-hour resolution, we used the shifts of the hospital personnel in our analysis. These are the morning shift (8:00 - 15:00), the afternoon shift (15:00 - 21:00), and the night shift (21:00 - 8:00).

The 15 columns for each entry in the original dataset (see \ref{si:original-cols}) were processed into the following 6:

\begin{itemize}
 \item \textbf{Date and shift of entry} at the ED.
 \item \textbf{Date and shift of exit} from the ED.
 \item \textbf{Sex} of the patient at the time of the visit. 
 \item \textbf{Age} of the patient at the time of the visit.
 \item If the patient was a \textbf{resident} in the Balearic Islands or a \textbf{non-resident}. To be identified as a resident, they need to have both Spain as the country of residence (\textit{España} as \textit{Pais residencia}) and the Balearic Islands as province of residence (\textit{Baleares} as \textit{Provincia residencia}).
 \item Whether the patient was \textbf{hospitalized} immediately from the ED, or not. To do this, we checked if the entry for ``reason of discharge from the ED" was hospitalization (\textit{Motivo alta} was \textit{Paso a hospitalizaci\'on}).
\end{itemize}

\begin{figure}[H]
\centering
\includegraphics[width=0.75\linewidth]{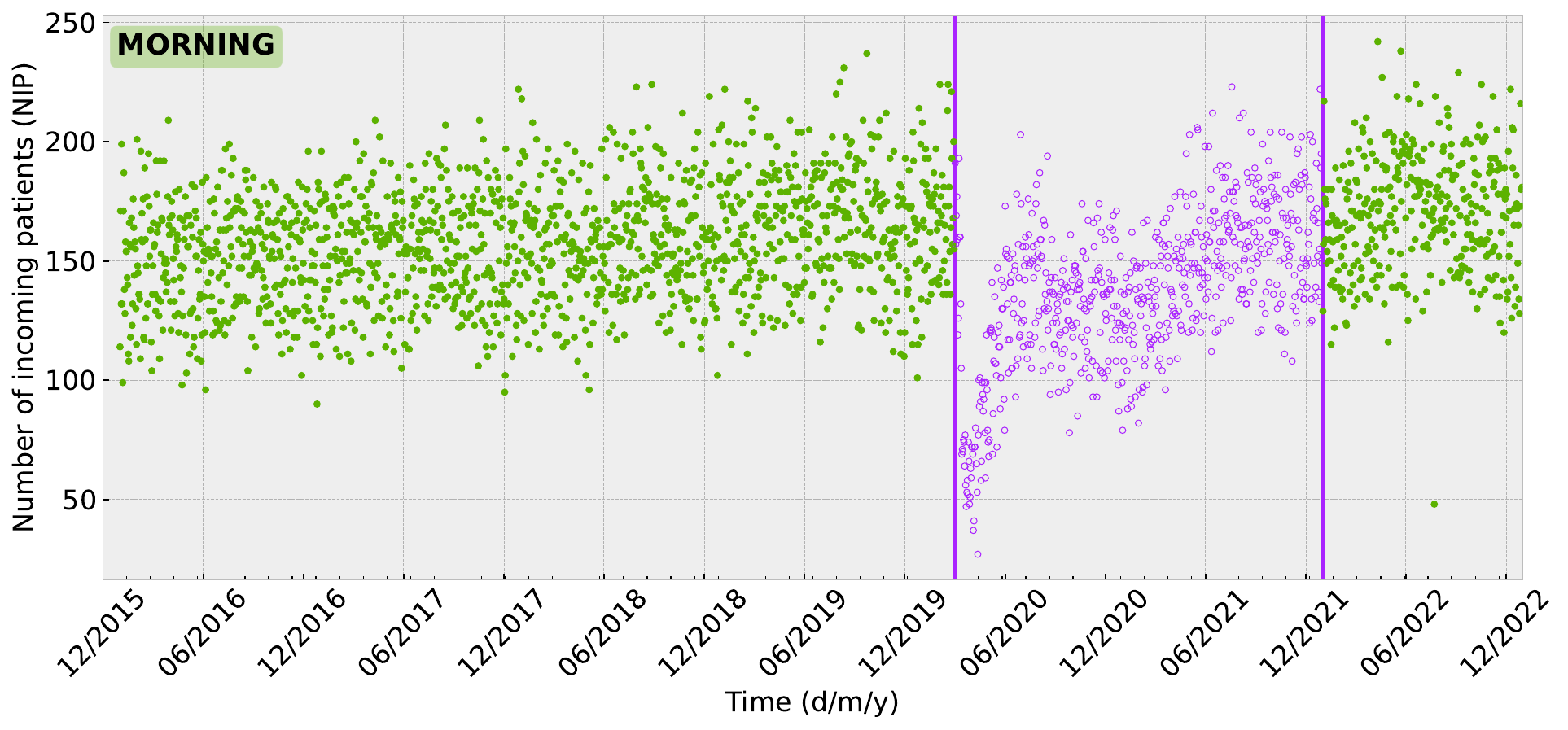}
\includegraphics[width=0.75\linewidth]{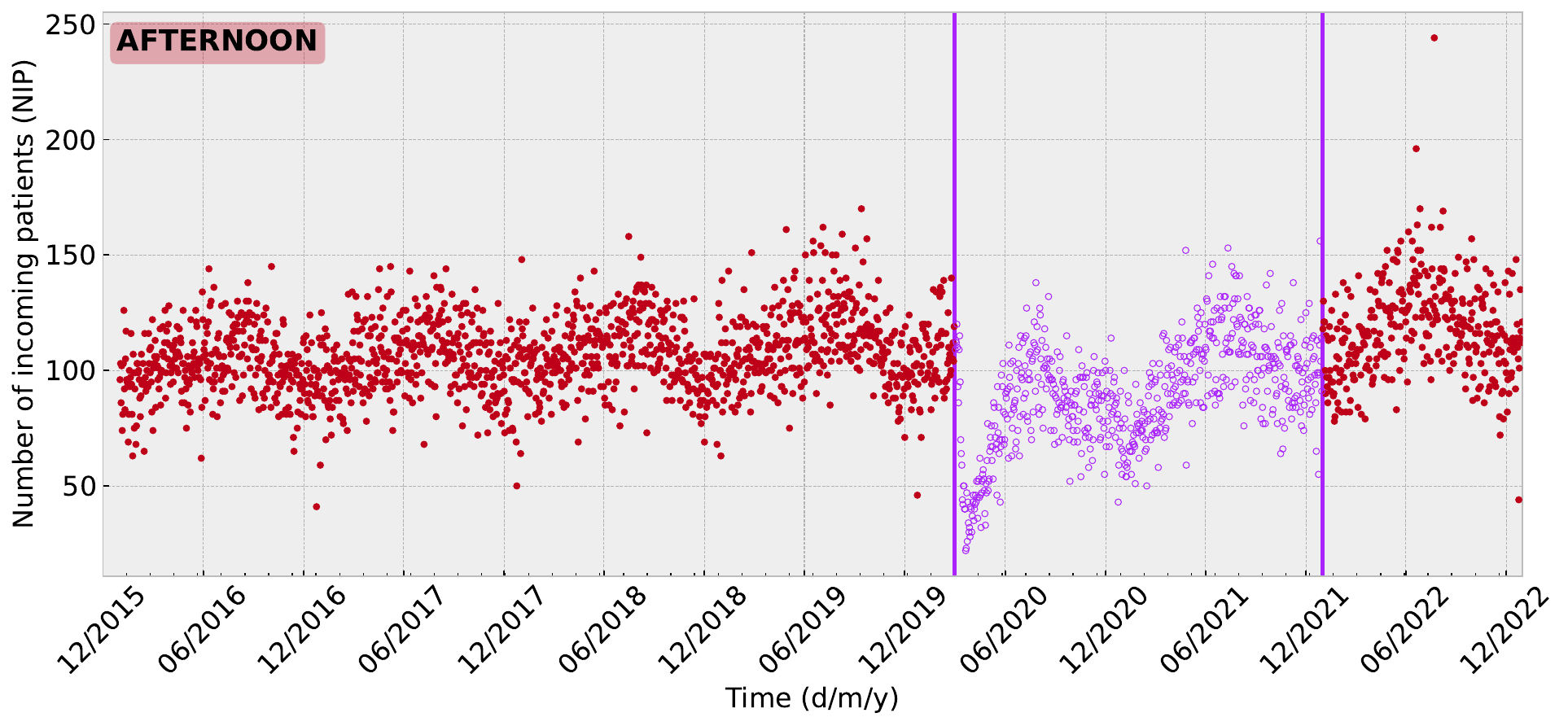}
\includegraphics[width=0.75\linewidth]{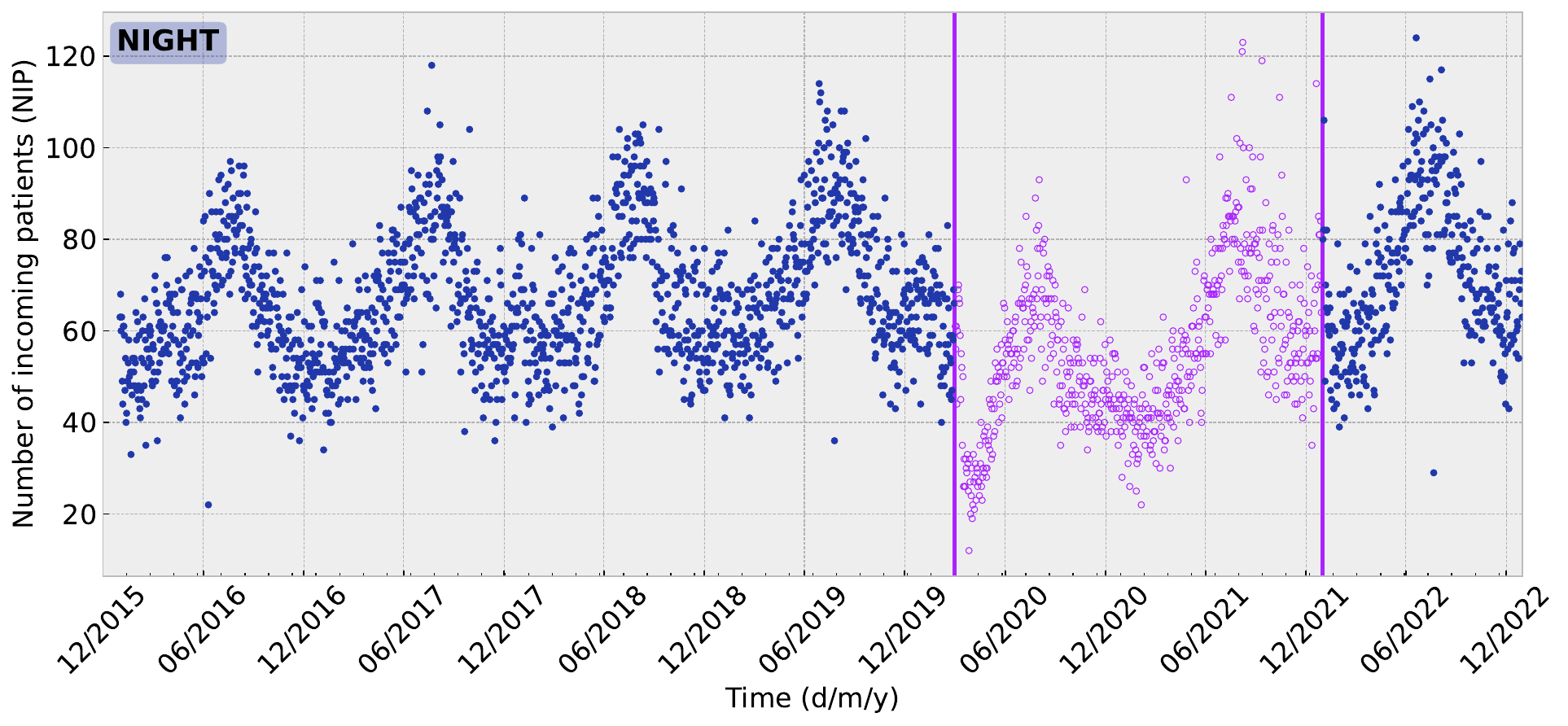}
\caption{\label{Fig1}Number of incoming patients (NIP) to the ED of Son Espases. Each point corresponds to the NIP for a specific day and shift. Each subfigure shows a different shift (morning in green, afternoon in red, night in blue). The purple points between the dates March 1, 2020 and December 31, 2021 are the values registered during the assumed pandemic period, and excluded from our analysis.}
\end{figure}
\medskip

\medskip

To explicitly avoid the impact of the COVID-19 pandemic, all entries from March 1, 2020, to December 31, 2021, were excluded. As seen in Figure~\ref{Fig1}, this period was characterized by a sharp decrease in the number of incoming patients (NIP), followed by a gradual recovery after the end of the pandemic. While we recognize a possible interest in these pandemic-related patterns, the analysis of this data is outside the scope of the current investigation. By excluding this period of time, the dataset was reduced to a total of 634,357 entries. We note that post-COVID patterns may differ from those of the pre-COVID era due to the lasting societal effects of the pandemic. For this reason, we divided our dataset into the following subsets:

\begin{itemize}
 \item A \textbf{training dataset} spanning December~26th, 2015 to February~28th, 2018, used to train all models.

 \item A \textbf{validation dataset} covering the period March 1st, 2018, to February 28th, 2019. This dataset was used to tune the hyperparameters of our models.

 \item A \textbf{test dataset} covering the period March 1st, 2019, to February 29th, 2020. This dataset was used to evaluate the performance of our models.

 \item A \textbf{post-COVID dataset} covering a time period after the pandemic (January 1st, 2022, to December 31st, 2022). This dataset was not used in our primary analysis, but as an additional test dataset to study the accuracy of the prediction model (trained and validated on pre-COVID data) in the post-COVID period.
\end{itemize}

We make the four datasets available on a GitHub repository \cite{repository}, along with the code used for our study.

The NIP exhibits significant variation across the three daily shifts, as can be seen by comparing the three panels of Fig 1. The most significant difference across shifts is in the seasonal behaviour of the data over the course of the calendar year. This seasonality is evident for the night shift, and almost absent for the morning shift. It is worth noting that despite the oscillations the numbers of incoming patients grow steadily over time (see also Fig. 1). We attribute this to an increase of the population.

The behavior of NIP for non-residents is very different than that for residents (\ref{si:nip-tourists}). Because of the low number of non-residents among the incoming patients (5.15\% of the total entries), we did not attempt to make predictions for non-residents as a separate class. Although Mallorca attracts a large number of tourists, the proportion of non-residents among the patients in the dataset is low, among other reasons, because non-residents tend to attend private hospitals.

As the data does not show any obvious differences between sexes in ED attendance patterns (\ref{si:nip-sex}), we do not attempt to predict sex-specific NIPs. 55.48\% of all patients attending the ED were female and 44.52\% male. The reasons for this imbalance are beyond the scope of this paper. Our dataset includes three patients without any information about their sex. We did not exclude these patients from the dataset as we did not perform any sex-specific analysis or prediction.

\subsection{Age cohorts and hospitalization risk}

\begin{figure}[!h]
\centering
\includegraphics[width=0.666\linewidth]{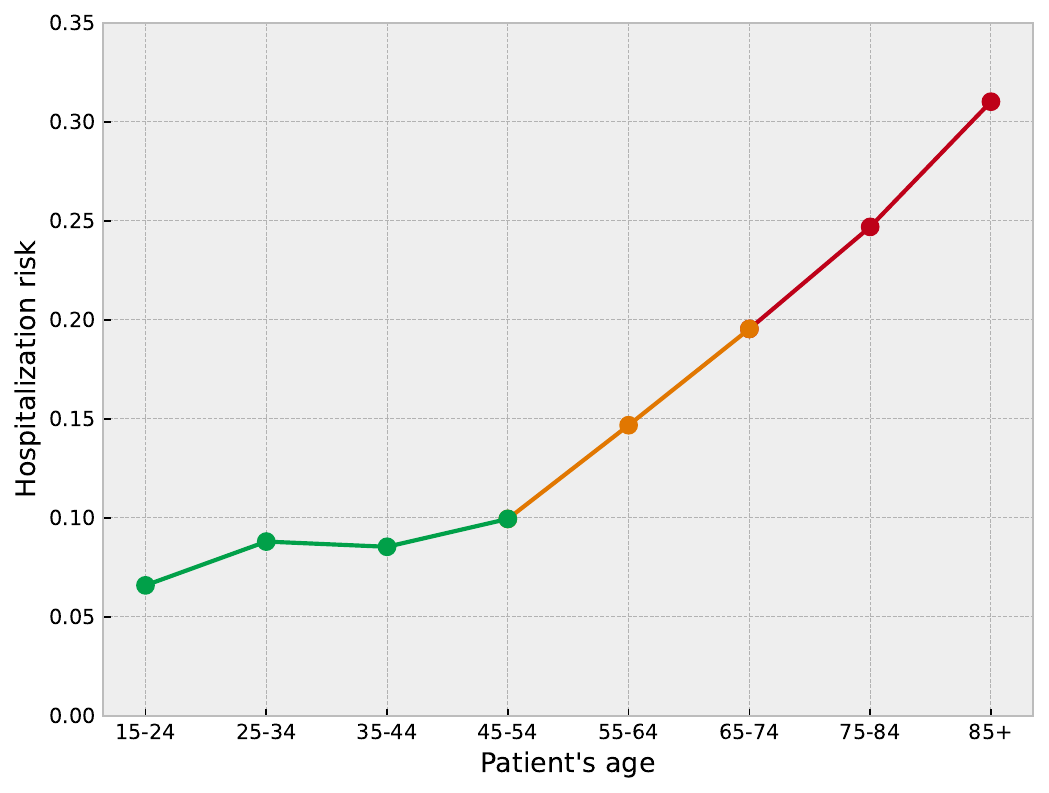}
\caption{\label{Fig2}Hospitalization risk $h$ as a function of the patient's age cohort. $h$ is the fraction of patients of a specific cohort who were hospitalized immediately after attending the ED. This is used to group patients in three different risk groups: low-risk group ($h<0.1$, green line), medium-risk group ($0.1<h<0.2$, orange line), and high-risk group ($h>0.2$, red line).}
\end{figure}

Table \ref{tab:motivo-alta} shows the most frequent reasons for a patient's discharge from the ED. The primary concern for hospital resource allocation is whether an emergency department visit leads to hospitalization. As hospitalization risk varies with patient age, we provide separate predictions for the different risk groups defined below.

We define the hospitalization risk $h$ for a given patient cohort as the fraction of patients in the dataset who are admitted to the hospital following an emergency department visit, rather than being discharged. Figure~\ref{Fig2} illustrates how $h$ increases with age. This age dependence motivates dividing patients into risk-based cohorts. We define the following classes:

\begin{itemize}
 \item \textbf{Low-risk group ($h\le 0.1$):} Ages 15 to 54.
 \item \textbf{Medium-risk group ($0.1<h\le 0.2$):} Ages 55 to 74.
 \item \textbf{High-risk group ($h>0.2$):} Ages 75 and above.
\end{itemize}

Data for admission into hospital beds from the ED at HUSE has a resolution of 24 hours. Therefore, we make daily predictions of NIP for the different risk groups and not shift-based predictions.

\begin{table}[!h]
\centering
\begin{tabular}{|l|l|l|}
\hline
\textbf{Reason of discharge} & \textbf{Counts} & \textbf{Percentage} \\ \hline
Full recovery/healing & 523968 & 82.60 \\ \hline
Hospitalization & 83247 & 13.12 \\ \hline
Failure to appear at the medical eval. & 7264 & 1.15 \\ \hline
Escaped (\textit{Fuga}) & 6719 & 1.06 \\ \hline
Voluntary discharge & 4226 & 0.67 \\ \hline
Failure to appear at the triage & 2845 & 0.45 \\ \hline
Moved to a socio-health center & 1641 & 0.26 \\ \hline
Moved to another hospital & 1302 & 0.21 \\ \hline
Moved to hospital of origin & 1144 & 0.18 \\ \hline
Unknown & 819 & 0.13 \\ \hline
Death & 633 & 0.10 \\ \hline
Home hospitalization & 532 & 0.08 \\ \hline
\end{tabular}
\caption{{\label{Table1}\bf Counts and percentages for reasons of discharge (\textit{Motivo alta}).}
This table includes only reasons for discharge that appear in the dataset} with a percentage higher than 0.01\%
\label{tab:motivo-alta}
\end{table}

\subsection{Input variables and models}

\begin{figure}[!h]
\centering
\includegraphics[width=0.49\linewidth]{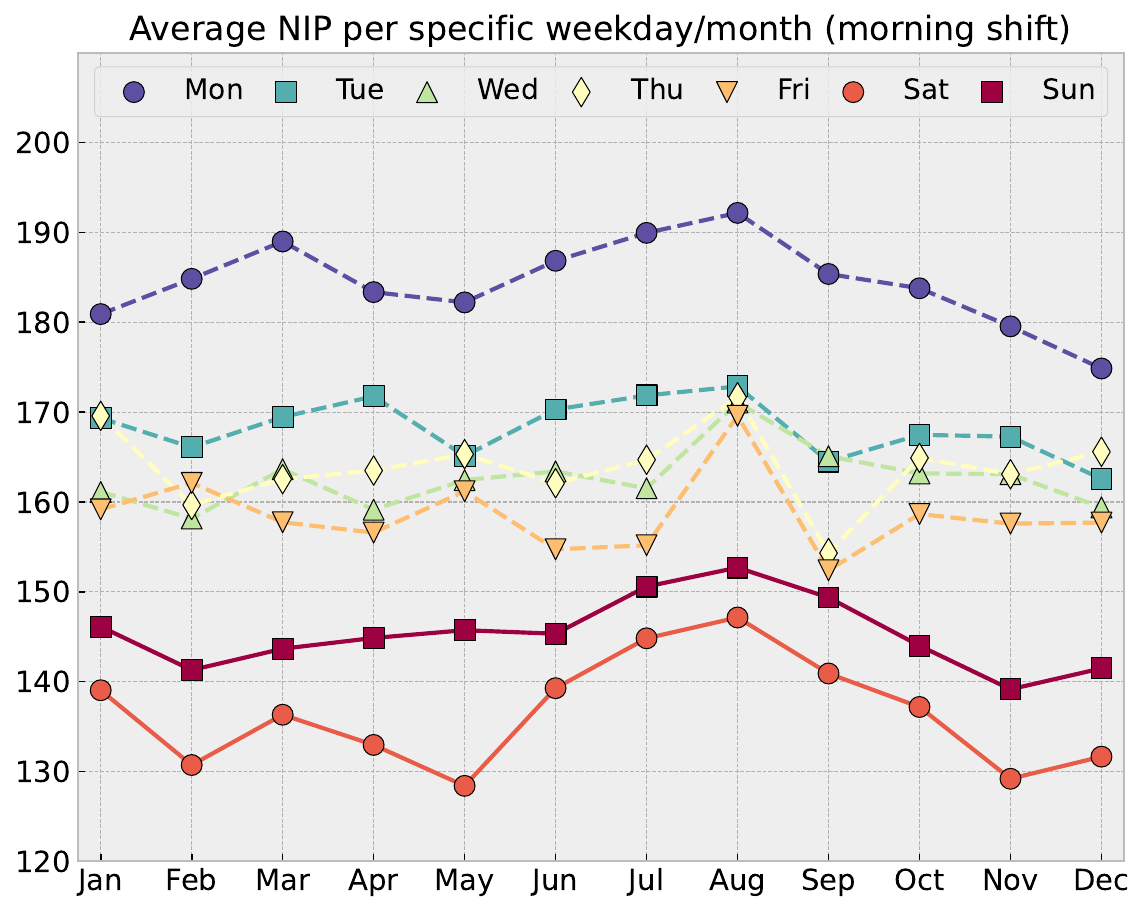}
\includegraphics[width=0.49\linewidth]{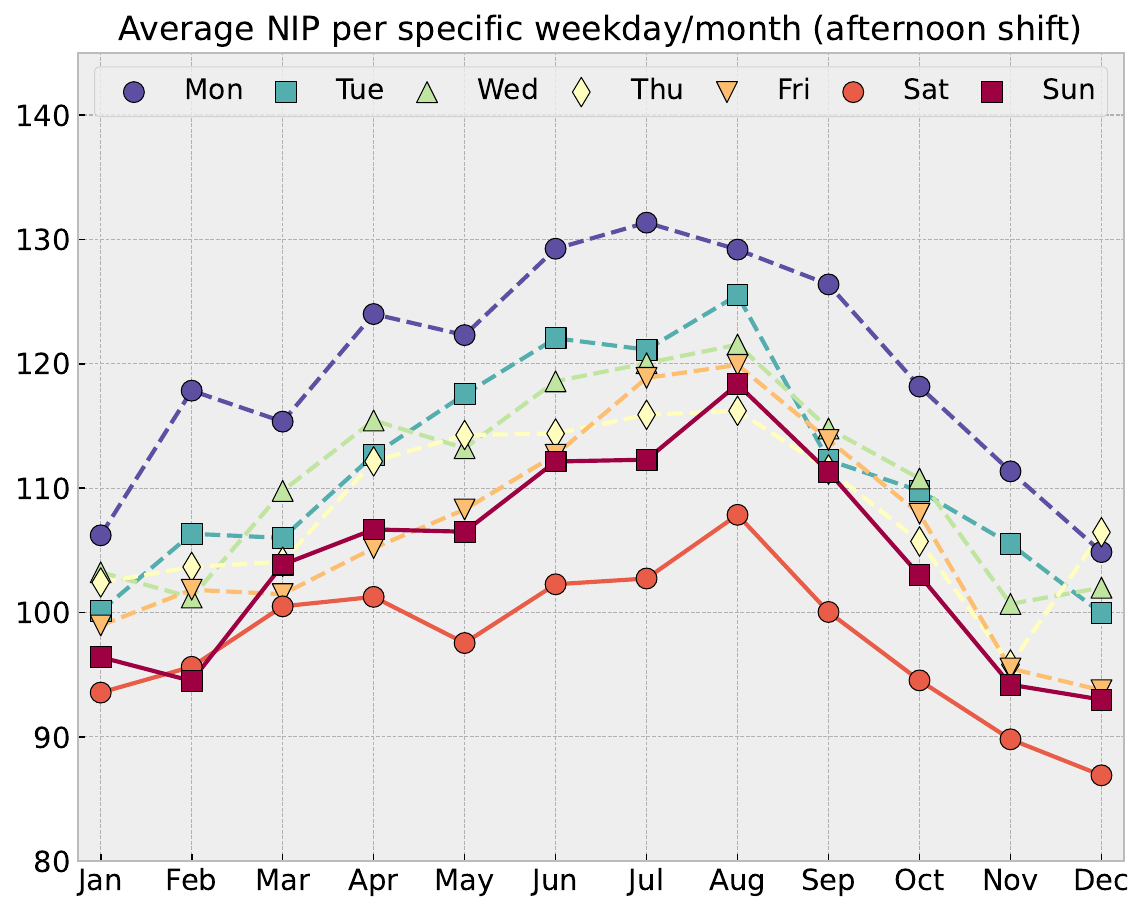}\\
\includegraphics[width=0.49\linewidth]{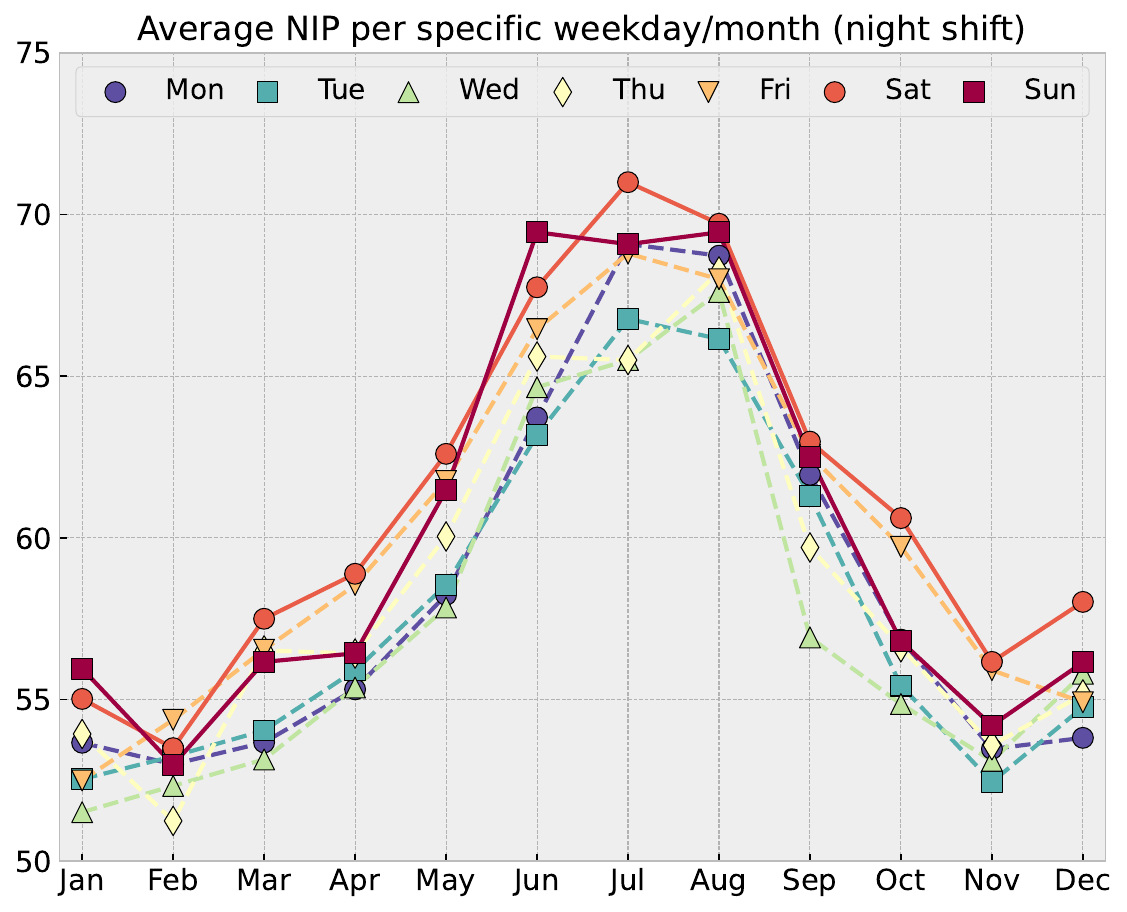}
\caption{Average NIP in the different shifts for given weekdays and months.
The panels respectively show morning, afternoon, and night shifts. Markers show the average of all NIPs for a specific shift, month, and weekday. For example, the lower plot shows that the average NIP during a night shift on Wednesdays in July is 65 (light green triangle point). As in Figure~\ref{Fig1}, the seasonal behavior of NIP is more pronounced for the night shift and almost absent for the morning shift.}
\end{figure}

Our choice of exogenous input variables is based on a number of observations. For example, the NIPs of the afternoon and night shifts exhibit considerable dependence on the time of the year (Fig. 3, upper right and bottom panels). We attribute this pattern to an increased overall population and engagement in risky activities during the summer months. On the contrary, arrivals during the mornings are largely independent of the time of the year (Fig. 3, upper left panel). Additionally, the NIP is different on different days of the week, particularly for the morning and afternoon shifts (Fig. 3, upper panels) being highest on Mondays, and lowest at the weekends. Based on these observations, we included calendar and population variables in the prediction models; additionally, we used weather data, following existing studies \cite{1, 2, 8, 10, 12, 14, 15, 18, 20, 21, 22}. More precisely, the inputs for our models are as follows:

\begin{itemize}
 \item \textbf{Calendar variables:} To predict outcomes on a specific date, our model uses the date itself (day, month and year) as input, along with the corresponding weekday. Additionally, we consider if any days within the five-day period surrounding the target date are designated holidays (national or local). Calendar variables were generated using the Python module \textit{workalendar} \cite{workalendar}, with the manual addition, when not already covered, of the local holidays in Palma or the Balearic Islands (in day/month format, these are 06/01, 20/01, 01/03, 01/05, 15/08, 12/10, 01/11, 06/12, 08/12, 26/12). The resulting input is therefore an array containing the following values: the number of days passed from the start of the dataset (an integer variable), the day of the year (integer from 0 to 365), the day of the week (integer from 0 to 6), and a bitstring of length 5 indicating which days on the five-day period surrounding the target date are holidays (5 values that are either 0 or 1).
 
 \item \textbf{Population variables:} We use the resident population and the number of tourists, both in the Balearic Islands. The data was obtained from the website of the \textit{Instituto Nacional de Estadística} \cite{population} (the national statistics agency in Spain). Resident data have a resolution of 6 months, while tourist data have a monthly resolution. As shown in the left-hand panel of Fig.~4, the resident population shows a monotonic growth in time. Daily estimates were obtained by performing a linear extrapolation of the past data. The number of tourists in the Balearic Islands shows strong periodicity (Fig.~4, right-hand panel), so a simple linear extrapolation of all past data is not appropriate. Instead, to estimate the number of tourists on a particular date, we performed a linear interpolation between the following two values: (i) the number of tourists in the month preceding or containing the target date, and (ii) the number of tourists in the calendar month following the target date, but from the year before.

 \begin{figure}[!h]
 \includegraphics[width=0.49\linewidth]{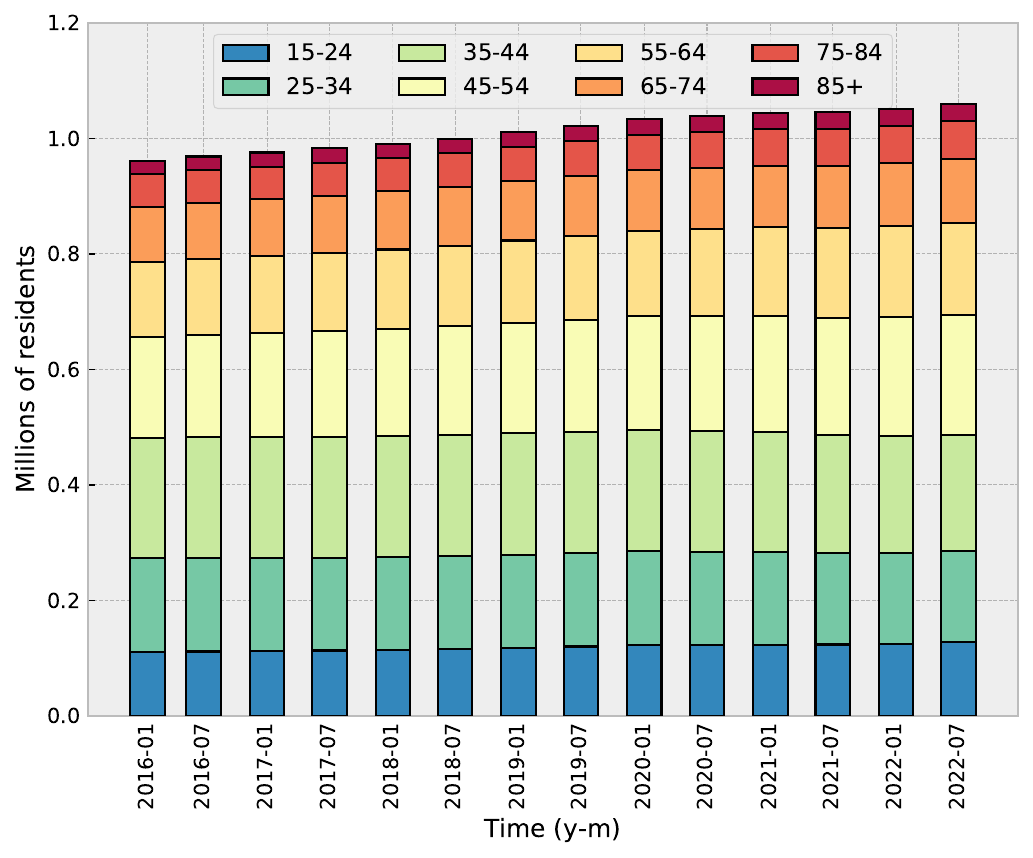}
 \includegraphics[width=0.49\linewidth]{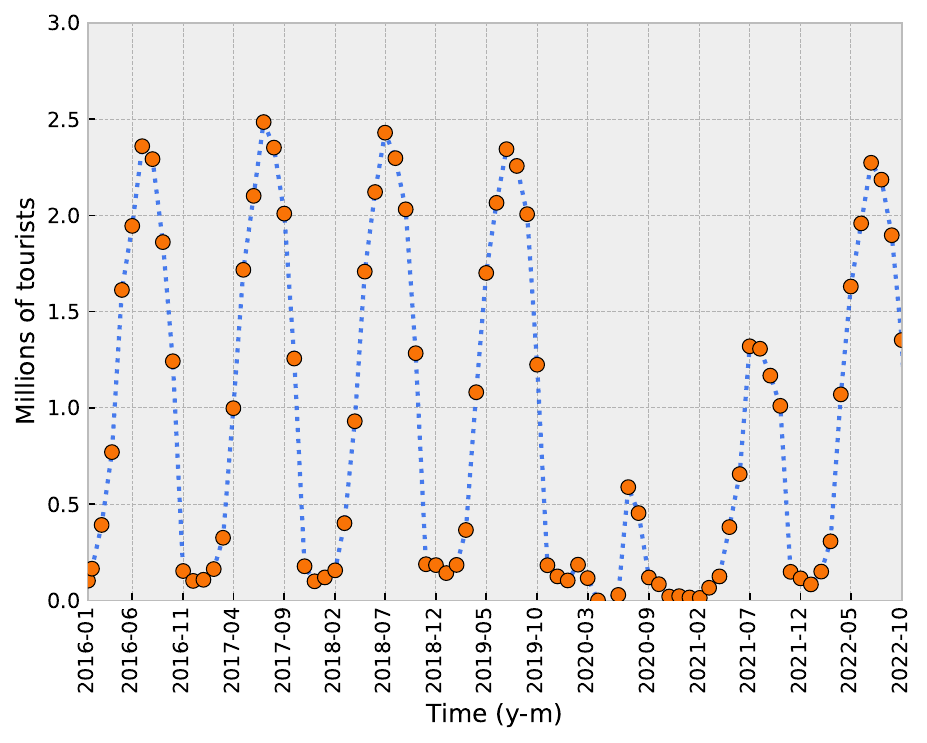}
 \caption{(a) Number of residents in the Balearic Islands per age-group. The data was sourced from the \textit{Instituto Nacional de Estadística} website and has a six-month resolution. (b) Number of monthly incoming tourists in the Balearic Islands, same source.}
 \end{figure}

\item \textbf{Weather variables:} For a given target date, we incorporate the weather forecast published the day before the target, sourced from the Spanish meteorology agency \textit{Agencia Estatal de Meteorología (AEMET)} \cite{weather}. The variables include maximum and minimum temperature, precipitation probability, and wind speed. The original data from AEMET consisted of text in natural language describing the predicted weather and integer numbers for the minimum and maximum temperatures. Precipitation probability and wind speed were extracted from the text as an intensity measure from 0 (absent) to 4 (very high intensity) by a local generative AI model. We employed the DeepSeek-R1 model \cite{deepseek} through a Python implementation. The prompt fed to the model can be found in the Appendix (\ref{si:ai-prompt}). This procedure does not introduce any target leakage. Since the data by AEMET already has a daily resolution, no interpolation was required.

\end{itemize}

The dataset of the preprocessed exogenous variables can be found in our repository \cite{repository}. 

Three machine-learning models were used to make predictions for the number of incoming patients. These were implemented using either \textit{PyTorch} \cite{pytorch} or \textit{scikit-learn} \cite{sklearn} Python libraries. All input data were preprocessed with scikit-learn's function \textit{StandardScaler}, which standardizes features by removing the mean and scaling to unit variance. We tested the following models on the data:

\begin{itemize}
 \item \textbf{Random forest (RF):} an ensemble learning method that constructs a number of decision trees during training, and, for a given input, returns the average prediction across trees. The term ``random" indicates the use of random subsets of the training data for the construction of each tree and the fact that random subsets of features are considered at each split in the trees. This enhances the model's generalization ability, accuracy, and robustness \cite{sklearn}. We set the parameter \textit{n\_estimators} of scikit-learn's function \textit{RandomForestRegressor}, which is the number of trees in the forest, to 100. To choose this value, we measured the performance of the RF model (see the Section \ref{sec:perf-boot} for more details about performance metrics) on the validation dataset for values of the \textit{n\_estimators} hyperparameter in the interval from 25 to 200 with a spacing of 25, and then selected the value for which the model performed best. The performance of the RF method for different values of \textit{n\_estimators} is shown in the Appendix (\ref{si:rf-svr-tuning}, left-hand panel).

 \item \textbf{Support vector regressors (SVR):} a type of support vector machine adapted for regression, whose goal is to predict continuous rather than categorical outputs. We set the parameter \textit{degree} of scikit-learn's function \textit{SVR}, representing the degree of the polynomial kernel function, to 3. As we did for the RF method, to choose this value, we measured the performance of the SVR model (see the Section \ref{sec:perf-boot} for more details about performance metrics) on the validation dataset for integer values of the \textit{degree} hyperparameter from one to ten, and then selected the value for which the model performed best. The performance of the SVR method for different values of \textit{degree} is shown in the Appendix (\ref{si:rf-svr-tuning}, right-hand panel). We left the default kernel set to \textit{RBF}.

 \item \textbf{Feedforward neural network (FNN):} a basic artificial neural network in which information flows in one direction, from input to output. It consists of interconnected nodes of consecutive layers, where each node applies an activation function. The network is trained to make predictions using backpropagation, adjusting weights to minimize prediction errors. FNNs are commonly used in machine learning for tasks such as classification and regression. For our models, we used a 14-layer structure, implemented with the \textit{PyTorch} module \cite{pytorch}. Details of the implementation can be found in \ref{si:fnn}. The FNN was trained for 130 epochs. To choose this value, we trained the model for 500 epochs, measuring at each epoch the loss function on both the training and the validation datasets. The minimum loss function on the validation dataset is obtained at 130 epochs. A figure showing the loss function curves can be found in the Appendix (see \ref{si:fnn-tuning}).
\end{itemize}

For all three models, we used four different combinations of input variables: (i) all variables (calendar variables, resident and tourist populations, weather forecasts), (ii) all variables, except the weather forecasts, (iii) all variables, except the tourist population, and (iv) only calendar variables and the resident population, i.e. excluding both weather forecasts and information on tourists. In the following text, the combination of input variables is denoted as a suffix following the name of the corresponding model. `All', `No W', `No T', and `No W-No T' indicate the combinations (i) to (iv) respectively. For example, the random forest method that uses all variables except weather forecasts is denoted as RF-No W, while the feedforward neural network including all input variables (calendar, weather forecasts, and tourist data) is denoted as FNN-All.

To compare our non-time-series models to commonly used time-series models, we implemented a seasonal autoregressive integrated moving average (SARIMA) model \cite{sarima} without exogenous variables. This model was parameterized as SARIMA$(1,1,1)(1,1,1)_7$. The non-seasonal orders $(p, d, q) = (1, 1, 1)$ were selected as a minimal configuration to account for a linear trend and short-term temporal dependencies. The seasonal orders $(P, D, Q) = (1, 1, 1)$ with a period of $m=7$ were specified to capture the weekly patterns evident from Figure 3.

We also implement a SARIMAX model with the same parametrization using calendar and population data as exogenous variables. We exclude weather data for this model for two reasons: (i) only 1-day weather forecasts are included in our study, thus it is not clear how to construct a meaningful SARIMAX model for long prediction horizons which would include weather data; (ii) as shown in the results section, weather data turned out not to be statistically significant to improve the quality of the predictions and can thus be discarded as an input variable.

Both the SARIMA and SARIMAX models were implemented using the Python module \textit{statsmodels} \cite{statsmodels}. In order to make a prediction for a specific date in the test period, the models were trained using the patient arrival numbers from the 365 days preceding this target date. The full implementation of these models, including seeds and version details, can be found in our repository \cite{repository}.

In the figures, tables, and the following text, we denote as SARIMA-$n$ and SARIMAX-$n$ the SARIMA and the SARIMAX models described in the previous section with a prediction horizon of $n$ days. We compare their predictions with our non-time-series models for $n=1, 7$, $14$, and $28$ days in the future, to test if our models perform better than SARIMA and SARIMAX as the prediction horizon increases. From existing literature \cite{1,2,3,4,5,6,7,8,9,10,11,12,13,14,15,16,17,18}, we expect the SARIMA model to outperform our three models for short prediction horizons, but to gradually lose accuracy as the horizon increases. Ultimately, we expect that the SARIMA models will be outperformed by the non-time-series models, or at the very least that the accuracy of both types of models will become comparable. We anticipate broadly similar behavior for SARIMAX; however, overfitting effects may adversely affect its predictive accuracy, even over short time horizons.

\subsection{Performance metrics and bootstrapping}
\label{sec:perf-boot}

 The accuracy of the predictive models is evaluated using the symmetric mean absolute percentage error (SMAPE):
\begin{equation}
 \text{SMAPE} = 100 \times \frac{2}{N} \sum_{i=1}^N \bigg|\frac{y_i - \hat{y}_i}{y_i + \hat{y}_i}\bigg|,
 \label{eq:mape}
\end{equation}
where $y_i$ is the actual NIP for a day and shift (or risk group), $\hat{y}_i$ is the predicted NIP for that shift (or risk group) on that day, and $N=365$ is the number days in the test dataset. SMAPE is a metric similar to MAPE that avoids issues of dividing by zero. This is suitable for our dataset as it is possible to have no incoming patients on a given day, especially for the night shift or if the analysis is restricted to cohort of high-risk patients. The SMAPE was calculated for 1000 bootstrap samples. Each of these bootstrap samples was generated by sampling $N$ random days from the test dataset with equal probability and with replacement. Details concerning random seeds and sampling functions can be found in our repository \cite{repository}. We also employ the root-mean-squared error (RMSE) and the mean-average-error (MAE) as secondary performance measures:

\begin{equation}
 \text{RMSE} = \sqrt{\frac{1}{N}\sum_{i=1}^N(y_i - \hat{y}_i)^2}, \quad\quad \text{MAE} = \frac{1}{N}\sum_{i=1}^N|y_i - \hat{y}_i|
 \label{eq:rmse-mae}
\end{equation}

For specific pairs of models, we perform a Diebold--Mariano (DM) test \cite{diebold} to determine if the costs associated with prediction errors from the two models are statistically different from one another. As the cost function we used the symmetric mean average percentage error (SMAPE) as it is the main performance metric used in the results section. The test provides a p‑value for the null hypothesis that the two forecast models have an equivalent average cost. If the null hypothesis is accepted at the significance level $0.05$ we say that the two models have equal predictive accuracy. Since we are testing a total of 243 null hypotheses with these DM tests, we apply a Bonferroni correction for multiple-hypothesis testing \cite{bonferroni-corr}. For clarity, we note that equal predictive accuracy as determined by the DM test is not transitive, that is to say if model A is identified as equivalent to model B by the test, and model B as equivalent to model C, then models A and C need not be equivalent to one another \cite{diebold}.

For a given pair of prediction models, DM tests are run for each bootstrap sample, returning a total of 1000 p-values for that pair of models. We measured the degree of equivalence between the two models by the fraction of the DM test results among those 1000 bootstrap samples that led to the acceptance of the null hypothesis of equal predictive accuracy between the models. In future sections, we state that two models have equal predictive accuracy if they satisfy the null hypothesis in more than 75\% of the samples. The Bonferroni-corrected Diebold–Mariano test enforces strong control of false positives within each sample. The additional requirement that the null be rejected in at least 75\% of resamples serves as a complementary robustness criterion, ensuring that statistical significance is not driven by sample-specific variability, while avoiding the overly restrictive requirement of near-uniform rejection that would be incompatible with the finite-sample power of the DM test.

\section{Results}

\begin{table}[h]
\centering
\begin{tabular}{cc|ccc|ccc|}
\cline{3-8}
\textbf{} & \textbf{} & \multicolumn{3}{c|}{\textbf{Shift-based predictions}} & \multicolumn{3}{c|}{\textbf{Risk-group-based predictions}} \\ \hline
\multicolumn{1}{|c|}{\textbf{Method}} & \textbf{\begin{tabular}[c]{@{}c@{}}Input\\ variables\end{tabular}} & \multicolumn{1}{c|}{\textbf{\begin{tabular}[c]{@{}c@{}}SMAPE\\ (Morning)\end{tabular}}} & \multicolumn{1}{c|}{\textbf{\begin{tabular}[c]{@{}c@{}}SMAPE\\ (Afternoon)\end{tabular}}} & \textbf{\begin{tabular}[c]{@{}c@{}}SMAPE\\ (Night)\end{tabular}} & \multicolumn{1}{c|}{\textbf{\begin{tabular}[c]{@{}c@{}}SMAPE\\ (Low)\end{tabular}}} & \multicolumn{1}{c|}{\textbf{\begin{tabular}[c]{@{}c@{}}SMAPE\\ (Medium)\end{tabular}}} & \textbf{\begin{tabular}[c]{@{}c@{}}SMAPE\\ (High)\end{tabular}} \\ \hline
\multicolumn{1}{|c|}{RF} & All & \multicolumn{1}{c|}{\cellcolor[HTML]{CBCEFB}9.21} & \multicolumn{1}{c|}{\cellcolor[HTML]{C6E7FF}10.02} & \cellcolor[HTML]{CBCEFB}11.73 & \multicolumn{1}{c|}{\cellcolor[HTML]{CBCEFB}8.59} & \multicolumn{1}{c|}{\cellcolor[HTML]{CBCEFB}10.21} & \cellcolor[HTML]{CBCEFB}14.31 \\ \hline
\multicolumn{1}{|c|}{RF} & No W & \multicolumn{1}{c|}{\cellcolor[HTML]{CBCEFB}9.03} & \multicolumn{1}{c|}{\cellcolor[HTML]{CBCEFB}10.29} & \cellcolor[HTML]{CBCEFB}12.08 & \multicolumn{1}{c|}{\cellcolor[HTML]{CBCEFB}8.91} & \multicolumn{1}{c|}{\cellcolor[HTML]{CBCEFB}10.37} & \cellcolor[HTML]{CBCEFB}14.15 \\ \hline
\multicolumn{1}{|c|}{RF} & No T & \multicolumn{1}{c|}{\cellcolor[HTML]{CBCEFB}9.48} & \multicolumn{1}{c|}{\cellcolor[HTML]{CBCEFB}10.78} & 13.65 & \multicolumn{1}{c|}{9.95} & \multicolumn{1}{c|}{\cellcolor[HTML]{CBCEFB}10.96} & \cellcolor[HTML]{C6E7FF}13.35 \\ \hline
\multicolumn{1}{|c|}{RF} & No W-No T & \multicolumn{1}{c|}{\cellcolor[HTML]{CBCEFB}9.43} & \multicolumn{1}{c|}{\cellcolor[HTML]{CBCEFB}10.79} & \cellcolor[HTML]{CBCEFB}11.90 & \multicolumn{1}{c|}{\cellcolor[HTML]{CBCEFB}9.06} & \multicolumn{1}{c|}{\cellcolor[HTML]{CBCEFB}10.81} & \cellcolor[HTML]{CBCEFB}14.65 \\ \hline
\multicolumn{1}{|c|}{SVR} & All & \multicolumn{1}{c|}{12.97} & \multicolumn{1}{c|}{13.33} & 18.53 & \multicolumn{1}{c|}{12.92} & \multicolumn{1}{c|}{15.58} & 21.69 \\ \hline
\multicolumn{1}{|c|}{SVR} & No W & \multicolumn{1}{c|}{13.28} & \multicolumn{1}{c|}{13.44} & 18.42 & \multicolumn{1}{c|}{13.40} & \multicolumn{1}{c|}{16.11} & 23.08 \\ \hline
\multicolumn{1}{|c|}{SVR} & No T & \multicolumn{1}{c|}{12.95} & \multicolumn{1}{c|}{13.74} & 21.25 & \multicolumn{1}{c|}{13.56} & \multicolumn{1}{c|}{16.11} & 21.11 \\ \hline
\multicolumn{1}{|c|}{SVR} & No W-No T & \multicolumn{1}{c|}{13.43} & \multicolumn{1}{c|}{14.58} & 24.76 & \multicolumn{1}{c|}{15.40} & \multicolumn{1}{c|}{16.92} & 20.99 \\ \hline
\multicolumn{1}{|c|}{FNN} & All & \multicolumn{1}{c|}{13.21} & \multicolumn{1}{c|}{12.95} & 17.84 & \multicolumn{1}{c|}{\cellcolor[HTML]{CBCEFB}9.25} & \multicolumn{1}{c|}{12.89} & 19.95 \\ \hline
\multicolumn{1}{|c|}{FNN} & No W & \multicolumn{1}{c|}{\cellcolor[HTML]{CBCEFB}9.53} & \multicolumn{1}{c|}{\cellcolor[HTML]{CBCEFB}10.37} & 17.49 & \multicolumn{1}{c|}{12.32} & \multicolumn{1}{c|}{14.16} & 21.07 \\ \hline
\multicolumn{1}{|c|}{FNN} & No T & \multicolumn{1}{c|}{13.09} & \multicolumn{1}{c|}{14.13} & 20.98 & \multicolumn{1}{c|}{65.67} & \multicolumn{1}{c|}{67.31} & 72.84 \\ \hline
\multicolumn{1}{|c|}{FNN} & No W-No T & \multicolumn{1}{c|}{11.65} & \multicolumn{1}{c|}{11.93} & 17.49 & \multicolumn{1}{c|}{10.73} & \multicolumn{1}{c|}{13.24} & 18.80 \\ \hline
\multicolumn{1}{|c|}{SARIMA-1} & & \multicolumn{1}{c|}{7.69} & \multicolumn{1}{c|}{\cellcolor[HTML]{C6E7FF}9.14} & 10.48 & \multicolumn{1}{c|}{6.92} & \multicolumn{1}{c|}{\cellcolor[HTML]{C6E7FF}9.30} & 12.52 \\ \hline
\multicolumn{1}{|c|}{SARIMAX-1} & & \multicolumn{1}{c|}{8.16} & \multicolumn{1}{c|}{10.99} & 12.23 & \multicolumn{1}{c|}{8.22} & \multicolumn{1}{c|}{10.25} & 14.58 \\ \hline
\multicolumn{1}{|c|}{SARIMA-7} & & \multicolumn{1}{c|}{7.84} & \multicolumn{1}{c|}{\cellcolor[HTML]{C6E7FF}9.47} & \cellcolor[HTML]{CBCEFB}11.39 & \multicolumn{1}{c|}{7.40} & \multicolumn{1}{c|}{\cellcolor[HTML]{C6E7FF}9.28} & \cellcolor[HTML]{C6E7FF}12.68 \\ \hline
\multicolumn{1}{|c|}{SARIMAX-7} & & \multicolumn{1}{c|}{12.02} & \multicolumn{1}{c|}{17.16} & 19.28 & \multicolumn{1}{c|}{13.50} & \multicolumn{1}{c|}{14.23} & 23.04 \\ \hline
\multicolumn{1}{|c|}{SARIMA-14} & & \multicolumn{1}{c|}{\cellcolor[HTML]{C6E7FF}8.12} & \multicolumn{1}{c|}{\cellcolor[HTML]{CBCEFB}9.74} & \cellcolor[HTML]{CBCEFB}12.28 & \multicolumn{1}{c|}{\cellcolor[HTML]{C6E7FF}7.77} & \multicolumn{1}{c|}{\cellcolor[HTML]{C6E7FF}9.47} & \cellcolor[HTML]{C6E7FF}12.77 \\ \hline
\multicolumn{1}{|c|}{SARIMAX-14} & & \multicolumn{1}{c|}{17.23} & \multicolumn{1}{c|}{26.66} & 31.02 & \multicolumn{1}{c|}{22.94} & \multicolumn{1}{c|}{19.62} & 36.45 \\ \hline
\multicolumn{1}{|c|}{SARIMA-28} & & \multicolumn{1}{c|}{\cellcolor[HTML]{CBCEFB}8.52} & \multicolumn{1}{c|}{\cellcolor[HTML]{CBCEFB}10.48} & \cellcolor[HTML]{CBCEFB}13.58 & \multicolumn{1}{c|}{\cellcolor[HTML]{CBCEFB}9.40} & \multicolumn{1}{c|}{\cellcolor[HTML]{CBCEFB}9.67} & \cellcolor[HTML]{CBCEFB}13.23 \\ \hline
\multicolumn{1}{|c|}{SARIMAX-28} & & \multicolumn{1}{c|}{30.22} & \multicolumn{1}{c|}{50.23} & 63.86 & \multicolumn{1}{c|}{47.65} & \multicolumn{1}{c|}{31.56} & 70.90 \\ \hline
\end{tabular}
\caption{{\bf SMAPE (Eq. \ref{eq:mape}) for the different prediction methods.} Each row corresponds to a specific method. The RF, SVR, and FNN models were run with different combinations of input data, as explained in the text. `All' indicates all input variables (calendar, resident and tourist population, weather forecast). `No W' means that weather data was excluded, `No T' means that resident and tourist populations were excluded. The last eight rows are for the SARIMA-$n$ and SARIMAX-$n$ time-series models, where the integer number (n=1, 7, 14, 28) indicates the prediction horizon in days. Columns 3 to 5 are the SMAPEs for the predictions of the total NIP (across all age groups) in the different shifts. The last three columns show the SMAPEs for the three age cohorts discussed in the section on Age-specific predictions. Blue cells indicate that, according to the DM test, a model has the same prediction accuracy as the RF-No W model for that shift or risk group in at least 90\% of the bootstrap samples. Purple cells indicate that the same condition is satisfied for both the RF-No W and RF-No W-No T models. The focus on these two models is justified in the Results section.
}
\label{tab:mape}
\end{table}

\begin{figure}[!h]
\includegraphics[width=0.49\linewidth]{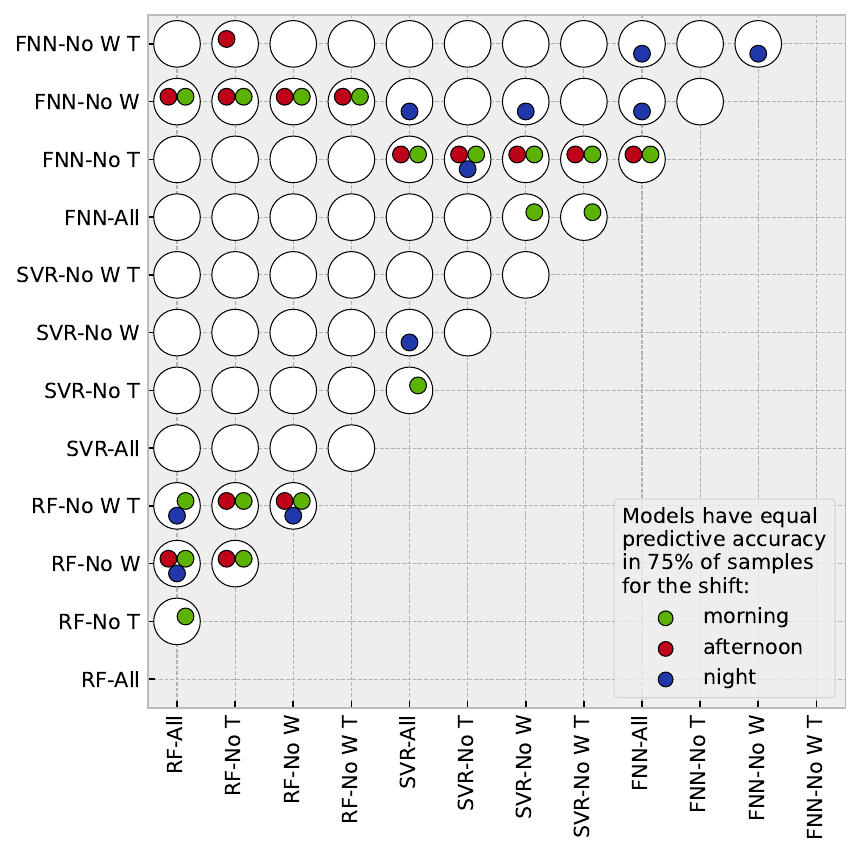}
\includegraphics[width=0.49\linewidth]{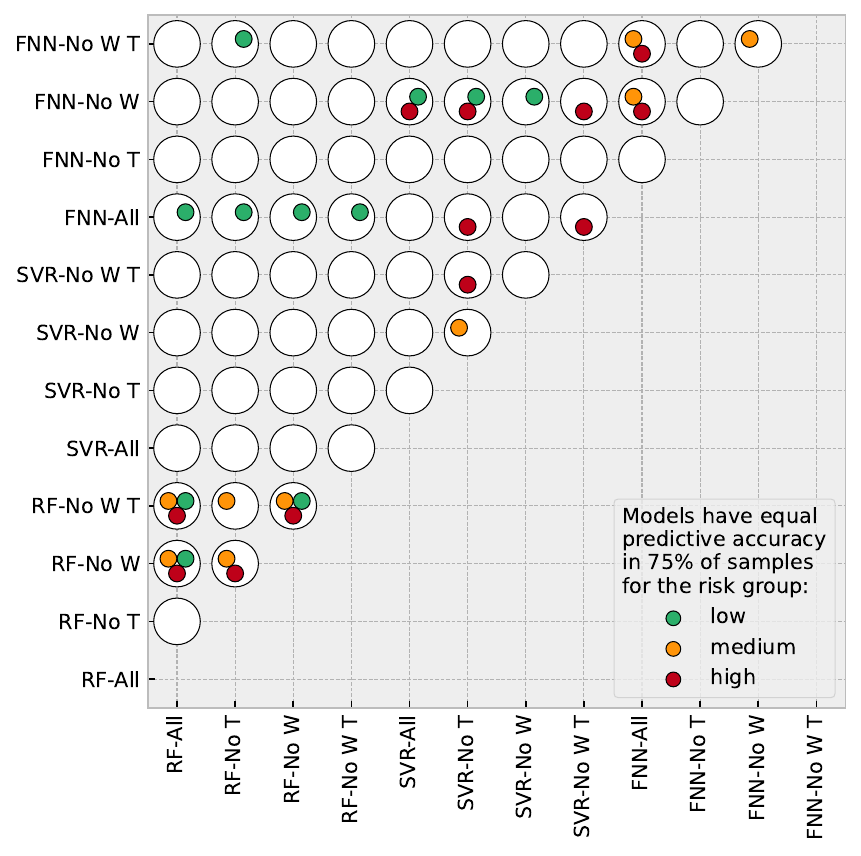}
\caption{This figure highlights which combinations of non-time-series models and input variables have equal predictive accuracy according to the DM test (shift-based predictions in panel (a) and risk-group-based predictions in panel (b)). Each white disk corresponds to a different combination, as indicated on the axes. Since the test is symmetric, we only show each comparison once and hence the lower-right part of each diagram is empty. The colored circles inside the white disks indicate for which shifts or risk groups the two models represented by the disk have equal predictive accuracy for at least 75\% of bootstrap samples.}
\end{figure}

\subsection{Non-time-series models}

Table \ref{tab:mape} shows the SMAPEs for each patient cohort, both for shift-based and risk-group-based predictions. The full set of performance metrics (SMAPE, RMSE, MAE) and the associated errors over the bootstrap samples can be found in the Appendix (\ref{si:metric-table}). The values in Table \ref{tab:mape} are the mean over the 1000 bootstrap samples. In this subsection, we focus on the upper half of the table, showing the SMAPEs for the non-time-series models. For each shift and risk group, the non-time-series model with the lowest SMAPE is always an RF model.

Figure 5 highlights which pairs of models have equal predictive accuracy according to the DM test among the non-time-series models. The full set of results of the DM tests, including the percentages of bootstrap samples that satisfy the null hypotheses, can be found in the Appendix (\ref{si:diebold-table}). Our goal is to select the simplest model, both in terms of input variables and method, that either has the lowest SMAPE for all shifts or has equal predictive accuracy to the one that satisfies this condition. SVRs always perform worse than RF models (SVRs are found to have a higher SMAPE and never have equal predictive accuracy as the RF with the lowest SMAPE). For some particular input combinations and shifts (or risk groups), FNNs have equal predictive accuracy as the RF with the lowest SMAPE. Nevertheless, FNNs are more computationally expensive than RFs, so we focus only on RF models for the rest of the subsection.
\medskip

In what follows, we analyse the results shift by shift and risk group by risk group, starting with the morning shift. For the latter, RF-No W yields the lowest SMAPE, suggesting that the inclusion of weather data induces overfitting. Furthermore, RF-No W exhibits equal predictive accuracy to that of all other RF models. In particular, further excluding tourist variables does not lead to a statistically significant loss of accuracy. Therefore, both sets of variables can be safely omitted.

For the afternoon shift, RF-All achieves the lowest SMAPE but has a predictive accuracy equal to that of RF-No W. This means that the increase in SMAPE caused by excluding weather information is not statistically relevant. Nevertheless, tourist data significantly improve the accuracy according to the DM test.

RF-All again achieves the lowest SMAPE for the night shift, although its predictive accuracy does not differ significantly from that of RF-No W and RF-No W-No T. We also find that RF-No T performs significantly worse, suggesting that weather data causes overfitting in the absence of tourist variables. By contrast, the inclusion of weather data slightly improves the SMAPE, but this improvement is not statistically significant according to the DM test.

We next focus on risk-group-based predictions. For both the low- and medium-risk groups, RF-All attains the lowest SMAPE. However, its predictive accuracy is not statistically different from that of the models excluding weather or tourist variables, mirroring the pattern observed for the morning and night shifts. Therefore, for these risk groups, both sets of variables can be excluded without a statistically significant loss in predictive performance.

For the high-risk group, the situation changes. RF-No T attains the lowest SMAPE. However, according to the DM test, its predictive accuracy is equal to that of RF-No W, and significantly smaller than that of RF-All (see panel b of Figure 5). This indicates that combining weather and tourist data (RF-All) leads to overfitting for this risk group. Since weather forecasts are available only one day ahead, whereas tourist data can be predicted on a monthly scale, RF-No W emerges as the most practical and robust choice for this group.

Summarising, for all shifts and risk groups, RF-No W proved to be either the model with the lowest SMAPE or to have equal predictive accuracy as the one with the lowest SMAPE. Therefore, the inclusion of weather data does not significantly improve prediction accuracy for any shift or risk group and can be discarded as an input variable.

Except for the afternoon shift and the high-risk group, RF-No W-No T also proved to be either the model with the lowest SMAPE or to have equal predictive accuracy as the one with the lowest SMAPE. Therefore, unlike weather data, the inclusion of tourist data significantly improves prediction accuracy for the afternoon shift and the high-risk group.

We also notice that the difference in SMAPE between the RF-No W models and RF-No W-No T is always lower than 0.5. Since the daily NIP for any shift or risk group is always lower than 250 patients, this difference in SMAPE corresponds to less than two patients per shift or risk group. This quantity, although statistically significant according to the DM test, is small in terms of resource allocation planning. This suggests that, while the inclusion of tourist data has a statistically significant effect, its overall impact is much weaker than implied by public perception and media narratives \cite{tourist1, tourist2}. This does not mean that tourists have no effect on patient inflow; rather, because of their strong seasonality, much of the information contained in tourist data is already captured by calendar variables.

Figure 6 illustrates how predictions of the RF-No W model align with shift-based data in two distinct time windows (May-June 2019 in panel (a) and November-December 2019 in panel (b)). The points represent the actual NIP, while the colored bars indicate the RMSE around the predicted value ($18.33$, $15.03$, $11.07$ for the morning, afternoon, and night shift respectively). Figure 7 illustrates how predictions of the RF-No W align with risk-group-based data in the same two time windows. The points represent again the actual NIP, while the colored bars indicate the RMSE around the predicted value ($23.82$, $10.88$, $8.80$ for the low, medium, and high risk groups respectively). The selected model, RF-No W, relies exclusively on calendar and population variables, both resident and tourist, which enables the prediction of NIP for any future date, provided that no major events disrupt tourism or population growth patterns, such as the COVID-19 pandemic.

\begin{figure}[!h]
\includegraphics[width=0.75\linewidth]{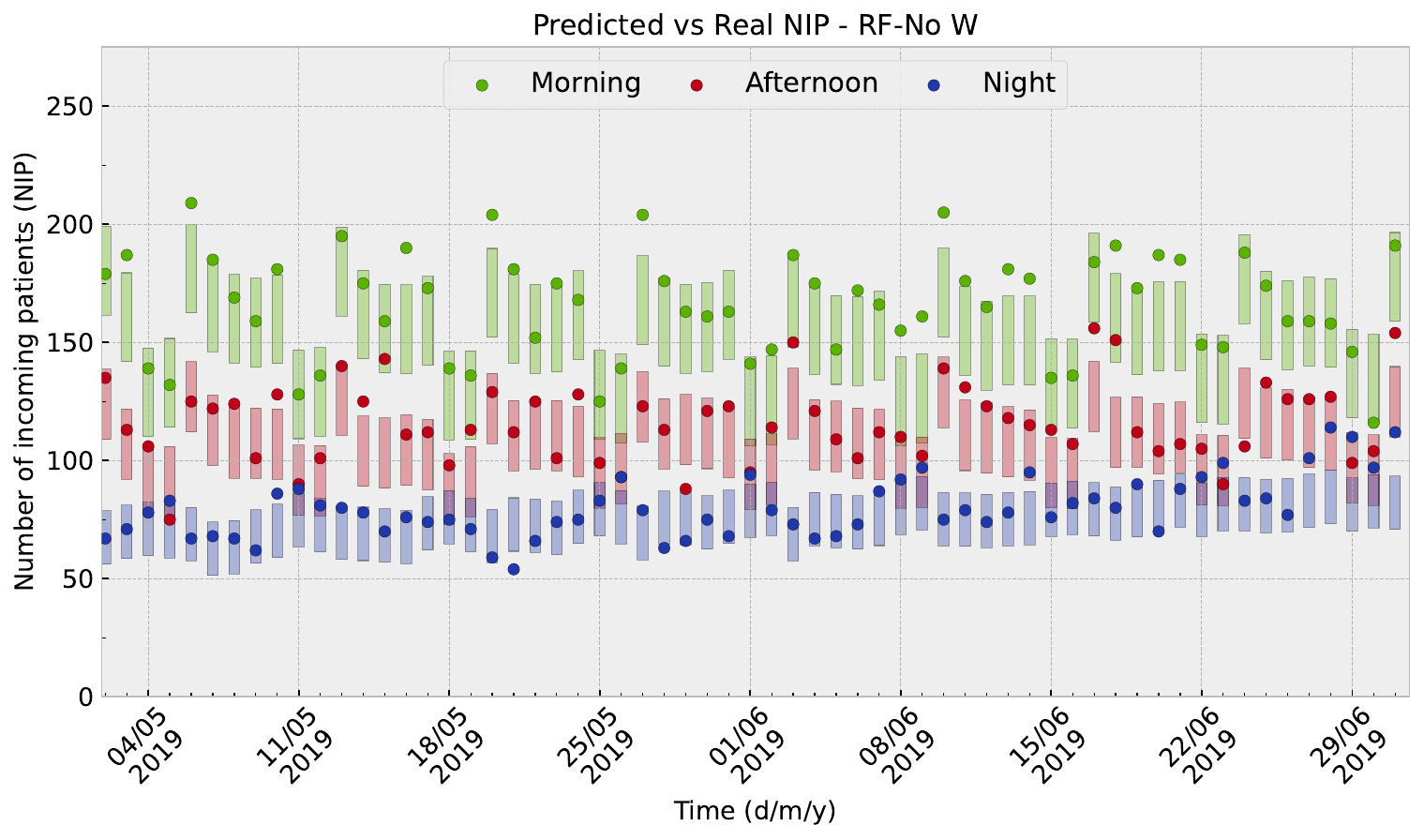}
\includegraphics[width=0.75\linewidth]{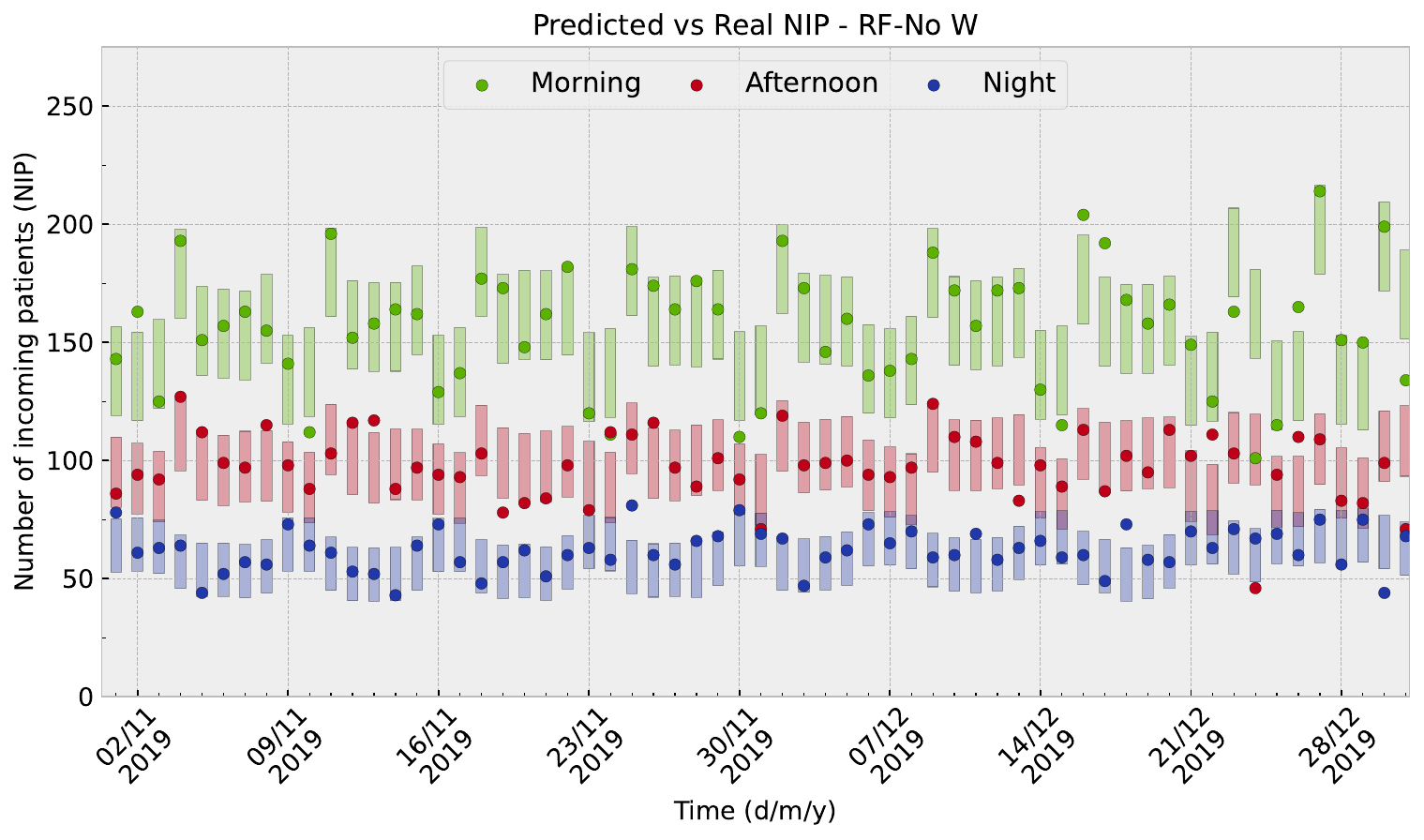}
\caption{The figures illustrate how predictions of the optimal model (RF-No W) align with true data in two distinct time windows (May-June in panel (a) and November-December in panel (b)). The points represent the actual NIP, while the colored bars indicate the RMSE around the predicted value.}
\end{figure}

\begin{figure}[!h]
\includegraphics[width=0.75\linewidth]{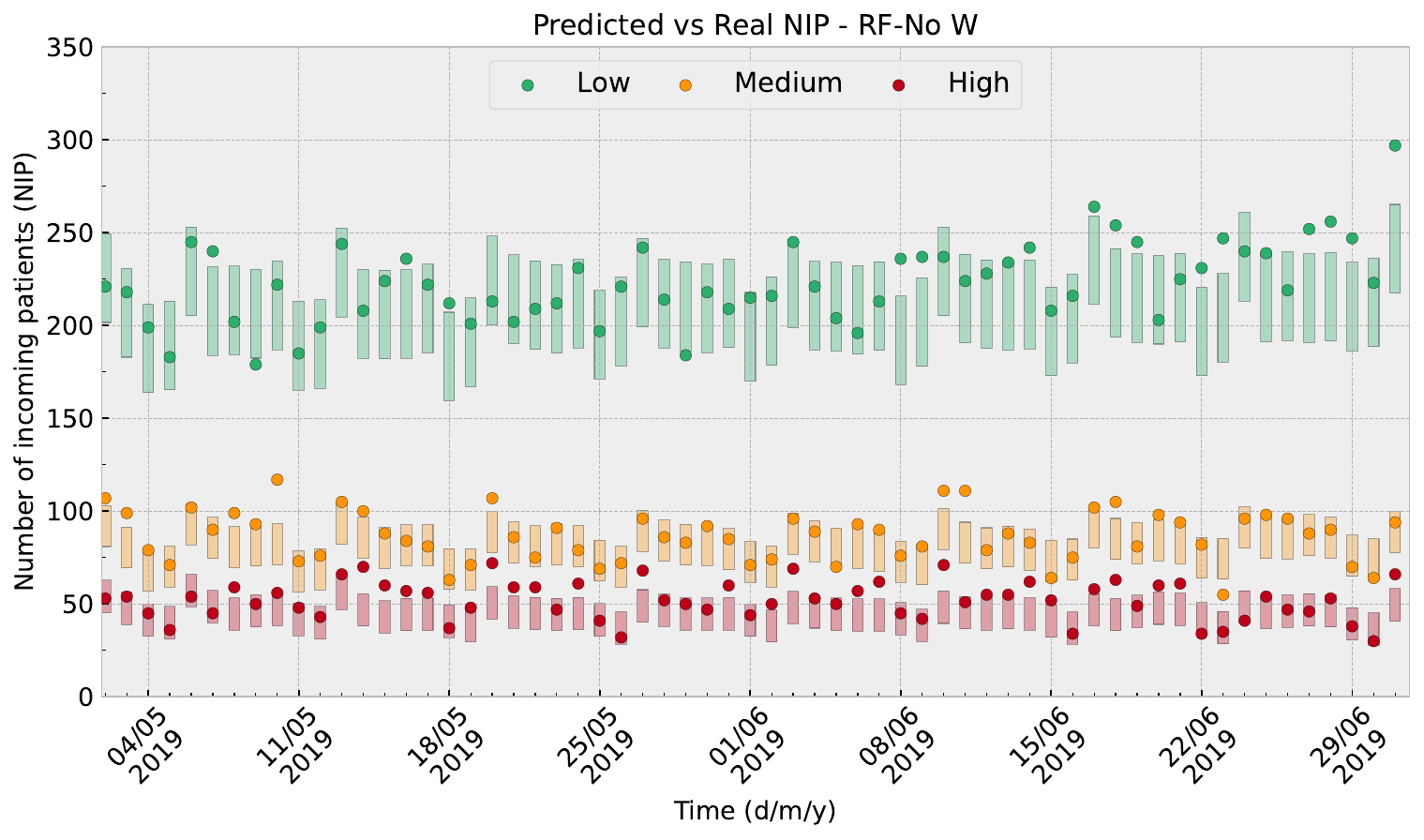}
\includegraphics[width=0.75\linewidth]{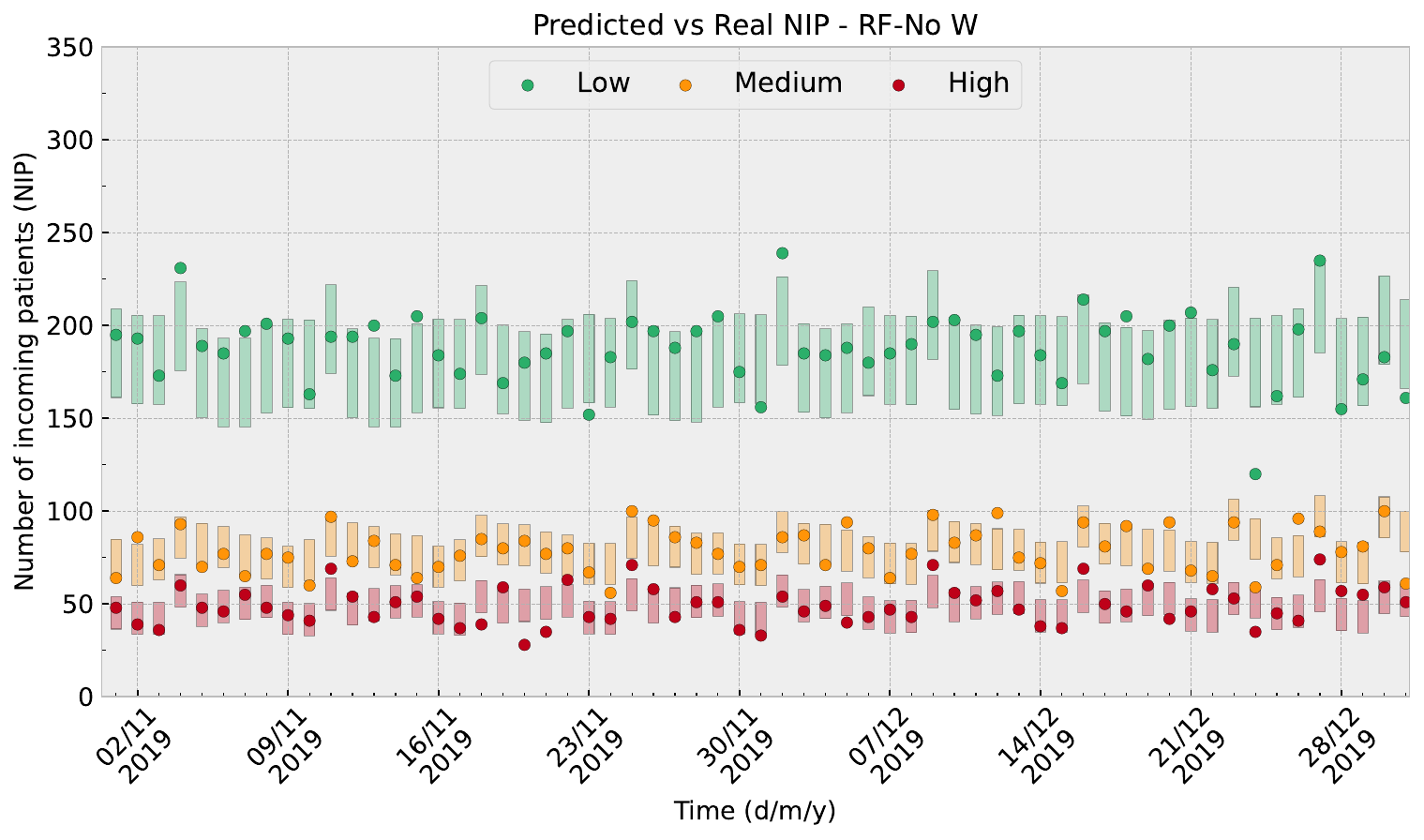}
\caption{These figures illustrate how predictions of the optimal model align with true data in two distinct time windows (May-June in panel (a) and November-December in panel (b)). The points represent the actual NIP, while the colored bars indicate the RMSE around the predicted value.}
\end{figure}

\subsection{Comparison with time-series models}

\begin{figure}[!h]
\includegraphics[width=0.49\linewidth]{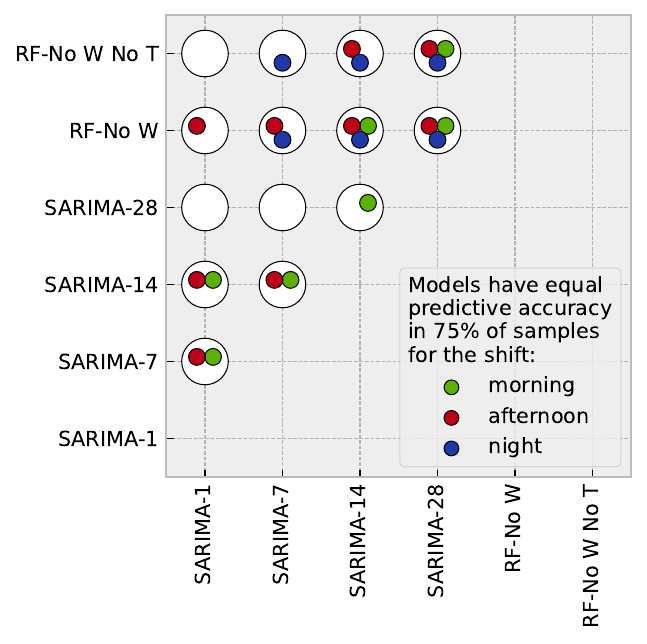}
\includegraphics[width=0.49\linewidth]{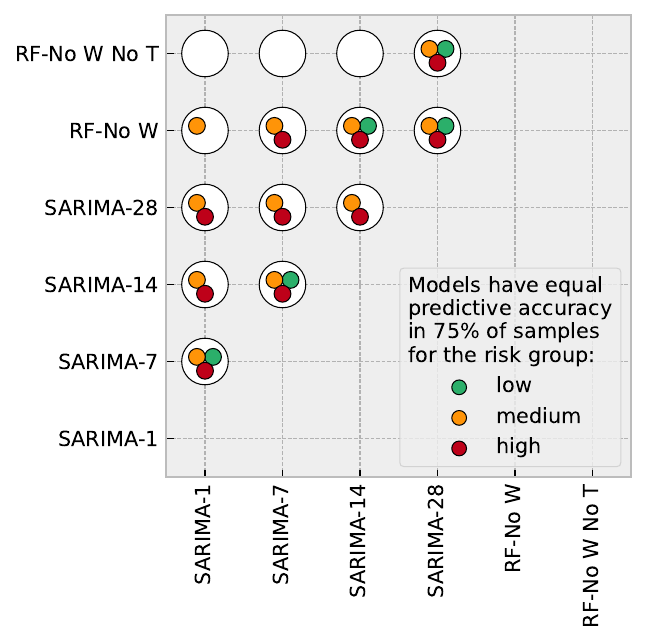}
\caption{This figure highlights which pairs of the following models have equal predictive accuracy according to the DM test (shift-based predictions in panel (a) and risk-group-based predictions in panel (b)): SARIMA with predictive horizons (1, 7, 14, 28), RF-No W, and RF-No W-No T. Each white disk corresponds to a different combination, as indicated on the axes. Since the test is symmetric, we only show each comparison once and hence the lower-right part of each diagram is empty. The colored circles inside the white disks indicate for which shifts or risk groups the two models represented by the disk have equal predictive accuracy for at least 75\% of the bootstrap samples.}
\end{figure}

The goal of this section is to compare the RF-No W and RF-No W-No T models with simple time-series models. As discussed in the Methods section, we compare our models with SARIMA and SARIMAX because these models are widely used in the relevant literature \cite{1, 2, 3, 4, 5, 6, 7, 8, 9, 10, 11, 12, 13, 14, 15, 16, 17, 18}. We parametrized these models as SARIMA$(1, 1, 1)(1, 1, 1)_7$ and SARIMAX$(1, 1, 1)(1, 1, 1)_7$ to represent a simple model that captures weekly oscillations in the NIP. Since the goal of this paper is to assess which external variables are relevant for non-time-series models for the prediction of NIP over long horizons we avoid a detailed search for the optimal model in the ARIMA family. We compare SARIMA and SARIMAX to both RF-No W and RF-No W-No T rather than only to RF-No W. This is because RF-No W-No T could also be a practical option for ED managers, as it avoids the use of the additional variable tourist data. We note though that the difference between their SMAPEs is statistically significant.

We now focus on the lower half of Table \ref{tab:mape}, showing the SMAPEs of time-series models. The full set of performance metrics (SMAPE, RMSE, MAE) and the associated errors over the bootstrap samples can be found in the Appendix (\ref{si:metric-table}). As expected, the SMAPEs of both SARIMA and SARIMAX increase with the prediction horizon. For all shifts and risk groups, the SMAPEs of SARIMAX are higher than the ones for SARIMA. This means that, for autoregressive models like SARIMAX, the inclusion of the exogenous variables lead to overfitting. Therefore, we will focus on SARIMA (no exogenous variables) for the rest of the study.

 Figure 8 highlights which pairs among the following models have equal predictive accuracy according to the DM test: SARIMA with predictive horizons (1, 7, 14, 28), RF-No W, and RF-No W-No T. The full set of results of the DM tests, including the exact percentages of samples that satisfy the null hypotheses, can be found in the Appendix (\ref{si:diebold-table}). Since the SMAPE of the SARIMA model increases with the prediction horizon, we determine from which horizon onward the predictive accuracy of SARIMA is equal to or below those of RF-No W and RF-No W-No T respectively.

 We start our discussion from the longest horizon. The predictive accuracy of SARIMA-28 is equal to that of both RF-No W and RF-No W-No T for any shift and risk group (Fig. 8). From this we expect that, for any longer horizon, the two RF models will have either equal or higher predictive accuracy than SARIMA.

 SARIMA-14 outperforms RF-No W-No T across all risk groups and for the morning shift, but has equal predictive accuracy as RF-No W for all cases. Thus, RF-No W already matches SARIMA in accuracy when forecasting 14 days ahead. This is also the case for SARIMA-7, excluding the morning shift and the low-risk group. This means that a simple model such as RF can achieve an equivalent accuracy as SARIMA, relying only on exogenous variables that are accessible or predictable.

\subsection{Testing the models on post-COVID data}

Making predictions of NIP in the post-COVID period would ideally involve retraining the models on post-COVID data. However, there is currently no sufficient post-COVID data (our dataset ends in December 2022, leaving us with only one year of post-COVID patient arrival numbers). For this reason, we limit ourselves to testing our models on post-COVID data without any retraining. In other words, we use the models trained on pre-COVID data as described above to make predictions for the post-COVID period. Our aim is to assess whether, and to what extent, model performance deteriorates as a result of the pandemic-induced changes in NIP patterns. Among the three non-time-series models, we only focus on the random forest model, since the latter outperformed both support vector regressors and feedforward neural networks in the pre-COVID period. For pre-COVID testing, the SARIMA model was trained on the year preceding the test dataset. This is impossible to do for post-COVID testing, since the year preceding the test dataset is the pandemic period. Thus, we tested the performance of the RF models and compared them with their performance during the pre-COVID period.

The full set of SMAPEs, RMSEs, and MAEs for the RF models tested on post-COVID data can be found in the Appendix (\ref{si:post-covid}). In particular, Table \ref{tab:post-covid-shift} shows the performance metrics for shift-based predictions, and Table \ref{tab:post-covid-risk} is for risk-group-based predictions. The SMAPEs worsen considerably compared to their pre-COVID values for all shifts and risk groups, excluding the night shift. For post-COVID data, we cannot identify a single optimal model. Different input combinations make the RF perform better on different shifts and risk groups, and all SMAPEs are compatible with each other according to their associated errors across the four input combinations.

In conclusion, our results highlight the importance of retraining the models using post COVID patient numbers as soon as enough data has been collected.

\section{Conclusions}

In this study, we have developed a non-time-series approach to predict patient volumes in emergency departments using exogenous variables and simple machine-learning techniques. At variance with time-series models, in the non-time-series models any increase in the prediction error arises solely from the inclusion of time-dependent exogenous variables, such as weather, or resident and tourist population. In contrast to weather, the resident and tourist populations are easily predictable on a fortnightly and monthly scale, a useful horizon for hospital resource allocation.

We found that random forests outperformed support vector regressors and feedforward neural networks for all shifts and risk groups. The RF-No W model, the random forest model using only calendar and population variables as inputs, was identified as the optimal model across the examined non-time-series models to predict the number of incoming patients for any shift or risk group. This choice was made by noting that RF-No W, for any shift or risk group, was either the model with the lowest SMAPE or had equivalent accuracy to the model with the lowest SMAPE.

The results allowed us to exclude weather forecasts as input variables. This is in contrast with a number of other studies \cite{1, 2, 8, 10, 12, 14, 15, 18, 20, 21, 22}, where weather proved to be a necessary exogenous variable for accurate forecasting of the number of incoming patients at emergency departments. We believe that our result is likely due to Mallorca's seasonal and mild weather, for which calendar data is already a good substitute as an input variable. 

We found the difference in SMAPE between the RF-No W and RF-No W-No T models always to be below 0.5. Since the daily NIP for each shift and risk group does exceeds 250 patients at the hospital we studied, this difference in SMAPE corresponds to fewer than two patients per shift or risk group per day. Although this difference is statistically significant according to the DM test, it is small from a practical point of view and has a limited impact on resource allocation planning. Therefore, the inclusion of the tourist population variable does not have the strong effect that might be expected given how the phenomenon is described by residents and mainstream media \cite{tourist1, tourist2}. This does not imply that tourism has no effect on patient arrivals; rather, due to its strong seasonal pattern, most of the information provided by tourist data is already captured by calendar variables. RF-No W-No T therefore remains a viable candidate for accurate predictions when tourist data is unavailable.

We also compared RF-No W and RF-No W-No T to time-series models at different predictive horizons ranging from 1 to 28 days. Because of the overfitting caused by calendar and tourist data, SARIMA proved to be more accurate than SARIMAX for each shift, risk group, and prediction horizon. RF-No W showed equal prediction accuracy than SARIMA for each shift and risk group for horizons of 14 and 28 days. This implies equal or better prediction accuracy for any horizon longer than 14 days. We acknowledge that the SARIMA model we used could be improved with a better choice of parameters, but our current analysis is sufficient to demonstrate that relatively simple non-time-series models show comparable performance as some of the time-series models frequently employed in the literature \cite{1, 2, 3, 4, 5, 6, 7, 8, 9, 10, 11, 12, 13, 14, 15, 16, 17, 18}. We also stress that any time-series model, including SARIMA, would frequently have to be fed with new data to maintain high accuracy. Non-time-series models, once fed with population and calendar variables (both predictable with low uncertainty), do not need to be retrained with new data unless major external events alter the NIP patterns (i.e. a pandemic or a long-lasting calamity). For this reason, while time-series models are preferable for day-to-day predictions, the optimal model we have identified (RF-No W) is a viable alternative for long-term resource and personnel allocation. 

A further conclusion of our work concerns the permanent impact of the pandemic on ED admission patterns. As shown in the Appendix (\ref{si:post-covid}), the SMAPEs of the Random Forest models, trained on pre-COVID data, increase significantly when tested on post-COVID data (night shift excluded). To overcome this issue, the models would have to be retrained on post-COVID data at a suitable point in the future, when there will be sufficient post-COVID data for adequate training, validation, and testing.

Although our methods can -- in principle -- be applied to other hospitals, we cannot expect our conclusions to be valid at all geographic locations. However, we believe that analogous results would likely be found in regions with similar weather and tourist-flow patterns (and provided hospital procedures are similar).

\subsection*{Acknowledgements}
We acknowledge Noemi P\'{e}rez Garc\'{\i}a and the Observatori de Dades Sanitàies, Serveis Centrals de Salut de les Illes Balears for providing the data used in this study. The funders played no role in the study design, data collection and analysis, decision to publish, or preparation of the manuscript. We also acknowledge the regional office of the State Meteorological Agency (AEMET) in the Balearic Islands for providing the weather forecast data.\\
Partial financial support has been received from Grants PID2021-122256NB-C21/C22 and PID2024-157493NB-C21/C22 funded by MICIU/AEI/10.13039/501100011033 and by “ERDF/EU”, and the María de Maeztu Program for units of Excellence in R\&D, grant CEX2021-001164-M. The project that gave rise to these results received the support of a fellowship from ``la Caixa'' Foundation (ID 100010434, https://lacaixafoundation.org/en/). The fellowship code is LCF/BQ/DI22/11940041. This is an award to PC
 
\subsection*{Author contributions statement}
PC, ARM, CRM, RT and TG designed the work. JJSS and ARM constructed the patient data. PC carried out the data analysis, and prepared all figures. ARM, CRM, RT and TG directed the work. PC wrote the first draft of the manuscript. PC, CRM, RT and TG revised and edited the manuscript. All authors reviewed the manuscript.
\subsection*{Data availability statement}
The data and code required to reproduce the study are available via a GitHub repository \url{https://github.com/complexParide/son_espases} \cite{repository}. This includes detailed shift-by-shift patient arrival numbers on each day in the training, validation and test periods, as well as daily patient arrivals in the low, medium and high risk groups. The input data for the prediction algorithm (calendar, population and weather) is also available in the repository.

\bibliography{bib}
\appendix
\makeatletter
\renewcommand{\thefigure}{\Alph{section}\arabic{figure}}
\renewcommand{\thetable}{A\arabic{table}} 
\@addtoreset{figure}{section}
\@addtoreset{table}{section}
\makeatother
\newpage

\section{}
\subsection{Original data columns}
\label{si:original-cols}
\noindent\textit{Año} (year of the visit),\\
\textit{Servicio alta} (service of registration. Most of the entries are \textit{urgencias}),\\
\textit{Sexo} (sex of the patient at the time of the visit)), \\\textit{Edad} (age of the patient at the time of the visit), \\\textit{Fecha y hora inicio urg} (date and time of entry at the ED), \\
\textit{Fecha y Hora Fin Urg} (date and time of exit from the ED.),\\
\textit{Tipo paciente} (if the patient is pediatric or not), \\
\textit{Procedencia} (why the patient attended the ED, i.e. their own decision, GP's decision, from another hospital...), \\
\textit{Motivo alta} (reason for discharge, i.e. hospitalized, dismissed, dead, run away...), \\
\textit{Triatge data hm} (time of triage, when applicable), \\
\textit{Nacionalidad}: (nationality of the patient), \\
\textit{Provincia residencia} (province of residence of the patient if they are a Spanish resident), \\
\textit{Pais residencia} (country of residence of the patient), \\
\textit{Centro salud} (health center associated with the patient)

\clearpage 

\subsection{ NIP as a function of time, residents vs non-residents}
\label{si:nip-tourists}

\begin{figure}[H]
\includegraphics[width=0.49\linewidth]{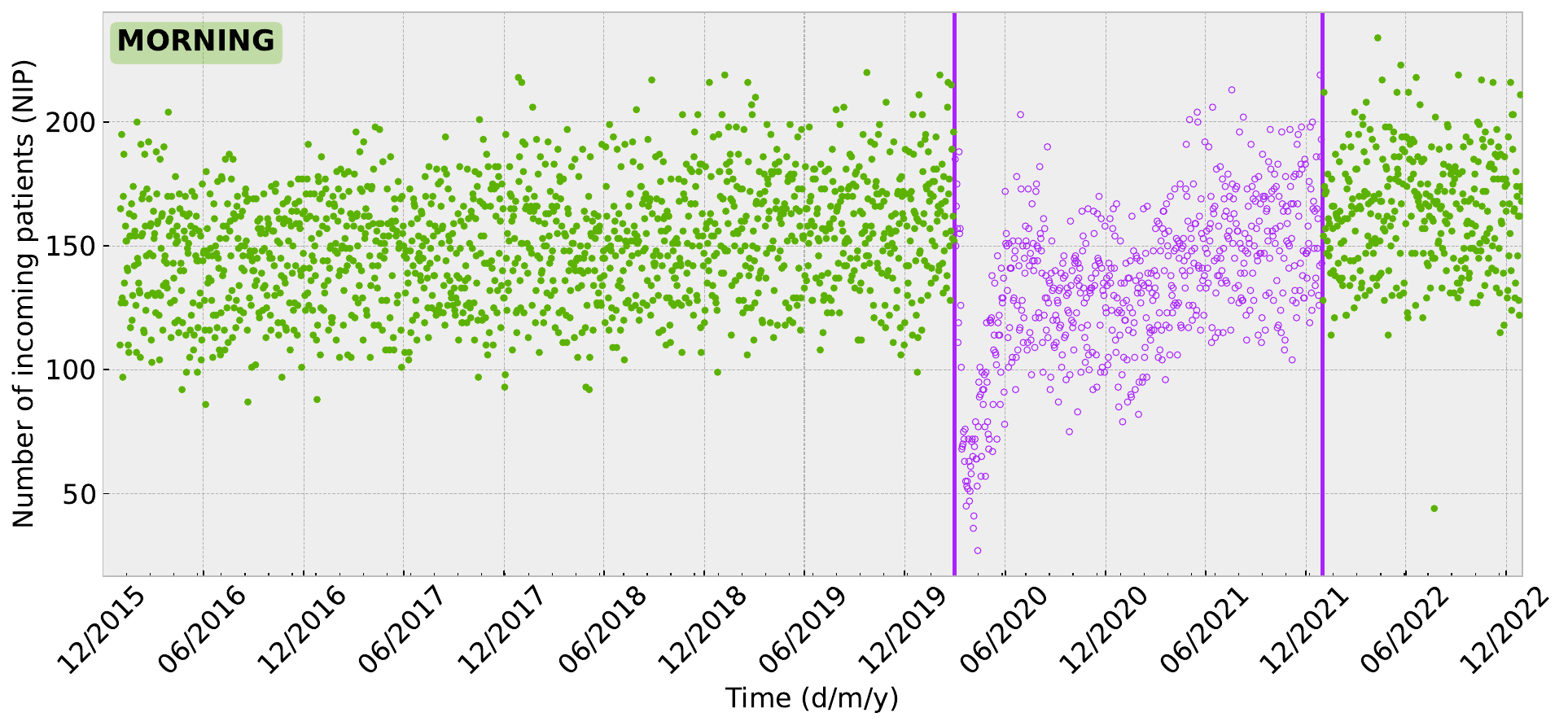}
\includegraphics[width=0.49\linewidth]{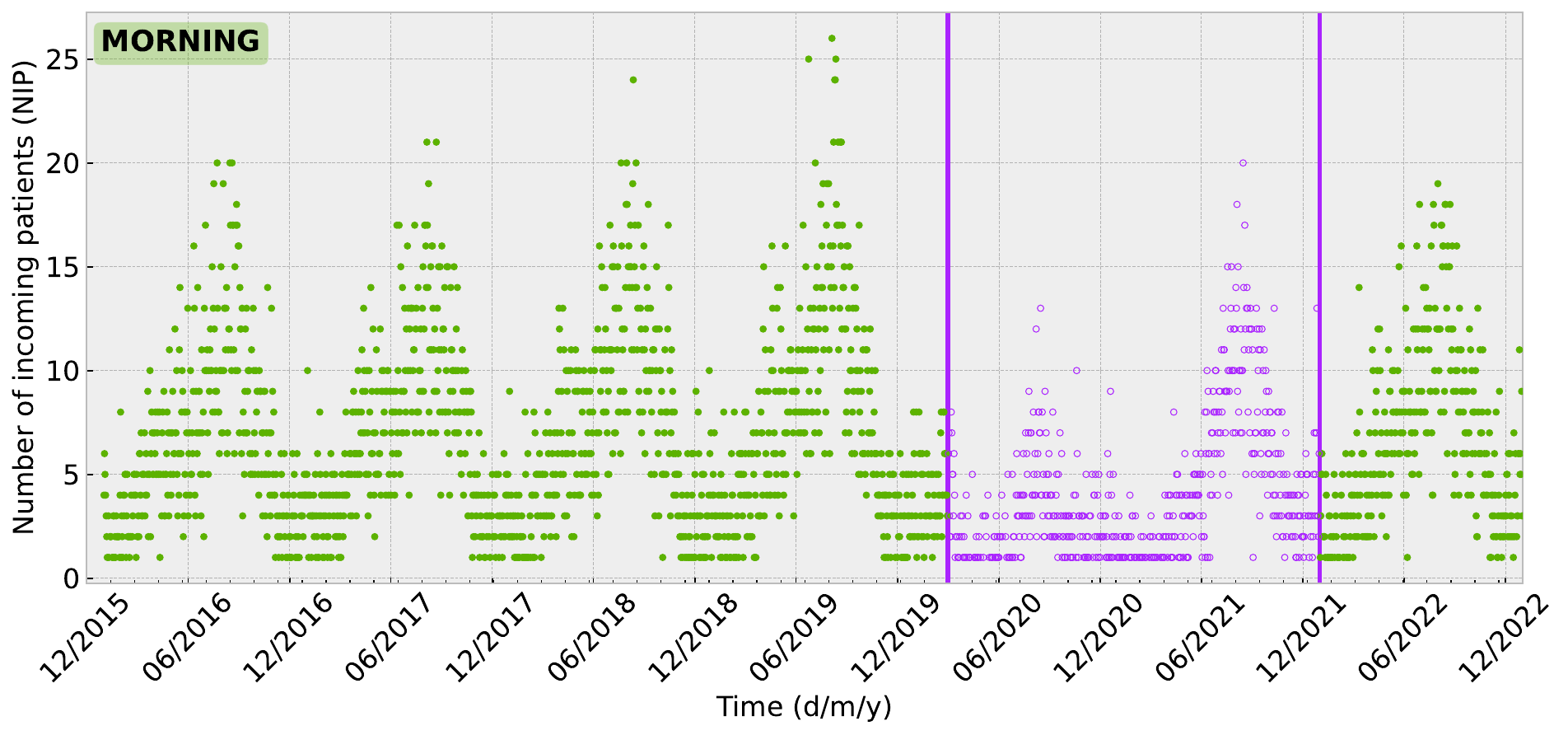}\\

\includegraphics[width=0.49\linewidth]{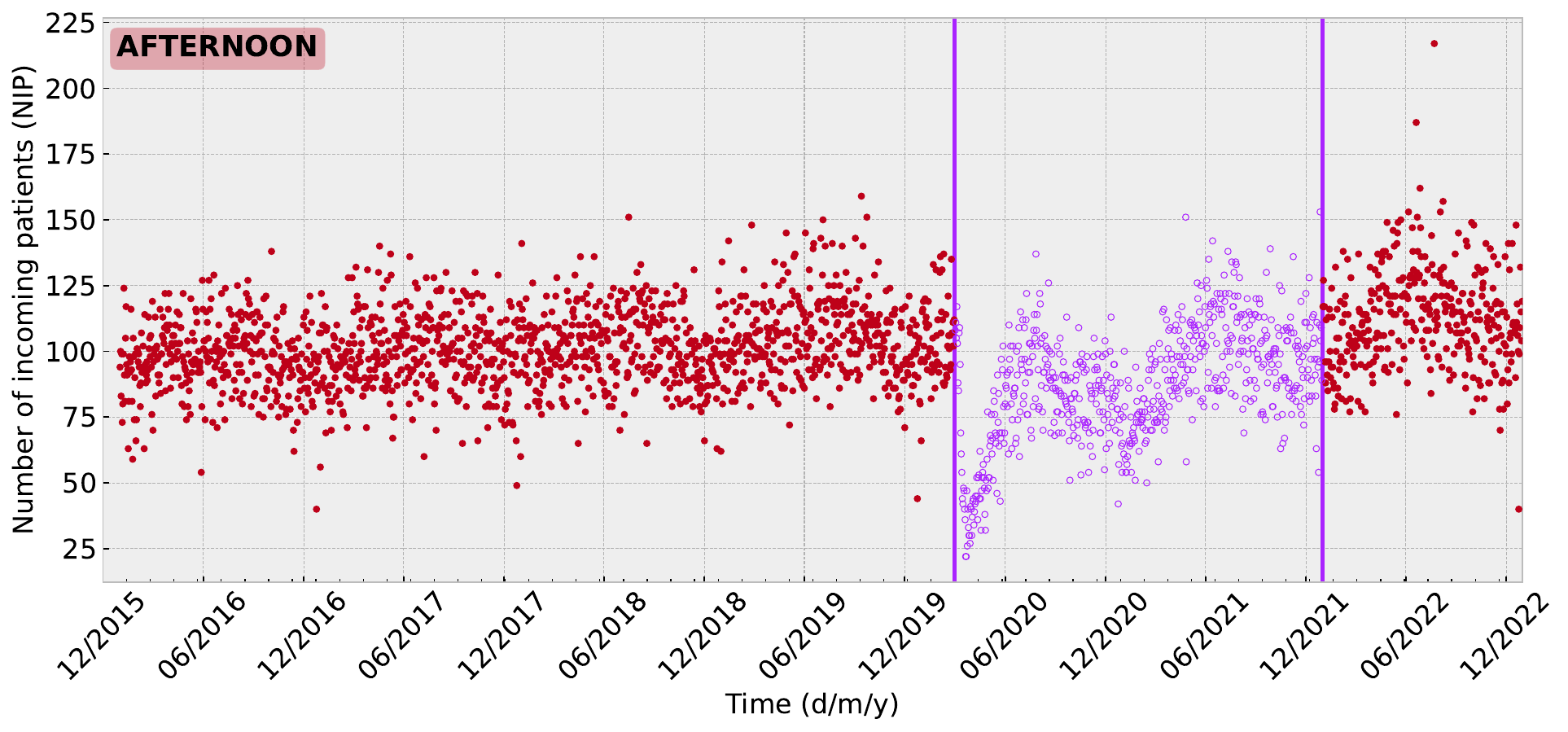}
\includegraphics[width=0.49\linewidth]{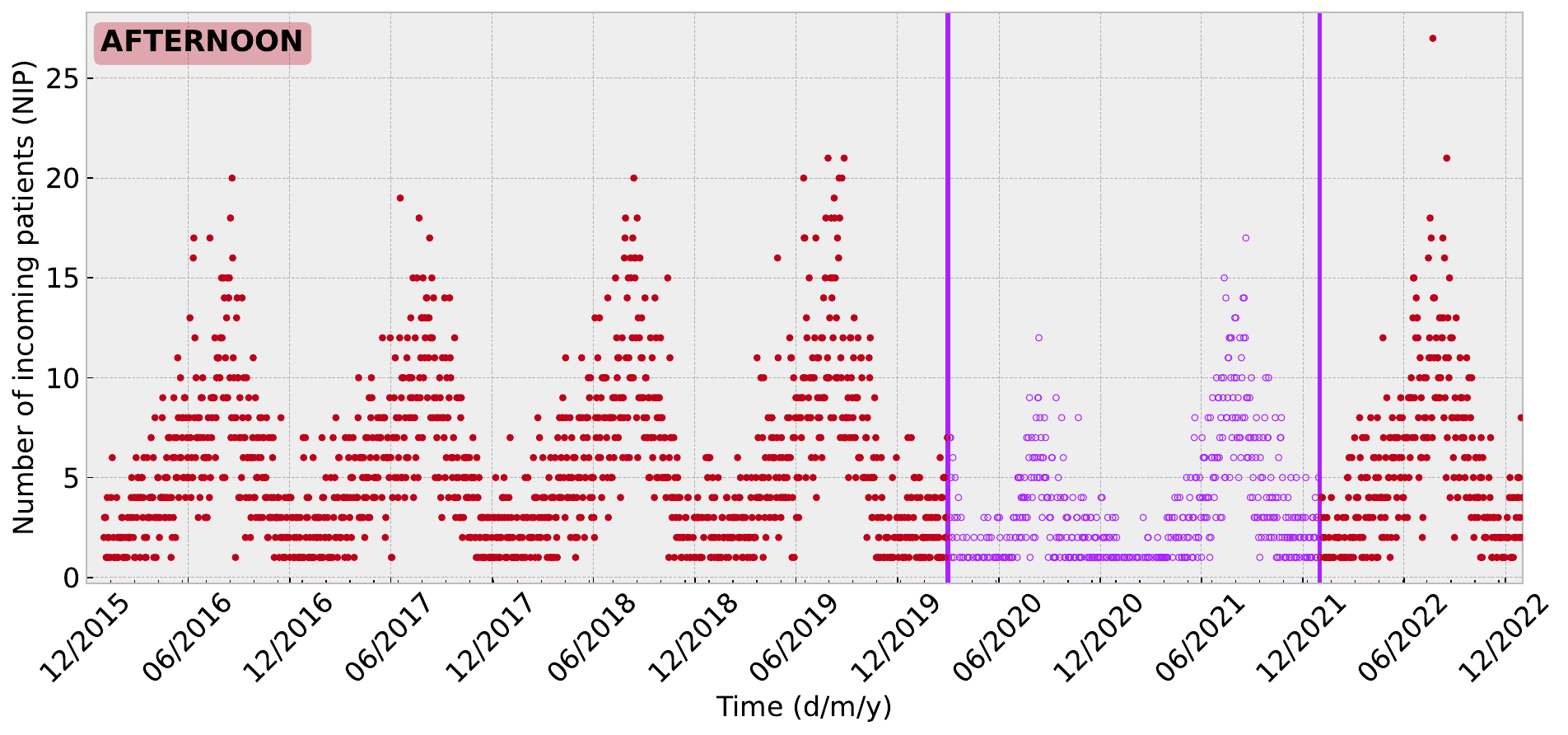}\\

\includegraphics[width=0.49\linewidth]{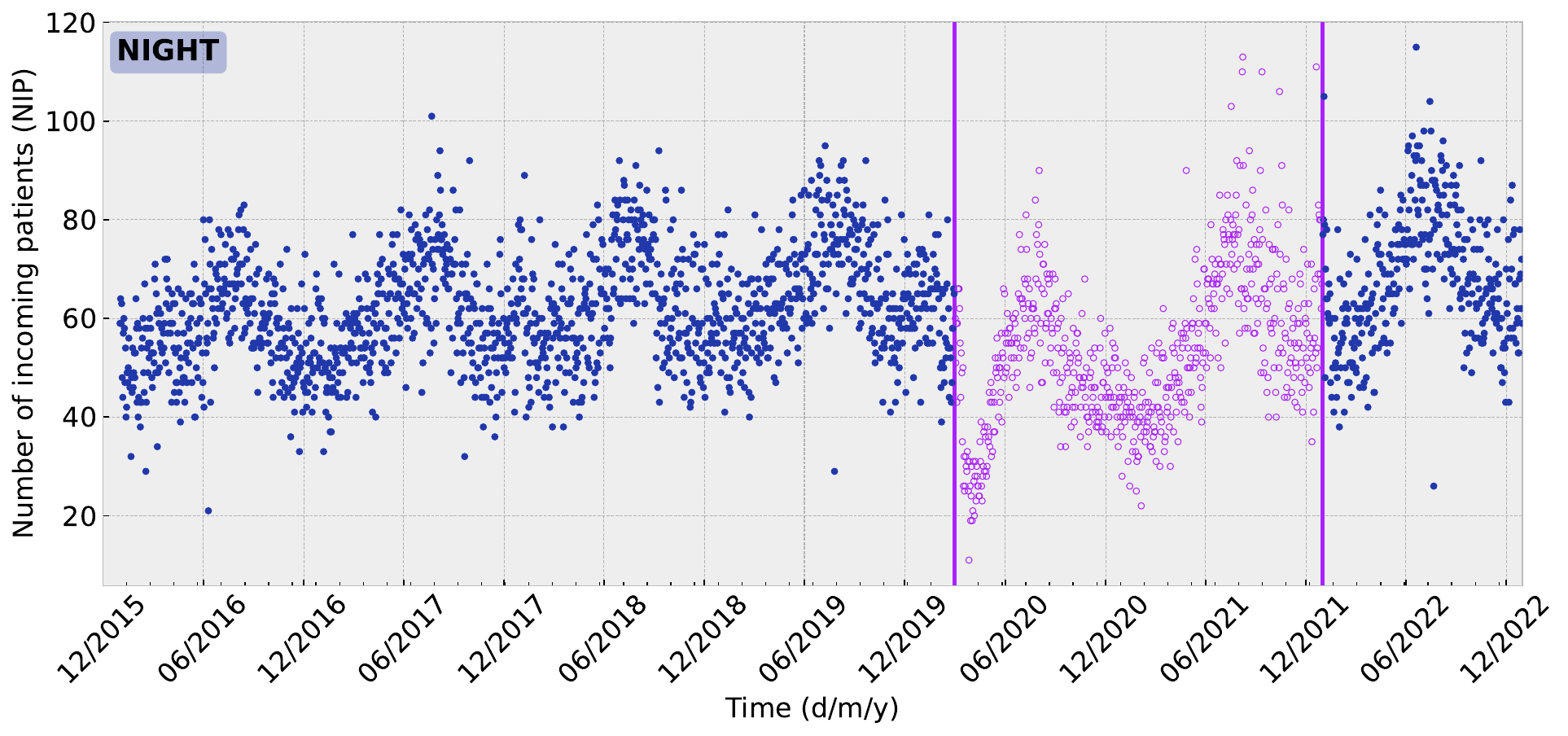}
\includegraphics[width=0.49\linewidth]{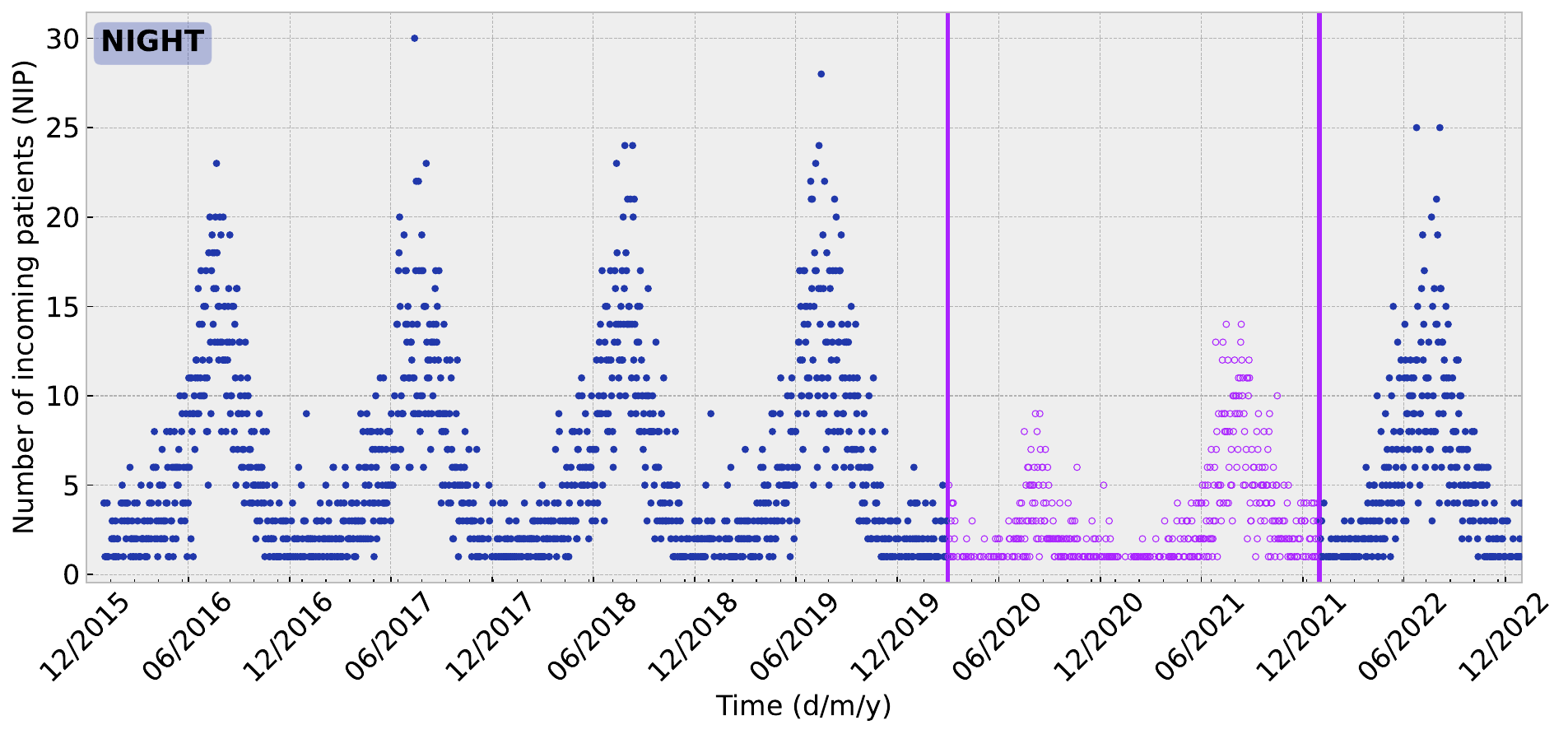}\\
\caption{NIP as a function of time, residents vs non-residents. Subfigures on the left show the NIP for residents, and subfigures on the right show the NIP for non-residents. Each point corresponds to the NIP for a specific day and shift. Each subfigure shows a different shift (morning in green, afternoon in red, night in blue). The purple points between the dates March 1, 2020 and December 31, 2021 are the values registered during the assumed pandemic period, and excluded from our analysis.}
\end{figure}

\clearpage 

\subsection{NIP as a function of time, females vs males}
\label{si:nip-sex}

\begin{figure}[H]
\includegraphics[width=0.49\linewidth]{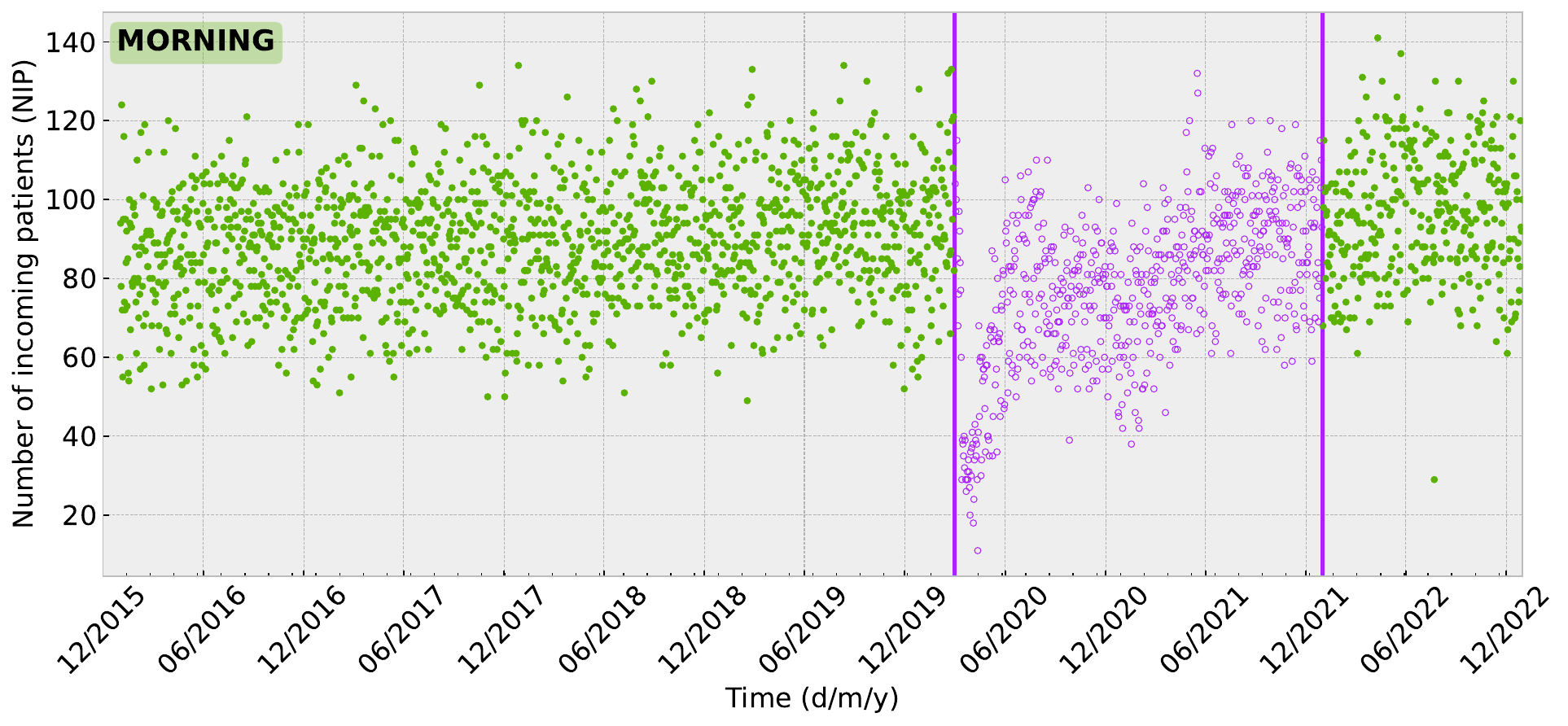}
\includegraphics[width=0.49\linewidth]{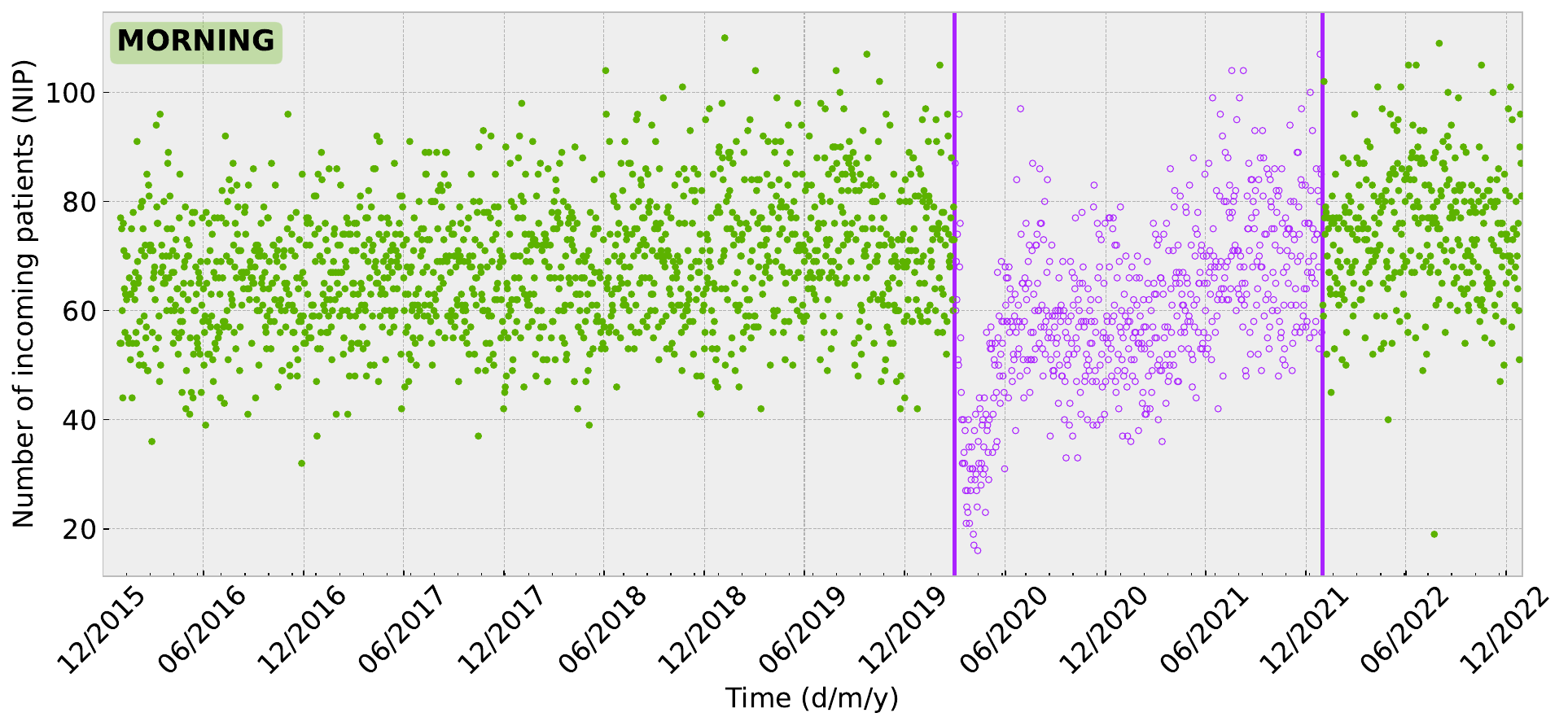}\\

\includegraphics[width=0.49\linewidth]{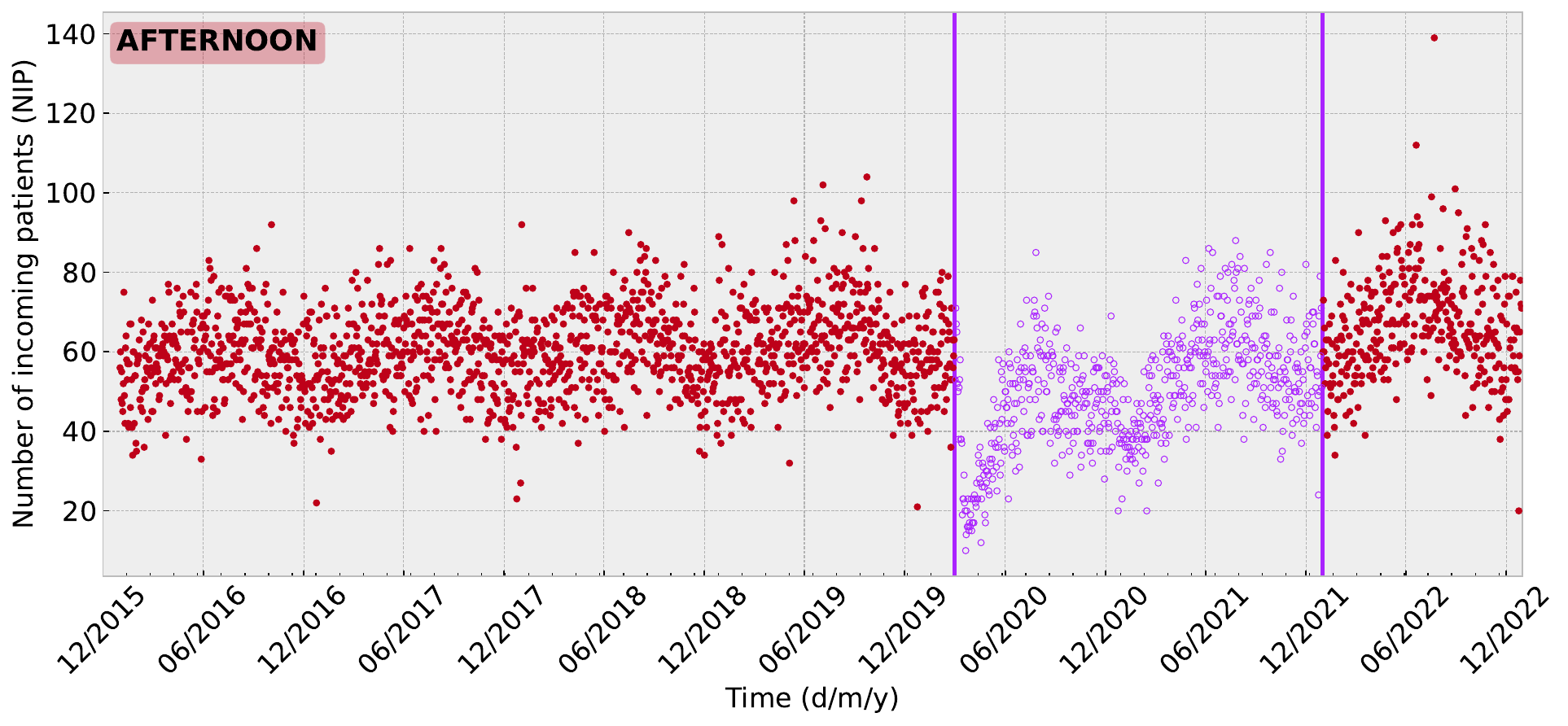}
\includegraphics[width=0.49\linewidth]{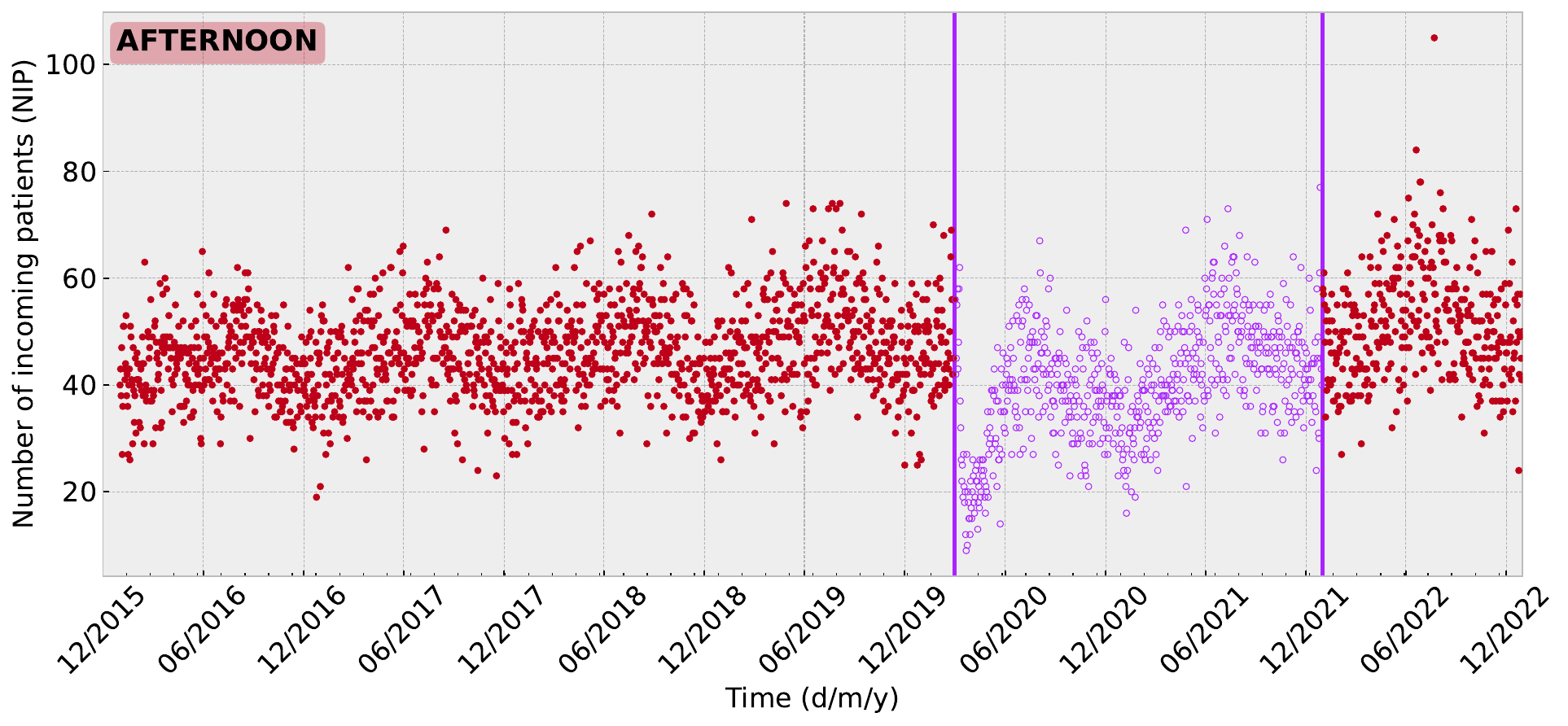}\\

\includegraphics[width=0.49\linewidth]{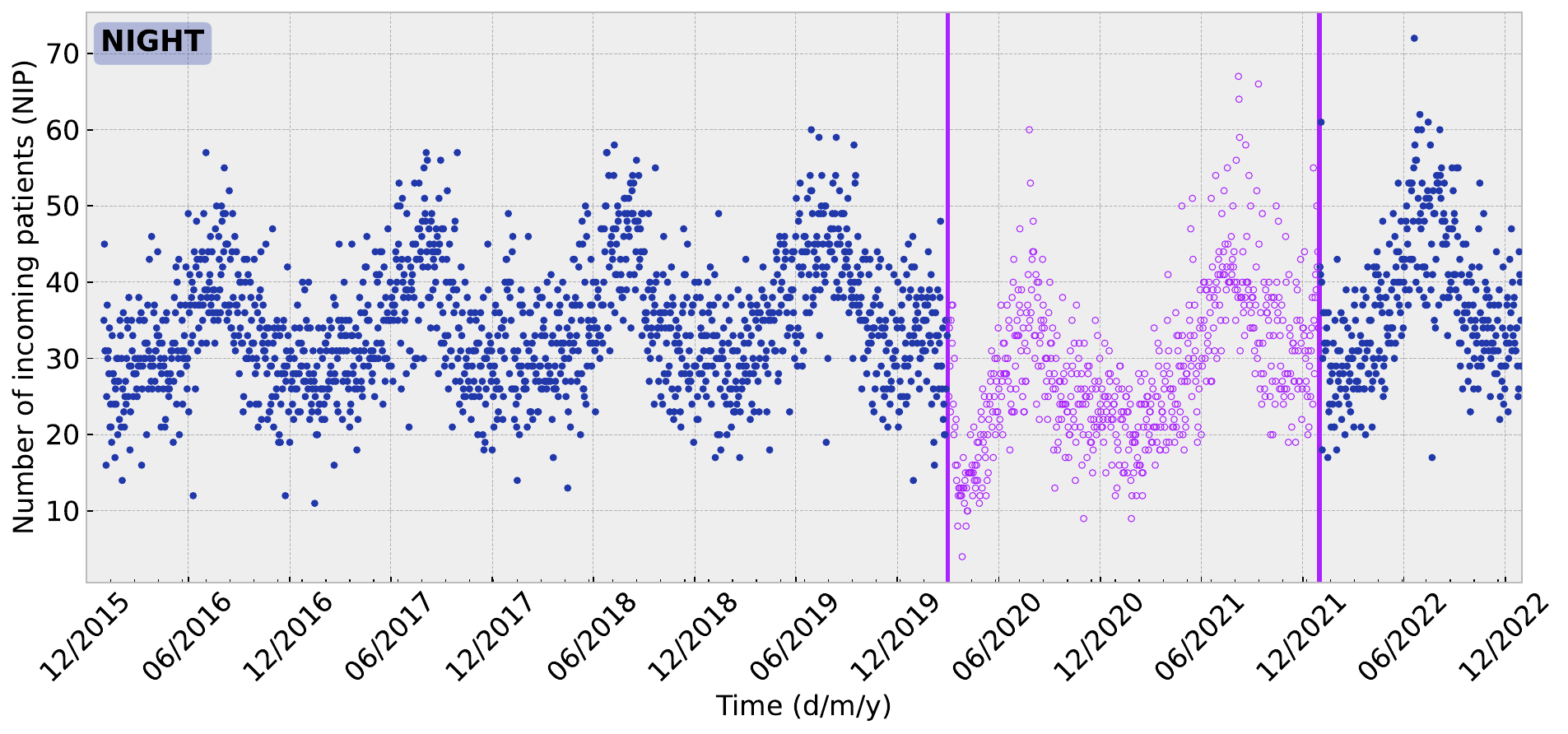}
\includegraphics[width=0.49\linewidth]{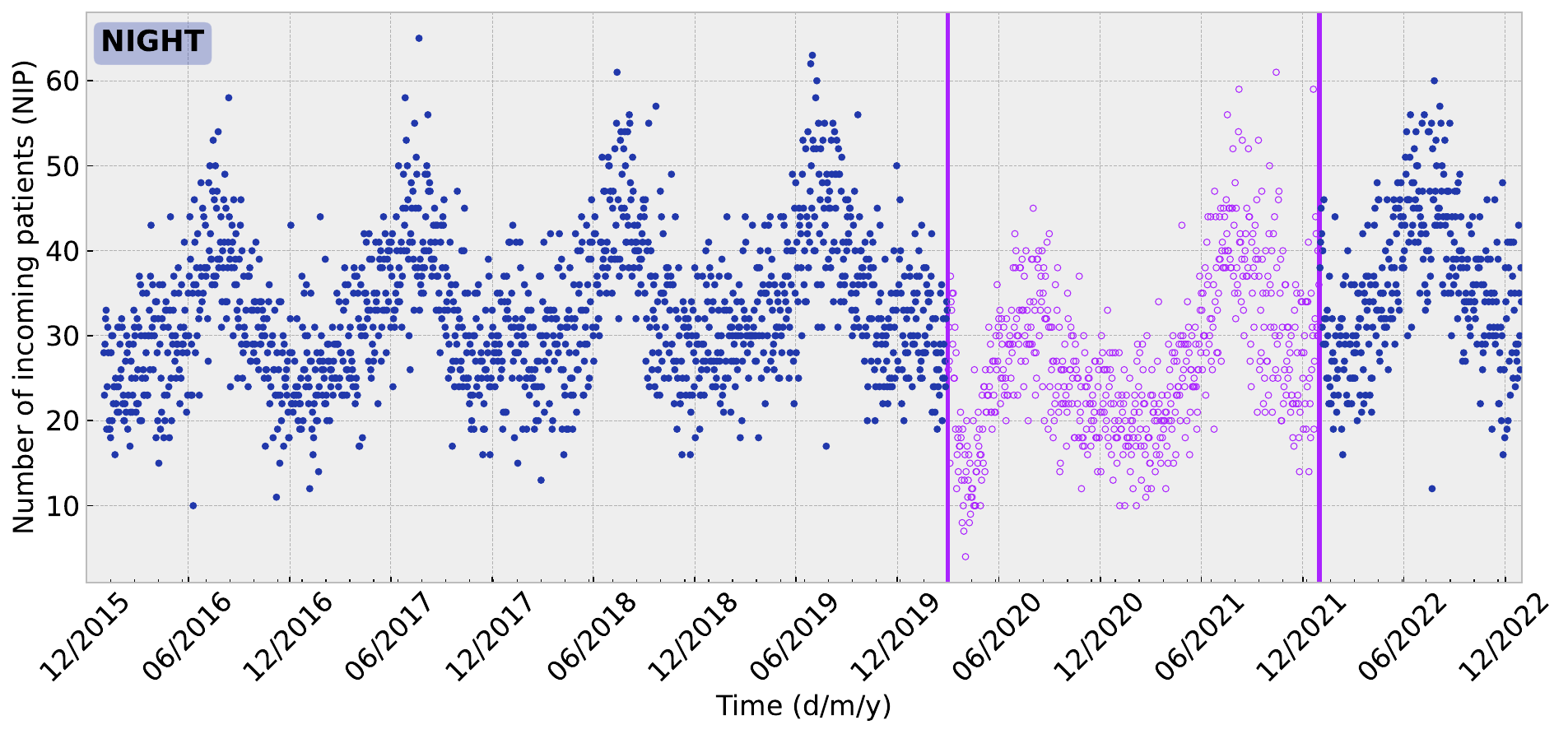}\\
\caption{NIP as a function of time, females vs males. Subfigures on the left show the NIP for female patients, while subfigures on the right show the NIP for male patients. Each point corresponds to the NIP for a specific day and shift. Each subfigure shows a different shift (morning in green, afternoon in red, night in blue). The purple points between the dates March 1, 2020 and December 31, 2021 are the values registered during the pandemic, and excluded from our analysis. The behaviour of the curves on the left and right respectively are very similar to one another, thus eliminating the need to develop a different model for each sex.}
\end{figure}

\subsection{DeepSeek prompt to preprocess weather data}
\label{si:ai-prompt}

You need to extract from the following spanish text information about the wind speed and the precipitation probability. You have to encode it as integer numbers from 0 to 4 as follows:
\medskip
\\
0 = none\\
1 = low\\
2 = medium\\
3 = high\\
4 = very high\\
\medskip

\noindent Reply just with two numbers separated by a single space (nothing else in your output message!) where the first one is the precipitation probability and the second one is the wind speed that you infer from the text, both of them from 0 to 4 as described above.

\subsection{Hyperparameter tuning for RF and SVR models}
\label{si:rf-svr-tuning}

\begin{figure}[H]
\includegraphics[width=\linewidth]{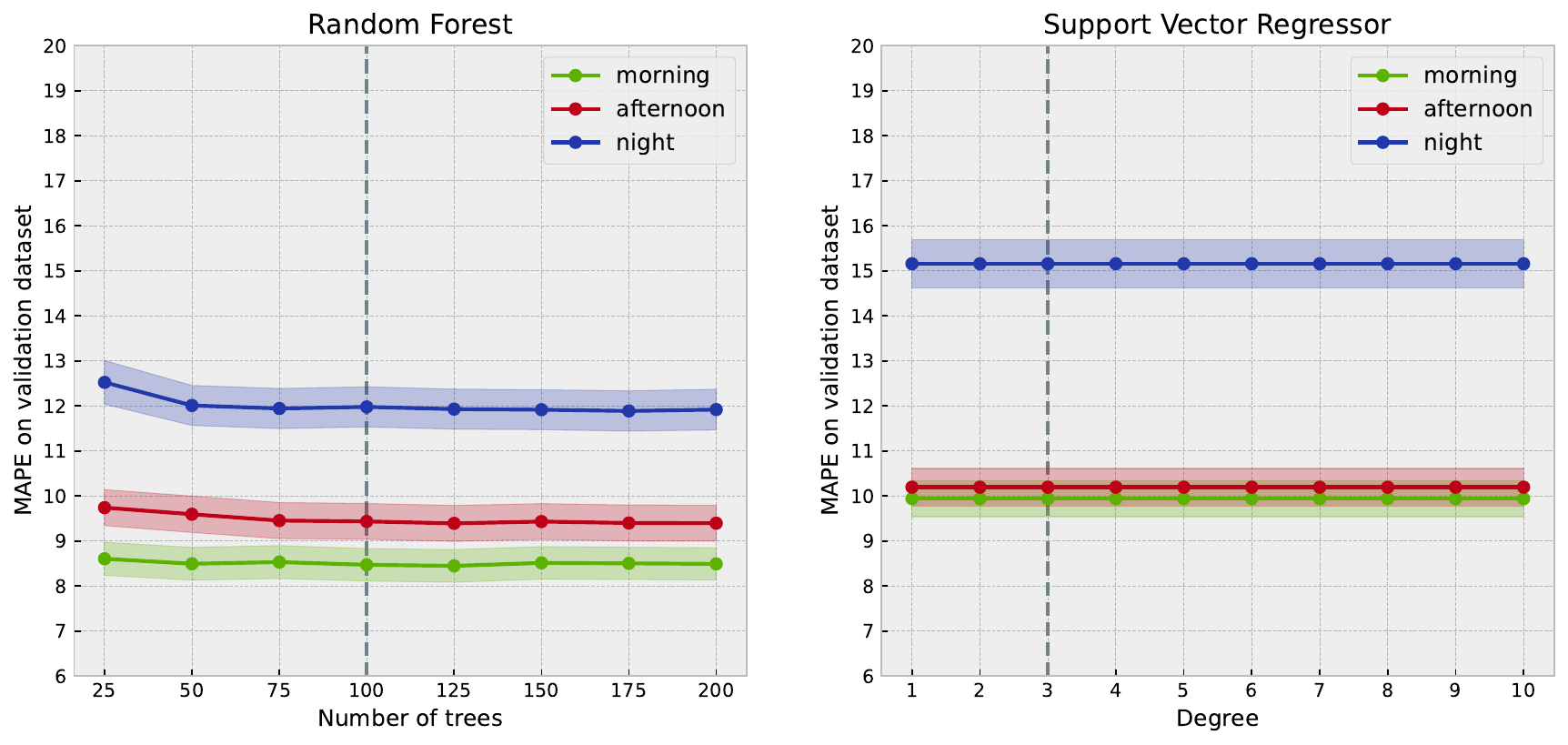}
\caption{Hyperparameter tuning for RF and SVR models. Subfigure on the left shows the SMAPE for the validation dataset obtained from the RF model for different values of the \textit{n\_estimators} hyperparameter. Subfigure on the right shows the SMAPE for the validation dataset obtained from the SVR model for different values of the \textit{degree} hyperparameter. The vertical grey lines indicate the optimal values.}
\end{figure}

\subsection{FNN detailed structure}
\label{si:fnn}
\begin{verbatim}
FNN = nn.Sequential(nn.Linear(X.shape[1], 32), nn.ReLU(),
 nn.Linear(32, 64), nn.ReLU(),
 nn.Linear(64, 128), nn.ReLU(),
 nn.Linear(128, 256), nn.ReLU(),
 nn.Linear(256, 512), nn.ReLU(),
 nn.Linear(512, 1024), nn.ReLU(),
 nn.Linear(1024, 2048), nn.ReLU(),
 nn.Linear(2048, 1024), nn.ReLU(),
 nn.Linear(1024, 512), nn.ReLU(),
 nn.Linear(512, 256), nn.ReLU(),
 nn.Linear(256, 128), nn.ReLU(),
 nn.Linear(128, 64), nn.ReLU(),
 nn.Linear(64, 32), nn.ReLU(),
 nn.Linear(32, Y.shape[1])
 )

criterion = nn.MSELoss()
optimizer = torch.optim.Adam(FNN.parameters(), lr=0.01)
\end{verbatim}

Here \texttt{nn} corresponds to the module \texttt{torch.nn}. The main hyperparameter is the number of training epochs. To tune this hyperparameter, we ran the model for 500 epochs and checked the loss function of both training and validation sets at each epoch. In order to minimize overfitting, we set the number of epochs to 130 (see \ref{si:fnn-tuning}).

\subsection{Hyperparameter tuning for the FNN model}
\label{si:fnn-tuning}

\begin{figure}[H]
\centering
\includegraphics[width=0.666\linewidth]{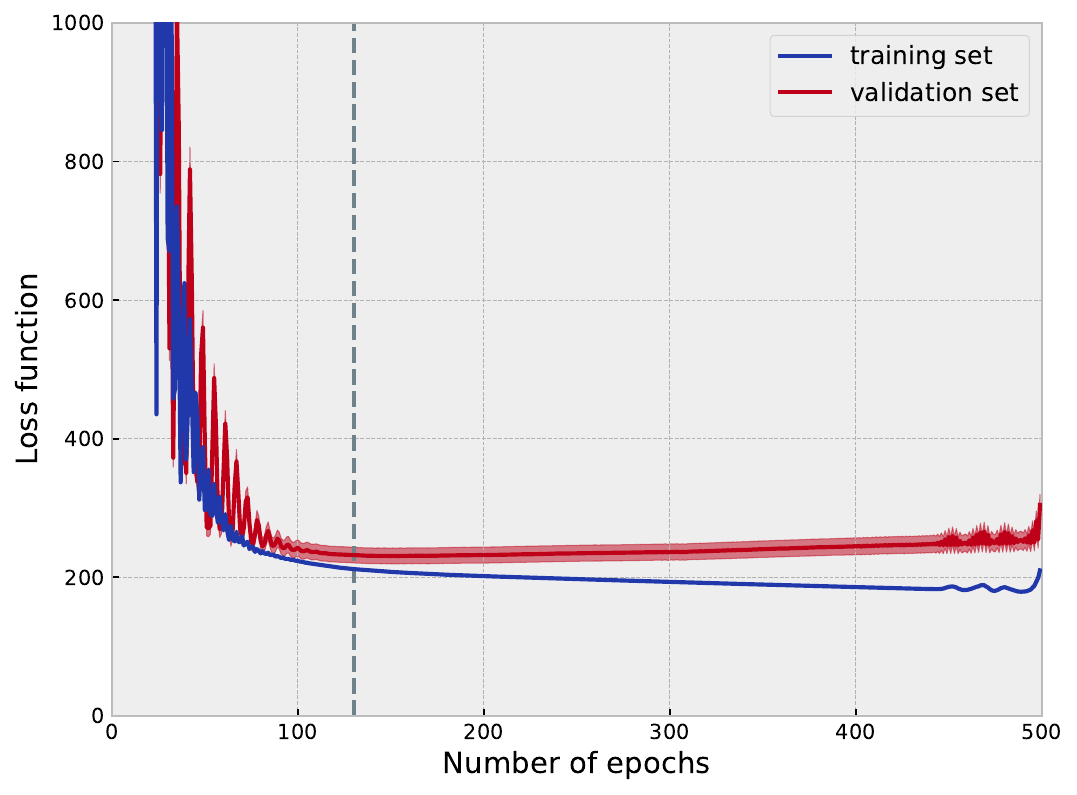}
\caption{Hyperparameter tuning for the FNN model. The figure shows the loss function at different training epochs for both the training (blue curve) and validation (red curve) dataset. The dashed vertical grey line indicates the optimal value, corresponding to the minimum of the validation curve.}
\end{figure}

\subsection{Full set of SMAPEs, RMSEs, and MAEs}
\label{si:metric-table}

\footnotesize
\begin{longtable}[c]{|c|c|c|c|c|c|c|c|c|}
\hline
\textbf{Method} & \textbf{\begin{tabular}[c]{@{}c@{}}Input\\ variable\end{tabular}} & \textbf{Shift} & \textbf{\begin{tabular}[c]{@{}c@{}}SMAPE\\ (mean)\end{tabular}} & \textbf{\begin{tabular}[c]{@{}c@{}}SMAPE\\ (st.dev.)\end{tabular}} & \textbf{\begin{tabular}[c]{@{}c@{}}RMSE\\ (mean)\end{tabular}} & \textbf{\begin{tabular}[c]{@{}c@{}}RMSE\\ (st.dev.)\end{tabular}} & \textbf{\begin{tabular}[c]{@{}c@{}}MAE\\ (mean)\end{tabular}} & \textbf{\begin{tabular}[c]{@{}c@{}}MAE\\ (st.dev.)\end{tabular}} \\ \hline
\endfirsthead
\endhead
SARIMA-1 & & Morning & 7.69 & 0.36 & 16.44 & 0.79 & 12.46 & 0.56 \\ \hline
SARIMA-1 & & Afternoon & 9.14 & 0.41 & 13.24 & 0.61 & 10.16 & 0.43 \\ \hline
SARIMA-1 & & Night & 10.48 & 0.48 & 9.70 & 0.57 & 7.47 & 0.33 \\ \hline
SARIMAX-1 & & Morning & 8.16 & 0.37 & 17.25 & 0.80 & 13.32 & 0.59 \\ \hline
SARIMAX-1 & & Afternoon & 10.99 & 0.47 & 15.54 & 0.67 & 12.15 & 0.49 \\ \hline
SARIMAX-1 & & Night & 12.23 & 0.51 & 10.91 & 0.49 & 8.66 & 0.35 \\ \hline
SARIMA-7 & & Morning & 7.84 & 0.37 & 16.74 & 0.78 & 12.73 & 0.57 \\ \hline
SARIMA-7 & & Afternoon & 9.47 & 0.41 & 13.58 & 0.59 & 10.48 & 0.44 \\ \hline
SARIMA-7 & & Night & 11.39 & 0.51 & 10.44 & 0.56 & 8.12 & 0.35 \\ \hline
SARIMAX-7 & & Morning & 12.02 & 0.57 & 29.79 & 2.60 & 20.49 & 1.11 \\ \hline
SARIMAX-7 & & Afternoon & 17.16 & 0.94 & 34.01 & 3.44 & 20.36 & 1.40 \\ \hline
SARIMAX-7 & & Night & 19.28 & 0.90 & 22.67 & 2.98 & 14.21 & 0.93 \\ \hline
SARIMA-14 & & Morning & 8.12 & 0.37 & 17.16 & 0.77 & 13.18 & 0.58 \\ \hline
SARIMA-14 & & Afternoon & 9.74 & 0.41 & 13.73 & 0.58 & 10.76 & 0.44 \\ \hline
SARIMA-14 & & Night & 12.28 & 0.51 & 11.10 & 0.57 & 8.79 & 0.36 \\ \hline
SARIMAX-14 & & Morning & 17.23 & 0.90 & 55.97 & 9.10 & 31.30 & 2.48 \\ \hline
SARIMAX-14 & & Afternoon & 26.66 & 1.75 & 75.35 & 9.76 & 35.42 & 3.53 \\ \hline
SARIMAX-14 & & Night & 31.02 & 1.49 & 44.66 & 6.52 & 23.64 & 1.99 \\ \hline
SARIMA-28 & & Morning & 8.52 & 0.38 & 17.90 & 0.76 & 13.78 & 0.60 \\ \hline
SARIMA-28 & & Afternoon & 10.48 & 0.45 & 14.95 & 0.58 & 11.61 & 0.48 \\ \hline
SARIMA-28 & & Night & 13.58 & 0.59 & 12.73 & 0.61 & 9.83 & 0.43 \\ \hline
SARIMAX-28 & & Morning & 30.22 & 1.56 & 109.52 & 19.23 & 55.46 & 5.11 \\ \hline
SARIMAX-28 & & Afternoon & 50.23 & 2.81 & 174.03 & 21.72 & 78.49 & 8.19 \\ \hline
SARIMAX-28 & & Night & 63.86 & 3.07 & 95.48 & 12.80 & 48.09 & 4.23 \\ \hline
RF & All & Morning & 9.21 & 0.36 & 18.65 & 0.71 & 14.92 & 0.60 \\ \hline
RF & All & Afternoon & 10.02 & 0.44 & 14.77 & 0.70 & 11.08 & 0.49 \\ \hline
RF & All & Night & 11.73 & 0.50 & 10.88 & 0.51 & 8.27 & 0.36 \\ \hline
RF & No W & Morning & 9.03 & 0.36 & 18.33 & 0.71 & 14.58 & 0.60 \\ \hline
RF & No W & Afternoon & 10.29 & 0.45 & 15.03 & 0.71 & 11.35 & 0.50 \\ \hline
RF & No W & Night & 12.08 & 0.49 & 11.07 & 0.49 & 8.54 & 0.36 \\ \hline
RF & No T & Morning & 9.48 & 0.36 & 18.97 & 0.70 & 15.28 & 0.60 \\ \hline
RF & No T & Afternoon & 10.79 & 0.47 & 15.67 & 0.72 & 11.86 & 0.53 \\ \hline
RF & No T & Night & 13.65 & 0.55 & 12.38 & 0.53 & 9.57 & 0.40 \\ \hline
RF & No W-No T & Morning & 9.43 & 0.36 & 18.99 & 0.71 & 15.28 & 0.59 \\ \hline
RF & No W-No T & Afternoon & 10.79 & 0.47 & 15.56 & 0.72 & 11.88 & 0.52 \\ \hline
RF & No W-No T & Night & 11.90 & 0.48 & 10.87 & 0.48 & 8.43 & 0.34 \\ \hline
SVR & All & Morning & 12.97 & 0.47 & 25.91 & 0.95 & 21.10 & 0.82 \\ \hline
SVR & All & Afternoon & 13.33 & 0.53 & 19.32 & 0.81 & 14.75 & 0.62 \\ \hline
SVR & All & Night & 18.53 & 0.68 & 17.00 & 0.66 & 13.24 & 0.56 \\ \hline
SVR & No W & Morning & 13.28 & 0.48 & 26.45 & 0.96 & 21.57 & 0.83 \\ \hline
SVR & No W & Afternoon & 13.44 & 0.53 & 19.45 & 0.81 & 14.87 & 0.62 \\ \hline
SVR & No W & Night & 18.42 & 0.69 & 16.98 & 0.66 & 13.17 & 0.56 \\ \hline
SVR & No T & Morning & 12.95 & 0.47 & 25.87 & 0.95 & 21.07 & 0.82 \\ \hline
SVR & No T & Afternoon & 13.74 & 0.54 & 19.79 & 0.81 & 15.18 & 0.63 \\ \hline
SVR & No T & Night & 21.25 & 0.77 & 19.00 & 0.70 & 14.95 & 0.62 \\ \hline
SVR & No W-No T & Morning & 13.43 & 0.48 & 26.72 & 0.97 & 21.81 & 0.84 \\ \hline
SVR & No W-No T & Afternoon & 14.58 & 0.56 & 20.76 & 0.83 & 16.03 & 0.66 \\ \hline
SVR & No W-No T & Night & 24.76 & 0.85 & 21.26 & 0.73 & 17.04 & 0.67 \\ \hline
FNN & All & Morning & 13.21 & 0.48 & 26.26 & 0.96 & 21.47 & 0.83 \\ \hline
FNN & All & Afternoon & 12.95 & 0.52 & 18.89 & 0.80 & 14.36 & 0.61 \\ \hline
FNN & All & Night & 17.84 & 0.67 & 16.54 & 0.65 & 12.79 & 0.54 \\ \hline
FNN & No W & Morning & 9.53 & 0.40 & 19.44 & 0.78 & 15.52 & 0.65 \\ \hline
FNN & No W & Afternoon & 10.37 & 0.46 & 15.31 & 0.72 & 11.55 & 0.51 \\ \hline
FNN & No W & Night & 17.49 & 0.64 & 15.91 & 0.59 & 12.54 & 0.50 \\ \hline
FNN & No T & Morning & 13.09 & 0.47 & 25.90 & 0.98 & 21.08 & 0.82 \\ \hline
FNN & No T & Afternoon & 14.13 & 0.54 & 19.34 & 0.75 & 15.32 & 0.61 \\ \hline
FNN & No T & Night & 20.98 & 0.79 & 18.70 & 0.71 & 14.62 & 0.61 \\ \hline
FNN & No W-No T & Morning & 11.65 & 0.45 & 23.17 & 0.83 & 18.97 & 0.73 \\ \hline
FNN & No W-No T & Afternoon & 11.93 & 0.47 & 17.11 & 0.72 & 13.27 & 0.54 \\ \hline
FNN & No W-No T & Night & 17.49 & 0.64 & 16.05 & 0.62 & 12.56 & 0.51\\\hline

\caption{Performance metrics (SMAPE, RMSE, MAE) and associated standard deviations across models and input variables. Shift-based predictions, metrics averaged across 1000 bootstrap samples.}
\label{tab:metric-shift}\\
\end{longtable}

\footnotesize
\begin{longtable}[c]{|c|c|c|c|c|c|c|c|c|}
\hline
\textbf{Method} & \textbf{\begin{tabular}[c]{@{}c@{}}Input\\ variable\end{tabular}} & \textbf{\begin{tabular}[c]{@{}c@{}}Risk\\ group\end{tabular}} & \textbf{\begin{tabular}[c]{@{}c@{}}SMAPE\\ (mean)\end{tabular}} & \textbf{\begin{tabular}[c]{@{}c@{}}SMAPE\\ (st.dev.)\end{tabular}} & \textbf{\begin{tabular}[c]{@{}c@{}}RMSE\\ (mean)\end{tabular}} & \textbf{\begin{tabular}[c]{@{}c@{}}RMSE\\ (st.dev.)\end{tabular}} & \textbf{\begin{tabular}[c]{@{}c@{}}MAE\\ (mean)\end{tabular}} & \textbf{\begin{tabular}[c]{@{}c@{}}MAE\\ (st.dev.)\end{tabular}} \\ \hline
\endfirsthead
\endhead
SARIMA-1 & & Low & 6.92 & 0.30 & 18.86 & 0.81 & 14.81 & 0.63 \\ \hline
SARIMA-1 & & Medium & 9.30 & 0.38 & 9.82 & 0.37 & 7.76 & 0.32 \\ \hline
SARIMA-1 & & High & 12.52 & 0.50 & 7.71 & 0.28 & 6.15 & 0.24 \\ \hline
SARIMAX-1 & & Low & 8.22 & 0.31 & 21.70 & 0.83 & 17.54 & 0.67 \\ \hline
SARIMAX-1 & & Medium & 10.25 & 0.42 & 10.87 & 0.40 & 8.58 & 0.35 \\ \hline
SARIMAX-1 & & High & 14.58 & 0.61 & 8.99 & 0.33 & 7.04 & 0.29 \\ \hline
SARIMA-7 & & Low & 7.40 & 0.32 & 20.27 & 0.85 & 15.85 & 0.67 \\ \hline
SARIMA-7 & & Medium & 9.28 & 0.38 & 9.86 & 0.38 & 7.74 & 0.32 \\ \hline
SARIMA-7 & & High & 12.68 & 0.50 & 7.78 & 0.28 & 6.23 & 0.24 \\ \hline
SARIMAX-7 & & Low & 13.50 & 0.64 & 45.32 & 4.10 & 30.06 & 1.76 \\ \hline
SARIMAX-7 & & Medium & 14.23 & 0.67 & 20.54 & 2.84 & 12.78 & 0.82 \\ \hline
SARIMAX-7 & & High & 23.04 & 0.99 & 20.29 & 2.51 & 12.26 & 0.83 \\ \hline
SARIMA-14 & & Low & 7.77 & 0.33 & 21.52 & 0.87 & 16.70 & 0.71 \\ \hline
SARIMA-14 & & Medium & 9.47 & 0.39 & 10.03 & 0.37 & 7.91 & 0.32 \\ \hline
SARIMA-14 & & High & 12.77 & 0.52 & 7.87 & 0.29 & 6.26 & 0.25 \\ \hline
SARIMAX-14 & & Low & 22.94 & 1.16 & 94.73 & 10.84 & 53.81 & 4.14 \\ \hline
SARIMAX-14 & & Medium & 19.62 & 1.12 & 39.32 & 7.86 & 19.23 & 1.88 \\ \hline
SARIMAX-14 & & High & 36.45 & 1.74 & 42.00 & 5.81 & 20.81 & 1.90 \\ \hline
SARIMA-28 & & Low & 9.40 & 0.38 & 25.64 & 0.97 & 20.26 & 0.82 \\ \hline
SARIMA-28 & & Medium & 9.67 & 0.39 & 10.20 & 0.37 & 8.08 & 0.32 \\ \hline
SARIMA-28 & & High & 13.23 & 0.53 & 8.18 & 0.31 & 6.50 & 0.26 \\ \hline
SARIMAX-28 & & Low & 47.65 & 2.10 & 207.45 & 22.98 & 114.69 & 9.12 \\ \hline
SARIMAX-28 & & Medium & 31.56 & 1.58 & 79.84 & 16.31 & 34.58 & 3.86 \\ \hline
SARIMAX-28 & & High & 70.90 & 3.04 & 91.91 & 12.29 & 43.06 & 4.25 \\ \hline
RF & All & Low & 8.59 & 0.35 & 23.38 & 0.97 & 18.25 & 0.77 \\ \hline
RF & All & Medium & 10.21 & 0.40 & 10.73 & 0.42 & 8.47 & 0.35 \\ \hline
RF & All & High & 14.31 & 0.55 & 8.82 & 0.32 & 7.00 & 0.29 \\ \hline
RF & No W & Low & 8.91 & 0.34 & 23.82 & 0.94 & 18.86 & 0.75 \\ \hline
RF & No W & Medium & 10.37 & 0.39 & 10.88 & 0.43 & 8.62 & 0.34 \\ \hline
RF & No W & High & 14.15 & 0.56 & 8.80 & 0.32 & 6.93 & 0.29 \\ \hline
RF & No T & Low & 9.95 & 0.37 & 26.19 & 0.93 & 20.94 & 0.81 \\ \hline
RF & No T & Medium & 10.96 & 0.41 & 11.33 & 0.42 & 9.07 & 0.35 \\ \hline
RF & No T & High & 13.35 & 0.52 & 8.22 & 0.30 & 6.54 & 0.26 \\ \hline
RF & No W-No T & Low & 9.06 & 0.35 & 24.41 & 0.98 & 19.26 & 0.77 \\ \hline
RF & No W-No T & Medium & 10.81 & 0.44 & 11.67 & 0.47 & 9.03 & 0.38 \\ \hline
RF & No W-No T & High & 14.65 & 0.55 & 8.88 & 0.32 & 7.15 & 0.28 \\ \hline
SVR & All & Low & 12.92 & 0.51 & 35.30 & 1.41 & 27.47 & 1.18 \\ \hline
SVR & All & Medium & 15.58 & 0.52 & 15.85 & 0.52 & 12.84 & 0.47 \\ \hline
SVR & All & High & 21.69 & 0.73 & 12.83 & 0.43 & 10.38 & 0.39 \\ \hline
SVR & No W & Low & 13.40 & 0.53 & 36.41 & 1.43 & 28.42 & 1.21 \\ \hline
SVR & No W & Medium & 16.11 & 0.54 & 16.30 & 0.53 & 13.24 & 0.48 \\ \hline
SVR & No W & High & 23.08 & 0.75 & 13.46 & 0.44 & 10.96 & 0.41 \\ \hline
SVR & No T & Low & 13.56 & 0.53 & 36.64 & 1.42 & 28.72 & 1.21 \\ \hline
SVR & No T & Medium & 16.11 & 0.53 & 16.29 & 0.53 & 13.24 & 0.48 \\ \hline
SVR & No T & High & 21.11 & 0.71 & 12.56 & 0.42 & 10.13 & 0.39 \\ \hline
SVR & No W-No T & Low & 15.40 & 0.56 & 40.40 & 1.47 & 32.25 & 1.28 \\ \hline
SVR & No W-No T & Medium & 16.92 & 0.55 & 16.94 & 0.54 & 13.84 & 0.50 \\ \hline
SVR & No W-No T & High & 20.99 & 0.72 & 12.53 & 0.42 & 10.09 & 0.39 \\ \hline
FNN & All & Low & 9.25 & 0.40 & 26.04 & 1.12 & 19.76 & 0.90 \\ \hline
FNN & All & Medium & 12.89 & 0.46 & 13.20 & 0.46 & 10.61 & 0.40 \\ \hline
FNN & All & High & 19.95 & 0.67 & 11.80 & 0.39 & 9.53 & 0.36 \\ \hline
FNN & No W & Low & 12.32 & 0.50 & 32.93 & 1.30 & 25.73 & 1.09 \\ \hline
FNN & No W & Medium & 14.16 & 0.46 & 13.95 & 0.47 & 11.55 & 0.40 \\ \hline
FNN & No W & High & 21.07 & 0.69 & 12.23 & 0.40 & 9.99 & 0.37 \\ \hline
FNN & No T & Low & 65.67 & 0.99 & 109.74 & 1.55 & 105.44 & 1.64 \\ \hline
FNN & No T & Medium & 67.31 & 0.89 & 43.78 & 0.56 & 42.24 & 0.59 \\ \hline
FNN & No T & High & 72.84 & 1.00 & 28.12 & 0.47 & 26.66 & 0.46 \\ \hline
FNN & No W-No T & Low & 10.73 & 0.46 & 29.76 & 1.25 & 23.06 & 1.04 \\ \hline
FNN & No W-No T & Medium & 13.24 & 0.44 & 13.56 & 0.46 & 11.02 & 0.40 \\ \hline
FNN & No W-No T & High & 18.80 & 0.68 & 11.53 & 0.40 & 9.14 & 0.37\\\hline
\caption{Performance metrics (SMAPE, RMSE, MAE) and associated standard deviations across models and input variables. Risk-group-based predictions, metrics averaged across 1000 bootstrap samples.}
\label{tab:metric-risk}\\
\end{longtable}

\subsection{Full set of DM p-values}
\label{si:diebold-table}

\footnotesize
\begin{longtable}[c]{|c|c|c|c|c|c|}
\hline
\textbf{Method 1} & \textbf{\begin{tabular}[c]{@{}c@{}}Input\\ variable 1\end{tabular}} & \textbf{Method 2} & \textbf{\begin{tabular}[c]{@{}c@{}}Input\\ variable 2\end{tabular}} & \textbf{Shift} & \textbf{\begin{tabular}[c]{@{}c@{}}Fraction of\\ equivalent samples\end{tabular}} \\ \hline
\endfirsthead
\endhead
SARIMA-1 & & SARIMA-7 & & Morning & 0.988 \\ \hline
SARIMA-1 & & SARIMA-7 & & Afternoon & 0.924 \\ \hline
SARIMA-1 & & SARIMA-7 & & Night & 0.471 \\ \hline
SARIMA-1 & & SARIMA-14 & & Morning & 0.791 \\ \hline
SARIMA-1 & & SARIMA-14 & & Afternoon & 0.795 \\ \hline
SARIMA-1 & & SARIMA-14 & & Night & 0.035 \\ \hline
SARIMA-1 & & SARIMA-28 & & Morning & 0.267 \\ \hline
SARIMA-1 & & SARIMA-28 & & Afternoon & 0.127 \\ \hline
SARIMA-1 & & SARIMA-28 & & Night & 0.0 \\ \hline
SARIMA-7 & & SARIMA-14 & & Morning & 0.869 \\ \hline
SARIMA-7 & & SARIMA-14 & & Afternoon & 0.982 \\ \hline
SARIMA-7 & & SARIMA-14 & & Night & 0.643 \\ \hline
SARIMA-7 & & SARIMA-28 & & Morning & 0.296 \\ \hline
SARIMA-7 & & SARIMA-28 & & Afternoon & 0.359 \\ \hline
SARIMA-7 & & SARIMA-28 & & Night & 0.045 \\ \hline
SARIMA-14 & & SARIMA-28 & & Morning & 0.833 \\ \hline
SARIMA-14 & & SARIMA-28 & & Afternoon & 0.604 \\ \hline
SARIMA-14 & & SARIMA-28 & & Night & 0.443 \\ \hline
SARIMA-1 & & RF & No W & Morning & 0.583 \\ \hline
SARIMA-1 & & RF & No W & Afternoon & 0.826 \\ \hline
SARIMA-1 & & RF & No W & Night & 0.675 \\ \hline
SARIMA-1 & & RF &  No W-No T & Morning & 0.245 \\ \hline
SARIMA-1 & & RF &  No W-No T & Afternoon & 0.403 \\ \hline
SARIMA-1 & & RF &  No W-No T & Night & 0.748 \\ \hline
SARIMA-7 & & RF & No W & Morning & 0.722 \\ \hline
SARIMA-7 & & RF & No W & Afternoon & 0.966 \\ \hline
SARIMA-7 & & RF & No W & Night & 0.996 \\ \hline
SARIMA-7 & & RF &  No W-No T & Morning & 0.385 \\ \hline
SARIMA-7 & & RF &  No W-No T & Afternoon & 0.736 \\ \hline
SARIMA-7 & & RF &  No W-No T & Night & 0.998 \\ \hline
SARIMA-14 & & RF & No W & Morning & 0.897 \\ \hline
SARIMA-14 & & RF & No W & Afternoon & 0.996 \\ \hline
SARIMA-14 & & RF & No W & Night & 0.999 \\ \hline
SARIMA-14 & & RF &  No W-No T & Morning & 0.614 \\ \hline
SARIMA-14 & & RF &  No W-No T & Afternoon & 0.893 \\ \hline
SARIMA-14 & & RF &  No W-No T & Night & 1.0 \\ \hline
SARIMA-28 & & RF & No W & Morning & 0.986 \\ \hline
SARIMA-28 & & RF & No W & Afternoon & 0.999 \\ \hline
SARIMA-28 & & RF & No W & Night & 0.924 \\ \hline
SARIMA-28 & & RF &  No W-No T & Morning & 0.911 \\ \hline
SARIMA-28 & & RF &  No W-No T & Afternoon & 1.0 \\ \hline
SARIMA-28 & & RF &  No W-No T & Night & 0.809 \\ \hline
RF & All & RF & No T & Morning & 0.964 \\ \hline
RF & All & RF & No T & Afternoon & 0.349 \\ \hline
RF & All & RF & No T & Night & 0.065 \\ \hline
RF & All & RF & No W & Morning & 0.981 \\ \hline
RF & All & RF & No W & Afternoon & 0.954 \\ \hline
RF & All & RF & No W & Night & 0.962 \\ \hline
RF & All & RF &  No W-No T & Morning & 0.997 \\ \hline
RF & All & RF &  No W-No T & Afternoon & 0.568 \\ \hline
RF & All & RF &  No W-No T & Night & 0.999 \\ \hline
RF & No T & RF & No W & Morning & 0.876 \\ \hline
RF & No T & RF & No W & Afternoon & 0.949 \\ \hline
RF & No T & RF & No W & Night & 0.453 \\ \hline
RF & No T & RF &  No W-No T & Morning & 1.0 \\ \hline
RF & No T & RF &  No W-No T & Afternoon & 1.0 \\ \hline
RF & No T & RF &  No W-No T & Night & 0.255 \\ \hline
RF & No W & RF &  No W-No T & Morning & 0.959 \\ \hline
RF & No W & RF &  No W-No T & Afternoon & 0.893 \\ \hline
RF & No W & RF &  No W-No T & Night & 0.996 \\ \hline
RF & All & SVR & All & Morning & 0.0 \\ \hline
RF & All & SVR & All & Afternoon & 0.0 \\ \hline
RF & All & SVR & All & Night & 0.0 \\ \hline
RF & All & SVR & No T & Morning & 0.0 \\ \hline
RF & All & SVR & No T & Afternoon & 0.0 \\ \hline
RF & All & SVR & No T & Night & 0.0 \\ \hline
RF & No T & SVR & All & Morning & 0.0 \\ \hline
RF & No T & SVR & All & Afternoon & 0.002 \\ \hline
RF & No T & SVR & All & Night & 0.0 \\ \hline
RF & All & SVR & No W & Morning & 0.0 \\ \hline
RF & All & SVR & No W & Afternoon & 0.0 \\ \hline
RF & All & SVR & No W & Night & 0.0 \\ \hline
RF & No W & SVR & All & Morning & 0.0 \\ \hline
RF & No W & SVR & All & Afternoon & 0.0 \\ \hline
RF & No W & SVR & All & Night & 0.0 \\ \hline
RF & All & SVR &  No W-No T & Morning & 0.0 \\ \hline
RF & All & SVR &  No W-No T & Afternoon & 0.0 \\ \hline
RF & All & SVR &  No W-No T & Night & 0.0 \\ \hline
RF &  No W-No T & SVR & All & Morning & 0.0 \\ \hline
RF &  No W-No T & SVR & All & Afternoon & 0.005 \\ \hline
RF &  No W-No T & SVR & All & Night & 0.0 \\ \hline
RF & No T & SVR & No T & Morning & 0.0 \\ \hline
RF & No T & SVR & No T & Afternoon & 0.0 \\ \hline
RF & No T & SVR & No T & Night & 0.0 \\ \hline
RF & No T & SVR & No W & Morning & 0.0 \\ \hline
RF & No T & SVR & No W & Afternoon & 0.0 \\ \hline
RF & No T & SVR & No W & Night & 0.001 \\ \hline
RF & No W & SVR & No T & Morning & 0.0 \\ \hline
RF & No W & SVR & No T & Afternoon & 0.0 \\ \hline
RF & No W & SVR & No T & Night & 0.0 \\ \hline
RF & No T & SVR &  No W-No T & Morning & 0.0 \\ \hline
RF & No T & SVR &  No W-No T & Afternoon & 0.0 \\ \hline
RF & No T & SVR &  No W-No T & Night & 0.0 \\ \hline
RF &  No W-No T & SVR & No T & Morning & 0.0 \\ \hline
RF &  No W-No T & SVR & No T & Afternoon & 0.001 \\ \hline
RF &  No W-No T & SVR & No T & Night & 0.0 \\ \hline
RF & No W & SVR & No W & Morning & 0.0 \\ \hline
RF & No W & SVR & No W & Afternoon & 0.0 \\ \hline
RF & No W & SVR & No W & Night & 0.0 \\ \hline
RF & No W & SVR &  No W-No T & Morning & 0.0 \\ \hline
RF & No W & SVR &  No W-No T & Afternoon & 0.0 \\ \hline
RF & No W & SVR &  No W-No T & Night & 0.0 \\ \hline
RF &  No W-No T & SVR & No W & Morning & 0.0 \\ \hline
RF &  No W-No T & SVR & No W & Afternoon & 0.004 \\ \hline
RF &  No W-No T & SVR & No W & Night & 0.0 \\ \hline
RF &  No W-No T & SVR &  No W-No T & Morning & 0.0 \\ \hline
RF &  No W-No T & SVR &  No W-No T & Afternoon & 0.0 \\ \hline
RF &  No W-No T & SVR &  No W-No T & Night & 0.0 \\ \hline
RF & All & FNN & All & Morning & 0.0 \\ \hline
RF & All & FNN & All & Afternoon & 0.0 \\ \hline
RF & All & FNN & All & Night & 0.0 \\ \hline
RF & All & FNN & No T & Morning & 0.0 \\ \hline
RF & All & FNN & No T & Afternoon & 0.0 \\ \hline
RF & All & FNN & No T & Night & 0.0 \\ \hline
RF & No T & FNN & All & Morning & 0.0 \\ \hline
RF & No T & FNN & All & Afternoon & 0.041 \\ \hline
RF & No T & FNN & All & Night & 0.006 \\ \hline
RF & All & FNN & No W & Morning & 0.999 \\ \hline
RF & All & FNN & No W & Afternoon & 0.993 \\ \hline
RF & All & FNN & No W & Night & 0.0 \\ \hline
RF & No W & FNN & All & Morning & 0.0 \\ \hline
RF & No W & FNN & All & Afternoon & 0.004 \\ \hline
RF & No W & FNN & All & Night & 0.0 \\ \hline
RF & All & FNN &  No W-No T & Morning & 0.012 \\ \hline
RF & All & FNN &  No W-No T & Afternoon & 0.093 \\ \hline
RF & All & FNN &  No W-No T & Night & 0.0 \\ \hline
RF &  No W-No T & FNN & All & Morning & 0.0 \\ \hline
RF &  No W-No T & FNN & All & Afternoon & 0.072 \\ \hline
RF &  No W-No T & FNN & All & Night & 0.0 \\ \hline
RF & No T & FNN & No T & Morning & 0.0 \\ \hline
RF & No T & FNN & No T & Afternoon & 0.0 \\ \hline
RF & No T & FNN & No T & Night & 0.0 \\ \hline
RF & No T & FNN & No W & Morning & 1.0 \\ \hline
RF & No T & FNN & No W & Afternoon & 0.994 \\ \hline
RF & No T & FNN & No W & Night & 0.055 \\ \hline
RF & No W & FNN & No T & Morning & 0.0 \\ \hline
RF & No W & FNN & No T & Afternoon & 0.0 \\ \hline
RF & No W & FNN & No T & Night & 0.0 \\ \hline
RF & No T & FNN &  No W-No T & Morning & 0.104 \\ \hline
RF & No T & FNN &  No W-No T & Afternoon & 0.756 \\ \hline
RF & No T & FNN &  No W-No T & Night & 0.024 \\ \hline
RF &  No W-No T & FNN & No T & Morning & 0.0 \\ \hline
RF &  No W-No T & FNN & No T & Afternoon & 0.0 \\ \hline
RF &  No W-No T & FNN & No T & Night & 0.0 \\ \hline
RF & No W & FNN & No W & Morning & 0.988 \\ \hline
RF & No W & FNN & No W & Afternoon & 1.0 \\ \hline
RF & No W & FNN & No W & Night & 0.0 \\ \hline
RF & No W & FNN &  No W-No T & Morning & 0.003 \\ \hline
RF & No W & FNN &  No W-No T & Afternoon & 0.365 \\ \hline
RF & No W & FNN &  No W-No T & Night & 0.0 \\ \hline
RF &  No W-No T & FNN & No W & Morning & 1.0 \\ \hline
RF &  No W-No T & FNN & No W & Afternoon & 0.994 \\ \hline
RF &  No W-No T & FNN & No W & Night & 0.0 \\ \hline
RF &  No W-No T & FNN &  No W-No T & Morning & 0.048 \\ \hline
RF &  No W-No T & FNN &  No W-No T & Afternoon & 0.748 \\ \hline
RF &  No W-No T & FNN &  No W-No T & Night & 0.0 \\ \hline
SVR & All & SVR & No T & Morning & 0.83 \\ \hline
SVR & All & SVR & No T & Afternoon & 0.0 \\ \hline
SVR & All & SVR & No T & Night & 0.0 \\ \hline
SVR & All & SVR & No W & Morning & 0.0 \\ \hline
SVR & All & SVR & No W & Afternoon & 0.035 \\ \hline
SVR & All & SVR & No W & Night & 0.925 \\ \hline
SVR & All & SVR &  No W-No T & Morning & 0.0 \\ \hline
SVR & All & SVR &  No W-No T & Afternoon & 0.0 \\ \hline
SVR & All & SVR &  No W-No T & Night & 0.0 \\ \hline
SVR & No T & SVR & No W & Morning & 0.0 \\ \hline
SVR & No T & SVR & No W & Afternoon & 0.003 \\ \hline
SVR & No T & SVR & No W & Night & 0.0 \\ \hline
SVR & No T & SVR &  No W-No T & Morning & 0.0 \\ \hline
SVR & No T & SVR &  No W-No T & Afternoon & 0.0 \\ \hline
SVR & No T & SVR &  No W-No T & Night & 0.0 \\ \hline
SVR & No W & SVR &  No W-No T & Morning & 0.0 \\ \hline
SVR & No W & SVR &  No W-No T & Afternoon & 0.0 \\ \hline
SVR & No W & SVR &  No W-No T & Night & 0.0 \\ \hline
SVR & All & FNN & All & Morning & 0.619 \\ \hline
SVR & All & FNN & All & Afternoon & 0.398 \\ \hline
SVR & All & FNN & All & Night & 0.061 \\ \hline
SVR & All & FNN & No T & Morning & 1.0 \\ \hline
SVR & All & FNN & No T & Afternoon & 0.965 \\ \hline
SVR & All & FNN & No T & Night & 0.011 \\ \hline
SVR & No T & FNN & All & Morning & 0.383 \\ \hline
SVR & No T & FNN & All & Afternoon & 0.007 \\ \hline
SVR & No T & FNN & All & Night & 0.0 \\ \hline
SVR & All & FNN & No W & Morning & 0.0 \\ \hline
SVR & All & FNN & No W & Afternoon & 0.0 \\ \hline
SVR & All & FNN & No W & Night & 0.86 \\ \hline
SVR & No W & FNN & All & Morning & 0.998 \\ \hline
SVR & No W & FNN & All & Afternoon & 0.021 \\ \hline
SVR & No W & FNN & All & Night & 0.009 \\ \hline
SVR & All & FNN &  No W-No T & Morning & 0.112 \\ \hline
SVR & All & FNN &  No W-No T & Afternoon & 0.188 \\ \hline
SVR & All & FNN &  No W-No T & Night & 0.576 \\ \hline
SVR &  No W-No T & FNN & All & Morning & 0.855 \\ \hline
SVR &  No W-No T & FNN & All & Afternoon & 0.0 \\ \hline
SVR &  No W-No T & FNN & All & Night & 0.0 \\ \hline
SVR & No T & FNN & No T & Morning & 1.0 \\ \hline
SVR & No T & FNN & No T & Afternoon & 1.0 \\ \hline
SVR & No T & FNN & No T & Night & 1.0 \\ \hline
SVR & No T & FNN & No W & Morning & 0.0 \\ \hline
SVR & No T & FNN & No W & Afternoon & 0.0 \\ \hline
SVR & No T & FNN & No W & Night & 0.002 \\ \hline
SVR & No W & FNN & No T & Morning & 0.998 \\ \hline
SVR & No W & FNN & No T & Afternoon & 0.988 \\ \hline
SVR & No W & FNN & No T & Night & 0.006 \\ \hline
SVR & No T & FNN &  No W-No T & Morning & 0.11 \\ \hline
SVR & No T & FNN &  No W-No T & Afternoon & 0.036 \\ \hline
SVR & No T & FNN &  No W-No T & Night & 0.0 \\ \hline
SVR &  No W-No T & FNN & No T & Morning & 0.993 \\ \hline
SVR &  No W-No T & FNN & No T & Afternoon & 0.997 \\ \hline
SVR &  No W-No T & FNN & No T & Night & 0.001 \\ \hline
SVR & No W & FNN & No W & Morning & 0.0 \\ \hline
SVR & No W & FNN & No W & Afternoon & 0.0 \\ \hline
SVR & No W & FNN & No W & Night & 0.892 \\ \hline
SVR & No W & FNN &  No W-No T & Morning & 0.014 \\ \hline
SVR & No W & FNN &  No W-No T & Afternoon & 0.101 \\ \hline
SVR & No W & FNN &  No W-No T & Night & 0.599 \\ \hline
SVR &  No W-No T & FNN & No W & Morning & 0.0 \\ \hline
SVR &  No W-No T & FNN & No W & Afternoon & 0.0 \\ \hline
SVR &  No W-No T & FNN & No W & Night & 0.0 \\ \hline
SVR &  No W-No T & FNN &  No W-No T & Morning & 0.007 \\ \hline
SVR &  No W-No T & FNN &  No W-No T & Afternoon & 0.0 \\ \hline
SVR &  No W-No T & FNN &  No W-No T & Night & 0.0 \\ \hline
FNN & All & FNN & No T & Morning & 1.0 \\ \hline
FNN & All & FNN & No T & Afternoon & 0.817 \\ \hline
FNN & All & FNN & No T & Night & 0.0 \\ \hline
FNN & All & FNN & No W & Morning & 0.0 \\ \hline
FNN & All & FNN & No W & Afternoon & 0.0 \\ \hline
FNN & All & FNN & No W & Night & 0.997 \\ \hline
FNN & All & FNN &  No W-No T & Morning & 0.004 \\ \hline
FNN & All & FNN &  No W-No T & Afternoon & 0.379 \\ \hline
FNN & All & FNN &  No W-No T & Night & 0.987 \\ \hline
FNN & No T & FNN & No W & Morning & 0.0 \\ \hline
FNN & No T & FNN & No W & Afternoon & 0.0 \\ \hline
FNN & No T & FNN & No W & Night & 0.0 \\ \hline
FNN & No T & FNN &  No W-No T & Morning & 0.749 \\ \hline
FNN & No T & FNN &  No W-No T & Afternoon & 0.217 \\ \hline
FNN & No T & FNN &  No W-No T & Night & 0.0 \\ \hline
FNN & No W & FNN &  No W-No T & Morning & 0.0 \\ \hline
FNN & No W & FNN &  No W-No T & Afternoon & 0.108 \\ \hline
FNN & No W & FNN &  No W-No T & Night & 1.0 \\ \hline
\caption{For each pair of models (with the respective input variables) in the table, we report the fraction of bootstrap samples for which the two modes have equivalent predictive accuracy according to the Diebold-Mariano test. This table is for shift-based predictions.}
\label{tab:diebold-shift}\\
\end{longtable}

\footnotesize
\begin{longtable}[c]{|c|c|c|c|c|c|}
\hline
\textbf{Method 1} & \textbf{\begin{tabular}[c]{@{}c@{}}Input\\ variable 1\end{tabular}} & \textbf{Method 2} & \textbf{\begin{tabular}[c]{@{}c@{}}Input\\ variable 2\end{tabular}} & \textbf{\begin{tabular}[c]{@{}c@{}}Risk\\ group\end{tabular}} & \textbf{\begin{tabular}[c]{@{}c@{}}Fraction of\\ equivalent samples\end{tabular}} \\ \hline
\endfirsthead
\endhead
SARIMA-1 & & SARIMA-7 & & Low & 0.817 \\ \hline
SARIMA-1 & & SARIMA-7 & & Medium & 1.0 \\ \hline
SARIMA-1 & & SARIMA-7 & & High & 0.997 \\ \hline
SARIMA-1 & & SARIMA-14 & & Low & 0.518 \\ \hline
SARIMA-1 & & SARIMA-14 & & Medium & 0.995 \\ \hline
SARIMA-1 & & SARIMA-14 & & High & 0.989 \\ \hline
SARIMA-1 & & SARIMA-28 & & Low & 0.0 \\ \hline
SARIMA-1 & & SARIMA-28 & & Medium & 0.963 \\ \hline
SARIMA-1 & & SARIMA-28 & & High & 0.902 \\ \hline
SARIMA-7 & & SARIMA-14 & & Low & 0.938 \\ \hline
SARIMA-7 & & SARIMA-14 & & Medium & 0.97 \\ \hline
SARIMA-7 & & SARIMA-14 & & High & 0.998 \\ \hline
SARIMA-7 & & SARIMA-28 & & Low & 0.0 \\ \hline
SARIMA-7 & & SARIMA-28 & & Medium & 0.935 \\ \hline
SARIMA-7 & & SARIMA-28 & & High & 0.95 \\ \hline
SARIMA-14 & & SARIMA-28 & & Low & 0.0 \\ \hline
SARIMA-14 & & SARIMA-28 & & Medium & 0.991 \\ \hline
SARIMA-14 & & SARIMA-28 & & High & 0.959 \\ \hline
SARIMA-1 & & RF & No W & Low & 0.062 \\ \hline
SARIMA-1 & & RF & No W & Medium & 0.796 \\ \hline
SARIMA-1 & & RF & No W & High & 0.747 \\ \hline
SARIMA-1 & & RF &  No W-No T & Low & 0.029 \\ \hline
SARIMA-1 & & RF &  No W-No T & Medium & 0.495 \\ \hline
SARIMA-1 & & RF &  No W-No T & High & 0.384 \\ \hline
SARIMA-7 & & RF & No W & Low & 0.416 \\ \hline
SARIMA-7 & & RF & No W & Medium & 0.787 \\ \hline
SARIMA-7 & & RF & No W & High & 0.824 \\ \hline
SARIMA-7 & & RF &  No W-No T & Low & 0.258 \\ \hline
SARIMA-7 & & RF &  No W-No T & Medium & 0.474 \\ \hline
SARIMA-7 & & RF &  No W-No T & High & 0.51 \\ \hline
SARIMA-14 & & RF & No W & Low & 0.798 \\ \hline
SARIMA-14 & & RF & No W & Medium & 0.91 \\ \hline
SARIMA-14 & & RF & No W & High & 0.869 \\ \hline
SARIMA-14 & & RF &  No W-No T & Low & 0.646 \\ \hline
SARIMA-14 & & RF &  No W-No T & Medium & 0.63 \\ \hline
SARIMA-14 & & RF &  No W-No T & High & 0.615 \\ \hline
SARIMA-28 & & RF & No W & Low & 0.996 \\ \hline
SARIMA-28 & & RF & No W & Medium & 0.97 \\ \hline
SARIMA-28 & & RF & No W & High & 0.979 \\ \hline
SARIMA-28 & & RF &  No W-No T & Low & 1.0 \\ \hline
SARIMA-28 & & RF &  No W-No T & Medium & 0.789 \\ \hline
SARIMA-28 & & RF &  No W-No T & High & 0.905 \\ \hline
RF & All & RF & No T & Low & 0.005 \\ \hline
RF & All & RF & No T & Medium & 0.192 \\ \hline
RF & All & RF & No T & High & 0.404 \\ \hline
RF & All & RF & No W & Low & 0.813 \\ \hline
RF & All & RF & No W & Medium & 0.999 \\ \hline
RF & All & RF & No W & High & 1.0 \\ \hline
RF & All & RF &  No W-No T & Low & 0.935 \\ \hline
RF & All & RF &  No W-No T & Medium & 0.948 \\ \hline
RF & All & RF &  No W-No T & High & 0.995 \\ \hline
RF & No T & RF & No W & Low & 0.349 \\ \hline
RF & No T & RF & No W & Medium & 0.889 \\ \hline
RF & No T & RF & No W & High & 0.871 \\ \hline
RF & No T & RF &  No W-No T & Low & 0.459 \\ \hline
RF & No T & RF &  No W-No T & Medium & 0.999 \\ \hline
RF & No T & RF &  No W-No T & High & 0.375 \\ \hline
RF & No W & RF &  No W-No T & Low & 0.998 \\ \hline
RF & No W & RF &  No W-No T & Medium & 0.985 \\ \hline
RF & No W & RF &  No W-No T & High & 0.959 \\ \hline
RF & All & SVR & All & Low & 0.0 \\ \hline
RF & All & SVR & All & Medium & 0.0 \\ \hline
RF & All & SVR & All & High & 0.0 \\ \hline
RF & All & SVR & No T & Low & 0.0 \\ \hline
RF & All & SVR & No T & Medium & 0.0 \\ \hline
RF & All & SVR & No T & High & 0.0 \\ \hline
RF & No T & SVR & All & Low & 0.002 \\ \hline
RF & No T & SVR & All & Medium & 0.0 \\ \hline
RF & No T & SVR & All & High & 0.0 \\ \hline
RF & All & SVR & No W & Low & 0.0 \\ \hline
RF & All & SVR & No W & Medium & 0.0 \\ \hline
RF & All & SVR & No W & High & 0.0 \\ \hline
RF & No W & SVR & All & Low & 0.0 \\ \hline
RF & No W & SVR & All & Medium & 0.0 \\ \hline
RF & No W & SVR & All & High & 0.0 \\ \hline
RF & All & SVR &  No W-No T & Low & 0.0 \\ \hline
RF & All & SVR &  No W-No T & Medium & 0.0 \\ \hline
RF & All & SVR &  No W-No T & High & 0.0 \\ \hline
RF &  No W-No T & SVR & All & Low & 0.0 \\ \hline
RF &  No W-No T & SVR & All & Medium & 0.0 \\ \hline
RF &  No W-No T & SVR & All & High & 0.0 \\ \hline
RF & No T & SVR & No T & Low & 0.0 \\ \hline
RF & No T & SVR & No T & Medium & 0.0 \\ \hline
RF & No T & SVR & No T & High & 0.0 \\ \hline
RF & No T & SVR & No W & Low & 0.0 \\ \hline
RF & No T & SVR & No W & Medium & 0.0 \\ \hline
RF & No T & SVR & No W & High & 0.0 \\ \hline
RF & No W & SVR & No T & Low & 0.0 \\ \hline
RF & No W & SVR & No T & Medium & 0.0 \\ \hline
RF & No W & SVR & No T & High & 0.0 \\ \hline
RF & No T & SVR &  No W-No T & Low & 0.0 \\ \hline
RF & No T & SVR &  No W-No T & Medium & 0.0 \\ \hline
RF & No T & SVR &  No W-No T & High & 0.0 \\ \hline
RF &  No W-No T & SVR & No T & Low & 0.0 \\ \hline
RF &  No W-No T & SVR & No T & Medium & 0.0 \\ \hline
RF &  No W-No T & SVR & No T & High & 0.0 \\ \hline
RF & No W & SVR & No W & Low & 0.0 \\ \hline
RF & No W & SVR & No W & Medium & 0.0 \\ \hline
RF & No W & SVR & No W & High & 0.0 \\ \hline
RF & No W & SVR &  No W-No T & Low & 0.0 \\ \hline
RF & No W & SVR &  No W-No T & Medium & 0.0 \\ \hline
RF & No W & SVR &  No W-No T & High & 0.0 \\ \hline
RF &  No W-No T & SVR & No W & Low & 0.0 \\ \hline
RF &  No W-No T & SVR & No W & Medium & 0.0 \\ \hline
RF &  No W-No T & SVR & No W & High & 0.0 \\ \hline
RF &  No W-No T & SVR &  No W-No T & Low & 0.0 \\ \hline
RF &  No W-No T & SVR &  No W-No T & Medium & 0.0 \\ \hline
RF &  No W-No T & SVR &  No W-No T & High & 0.0 \\ \hline
RF & All & FNN & All & Low & 0.936 \\ \hline
RF & All & FNN & All & Medium & 0.001 \\ \hline
RF & All & FNN & All & High & 0.0 \\ \hline
RF & All & FNN & No T & Low & 0.0 \\ \hline
RF & All & FNN & No T & Medium & 0.0 \\ \hline
RF & All & FNN & No T & High & 0.0 \\ \hline
RF & No T & FNN & All & Low & 0.973 \\ \hline
RF & No T & FNN & All & Medium & 0.222 \\ \hline
RF & No T & FNN & All & High & 0.0 \\ \hline
RF & All & FNN & No W & Low & 0.0 \\ \hline
RF & All & FNN & No W & Medium & 0.0 \\ \hline
RF & All & FNN & No W & High & 0.0 \\ \hline
RF & No W & FNN & All & Low & 0.996 \\ \hline
RF & No W & FNN & All & Medium & 0.015 \\ \hline
RF & No W & FNN & All & High & 0.0 \\ \hline
RF & All & FNN &  No W-No T & Low & 0.062 \\ \hline
RF & All & FNN &  No W-No T & Medium & 0.0 \\ \hline
RF & All & FNN &  No W-No T & High & 0.0 \\ \hline
RF &  No W-No T & FNN & All & Low & 1.0 \\ \hline
RF &  No W-No T & FNN & All & Medium & 0.382 \\ \hline
RF &  No W-No T & FNN & All & High & 0.0 \\ \hline
RF & No T & FNN & No T & Low & 0.0 \\ \hline
RF & No T & FNN & No T & Medium & 0.0 \\ \hline
RF & No T & FNN & No T & High & 0.0 \\ \hline
RF & No T & FNN & No W & Low & 0.043 \\ \hline
RF & No T & FNN & No W & Medium & 0.0 \\ \hline
RF & No T & FNN & No W & High & 0.0 \\ \hline
RF & No W & FNN & No T & Low & 0.0 \\ \hline
RF & No W & FNN & No T & Medium & 0.0 \\ \hline
RF & No W & FNN & No T & High & 0.0 \\ \hline
RF & No T & FNN &  No W-No T & Low & 0.97 \\ \hline
RF & No T & FNN &  No W-No T & Medium & 0.035 \\ \hline
RF & No T & FNN &  No W-No T & High & 0.0 \\ \hline
RF &  No W-No T & FNN & No T & Low & 0.0 \\ \hline
RF &  No W-No T & FNN & No T & Medium & 0.0 \\ \hline
RF &  No W-No T & FNN & No T & High & 0.0 \\ \hline
RF & No W & FNN & No W & Low & 0.0 \\ \hline
RF & No W & FNN & No W & Medium & 0.0 \\ \hline
RF & No W & FNN & No W & High & 0.0 \\ \hline
RF & No W & FNN &  No W-No T & Low & 0.244 \\ \hline
RF & No W & FNN &  No W-No T & Medium & 0.0 \\ \hline
RF & No W & FNN &  No W-No T & High & 0.0 \\ \hline
RF &  No W-No T & FNN & No W & Low & 0.0 \\ \hline
RF &  No W-No T & FNN & No W & Medium & 0.001 \\ \hline
RF &  No W-No T & FNN & No W & High & 0.0 \\ \hline
RF &  No W-No T & FNN &  No W-No T & Low & 0.207 \\ \hline
RF &  No W-No T & FNN &  No W-No T & Medium & 0.044 \\ \hline
RF &  No W-No T & FNN &  No W-No T & High & 0.006 \\ \hline
SVR & All & SVR & No T & Low & 0.0 \\ \hline
SVR & All & SVR & No T & Medium & 0.0 \\ \hline
SVR & All & SVR & No T & High & 0.0 \\ \hline
SVR & All & SVR & No W & Low & 0.0 \\ \hline
SVR & All & SVR & No W & Medium & 0.0 \\ \hline
SVR & All & SVR & No W & High & 0.0 \\ \hline
SVR & All & SVR &  No W-No T & Low & 0.0 \\ \hline
SVR & All & SVR &  No W-No T & Medium & 0.0 \\ \hline
SVR & All & SVR &  No W-No T & High & 0.0 \\ \hline
SVR & No T & SVR & No W & Low & 0.01 \\ \hline
SVR & No T & SVR & No W & Medium & 1.0 \\ \hline
SVR & No T & SVR & No W & High & 0.0 \\ \hline
SVR & No T & SVR &  No W-No T & Low & 0.0 \\ \hline
SVR & No T & SVR &  No W-No T & Medium & 0.0 \\ \hline
SVR & No T & SVR &  No W-No T & High & 0.977 \\ \hline
SVR & No W & SVR &  No W-No T & Low & 0.0 \\ \hline
SVR & No W & SVR &  No W-No T & Medium & 0.0 \\ \hline
SVR & No W & SVR &  No W-No T & High & 0.0 \\ \hline
SVR & All & FNN & All & Low & 0.0 \\ \hline
SVR & All & FNN & All & Medium & 0.0 \\ \hline
SVR & All & FNN & All & High & 0.4 \\ \hline
SVR & All & FNN & No T & Low & 0.0 \\ \hline
SVR & All & FNN & No T & Medium & 0.0 \\ \hline
SVR & All & FNN & No T & High & 0.0 \\ \hline
SVR & No T & FNN & All & Low & 0.0 \\ \hline
SVR & No T & FNN & All & Medium & 0.0 \\ \hline
SVR & No T & FNN & All & High & 0.835 \\ \hline
SVR & All & FNN & No W & Low & 0.986 \\ \hline
SVR & All & FNN & No W & Medium & 0.599 \\ \hline
SVR & All & FNN & No W & High & 0.989 \\ \hline
SVR & No W & FNN & All & Low & 0.0 \\ \hline
SVR & No W & FNN & All & Medium & 0.0 \\ \hline
SVR & No W & FNN & All & High & 0.0 \\ \hline
SVR & All & FNN &  No W-No T & Low & 0.0 \\ \hline
SVR & All & FNN &  No W-No T & Medium & 0.0 \\ \hline
SVR & All & FNN &  No W-No T & High & 0.0 \\ \hline
SVR &  No W-No T & FNN & All & Low & 0.0 \\ \hline
SVR &  No W-No T & FNN & All & Medium & 0.0 \\ \hline
SVR &  No W-No T & FNN & All & High & 0.884 \\ \hline
SVR & No T & FNN & No T & Low & 0.0 \\ \hline
SVR & No T & FNN & No T & Medium & 0.0 \\ \hline
SVR & No T & FNN & No T & High & 0.0 \\ \hline
SVR & No T & FNN & No W & Low & 0.826 \\ \hline
SVR & No T & FNN & No W & Medium & 0.155 \\ \hline
SVR & No T & FNN & No W & High & 0.999 \\ \hline
SVR & No W & FNN & No T & Low & 0.0 \\ \hline
SVR & No W & FNN & No T & Medium & 0.0 \\ \hline
SVR & No W & FNN & No T & High & 0.0 \\ \hline
SVR & No T & FNN &  No W-No T & Low & 0.0 \\ \hline
SVR & No T & FNN &  No W-No T & Medium & 0.0 \\ \hline
SVR & No T & FNN &  No W-No T & High & 0.0 \\ \hline
SVR &  No W-No T & FNN & No T & Low & 0.0 \\ \hline
SVR &  No W-No T & FNN & No T & Medium & 0.0 \\ \hline
SVR &  No W-No T & FNN & No T & High & 0.0 \\ \hline
SVR & No W & FNN & No W & Low & 0.914 \\ \hline
SVR & No W & FNN & No W & Medium & 0.2 \\ \hline
SVR & No W & FNN & No W & High & 0.202 \\ \hline
SVR & No W & FNN &  No W-No T & Low & 0.0 \\ \hline
SVR & No W & FNN &  No W-No T & Medium & 0.0 \\ \hline
SVR & No W & FNN &  No W-No T & High & 0.0 \\ \hline
SVR &  No W-No T & FNN & No W & Low & 0.0 \\ \hline
SVR &  No W-No T & FNN & No W & Medium & 0.001 \\ \hline
SVR &  No W-No T & FNN & No W & High & 0.999 \\ \hline
SVR &  No W-No T & FNN &  No W-No T & Low & 0.0 \\ \hline
SVR &  No W-No T & FNN &  No W-No T & Medium & 0.0 \\ \hline
SVR &  No W-No T & FNN &  No W-No T & High & 0.0 \\ \hline
FNN & All & FNN & No T & Low & 0.0 \\ \hline
FNN & All & FNN & No T & Medium & 0.0 \\ \hline
FNN & All & FNN & No T & High & 0.0 \\ \hline
FNN & All & FNN & No W & Low & 0.0 \\ \hline
FNN & All & FNN & No W & Medium & 0.762 \\ \hline
FNN & All & FNN & No W & High & 0.899 \\ \hline
FNN & All & FNN &  No W-No T & Low & 0.398 \\ \hline
FNN & All & FNN &  No W-No T & Medium & 0.999 \\ \hline
FNN & All & FNN &  No W-No T & High & 0.828 \\ \hline
FNN & No T & FNN & No W & Low & 0.0 \\ \hline
FNN & No T & FNN & No W & Medium & 0.0 \\ \hline
FNN & No T & FNN & No W & High & 0.0 \\ \hline
FNN & No T & FNN &  No W-No T & Low & 0.0 \\ \hline
FNN & No T & FNN &  No W-No T & Medium & 0.0 \\ \hline
FNN & No T & FNN &  No W-No T & High & 0.0 \\ \hline
FNN & No W & FNN &  No W-No T & Low & 0.468 \\ \hline
FNN & No W & FNN &  No W-No T & Medium & 0.927 \\ \hline
FNN & No W & FNN &  No W-No T & High & 0.061 \\ \hline
\caption{For each pair of models (with the respective input variables) in the table, we report the fraction of bootstrap samples for which the two modes have equivalent predictive accuracy according to the Diebold-Mariano test. This table is for risk-group-based predictions.}
\label{tab:diebold-risk}\\
\end{longtable}

\subsection{Full set of SMAPEs, RMSEs, and MAEs (post-COVID)}
\label{si:post-covid}
\begin{table}[h]
\centering
\begin{tabular}{|c|c|c|c|c|c|c|c|c|}
\hline
\textbf{Method} & \textbf{\begin{tabular}[c]{@{}c@{}}Input\\ variables\end{tabular}} & \textbf{Shift} & \textbf{\begin{tabular}[c]{@{}c@{}}SMAPE\\ (mean)\end{tabular}} & \textbf{\begin{tabular}[c]{@{}c@{}}SMAPE\\ (st.dev.)\end{tabular}} & \textbf{\begin{tabular}[c]{@{}c@{}}RMSE\\ (mean)\end{tabular}} & \textbf{\begin{tabular}[c]{@{}c@{}}RMSE\\ (st.dev.)\end{tabular}} & \textbf{\begin{tabular}[c]{@{}c@{}}MAE\\ (mean)\end{tabular}} & \textbf{\begin{tabular}[c]{@{}c@{}}MAE\\ (st.dev.)\end{tabular}} \\ \hline

RF & All & Morning & 15.19 & 0.72 & 33.45 & 1.56 & 24.88 & 1.18 \\ \hline
 
RF & All & Afternoon & 15.42 & 0.69 & 23.88 & 1.40 & 17.61 & 0.86 \\ \hline
 
RF & All & Night & 14.76 & 0.75 & 13.35 & 0.66 & 10.28 & 0.45 \\ \hline

RF & No W & Morning & 14.86 & 0.69 & 32.64 & 1.51 & 24.47 & 1.14 \\ \hline
 
RF & No W & Afternoon & 15.82 & 0.70 & 24.21 & 1.40 & 17.99 & 0.87 \\ \hline
 
RF & No W & Night & 15.31 & 0.74 & 13.60 & 0.64 & 10.63 & 0.44 \\ \hline

RF & No T & Morning & 14.96 & 0.71 & 33.07 & 1.52 & 24.54 & 1.16 \\ \hline
 
RF & No T & Afternoon & 16.25 & 0.72 & 24.76 & 1.43 & 18.39 & 0.89 \\ \hline
 
RF & No T & Night & 15.79 & 0.77 & 14.21 & 0.70 & 10.96 & 0.47 \\ \hline

RF & No W-No T & Morning & 15.16 & 0.70 & 33.32 & 1.51 & 24.94 & 1.15 \\ \hline
 
RF & No W-No T & Afternoon & 16.13 & 0.77 & 24.98 & 1.47 & 18.26 & 0.93 \\ \hline
 
RF & No W-No T & Night & 14.70 & 0.74 & 13.14 & 0.73 & 10.32 & 0.44 \\ \hline
\end{tabular}
\caption{Model performance on the post-COVID dataset for shift-based predictions. The model tested is the RF model for all four input combinations.}
\label{tab:post-covid-shift}
\end{table}

\begin{table}[h]
\centering
\begin{tabular}{|c|c|c|c|c|c|c|c|c|}
\hline
\textbf{Method} & \textbf{\begin{tabular}[c]{@{}c@{}}Input\\ variables\end{tabular}} & \textbf{\begin{tabular}[c]{@{}c@{}}Risk\\ group\end{tabular}} & \textbf{\begin{tabular}[c]{@{}c@{}}SMAPE\\ (mean)\end{tabular}} & \textbf{\begin{tabular}[c]{@{}c@{}}SMAPE\\ (st.dev.)\end{tabular}} & \textbf{\begin{tabular}[c]{@{}c@{}}RMSE\\ (mean)\end{tabular}} & \textbf{\begin{tabular}[c]{@{}c@{}}RMSE\\ (st.dev.)\end{tabular}} & \textbf{\begin{tabular}[c]{@{}c@{}}MAE\\ (mean)\end{tabular}} & \textbf{\begin{tabular}[c]{@{}c@{}}MAE\\ (st.dev.)\end{tabular}} \\ \hline

RF & All & Low & 11.63 & 0.48 & 32.60 & 1.39 & 25.12 & 1.10 \\ \hline
 
RF & All & Medium & 15.73 & 0.73 & 18.38 & 0.94 & 13.37 & 0.66 \\ \hline
 
RF & All & High & 20.47 & 0.80 & 13.09 & 0.56 & 10.25 & 0.43 \\ \hline

RF & No W & Low & 12.25 & 0.50 & 33.99 & 1.43 & 26.37 & 1.13 \\ \hline
 
RF & No W & Medium & 15.27 & 0.71 & 17.91 & 0.92 & 13.04 & 0.64 \\ \hline
 
RF & No W & High & 19.85 & 0.78 & 12.74 & 0.53 & 10.00 & 0.42 \\ \hline

RF & No T & Low & 12.61 & 0.50 & 34.37 & 1.34 & 27.00 & 1.12 \\ \hline
 
RF & No T & Medium & 15.54 & 0.72 & 18.09 & 0.91 & 13.20 & 0.65 \\ \hline
 
RF & No T & High & 20.41 & 0.80 & 13.09 & 0.57 & 10.22 & 0.42 \\ \hline

RF & No W-No T & Low & 12.48 & 0.50 & 34.38 & 1.44 & 26.86 & 1.12 \\ \hline
 
RF & No W-No T & Medium & 15.87 & 0.70 & 18.33 & 0.88 & 13.61 & 0.63 \\ \hline
 
RF & No W-No T & High & 19.86 & 0.79 & 12.78 & 0.54 & 9.99 & 0.42 \\ \hline
\end{tabular}
\caption{Model performance on the post-COVID dataset for risk-group-based predictions. The model tested is the RF model for all four input combinations.}
\label{tab:post-covid-risk}
\end{table}

\end{document}